\documentclass[a4paper, 11pt]{article}
\usepackage{jheppub}

\pdfoutput=1

\usepackage[utf8]{inputenc}
\usepackage{epsf, array, xcolor}
\usepackage{graphicx}
\usepackage{bbold}
\usepackage{amsmath, amssymb, mathtools}
\usepackage{epigraph}
\usepackage{textcomp}
\usepackage{color}
\usepackage{slashed}
\usepackage{graphics,psfrag}
\usepackage{graphicx,psfrag}
\usepackage{cancel}
\usepackage{float}
\usepackage{url}
\usepackage{pbox}

\usepackage{etoolbox}
\AtBeginEnvironment{tabular}{\scriptsize}

\title{\boldmath The \texorpdfstring{$e^+ e^- \rightarrow Z H$}{e+ e- -> Z H} Process in the SMEFT Beyond Leading Order}

\author[a]{Konstantin Asteriadis,}
\author[b]{Sally Dawson,}
\author[c]{Pier Paolo Giardino}
\author[b]{and Robert Szafron}

\emailAdd{konstantin.asteriadis@ur.de}
\emailAdd{dawson@bnl.gov}
\emailAdd{pier.giardino@uam.es}
\emailAdd{rszafron@bnl.gov}

\affiliation[a]{Institute for Theoretical Physics, 
    University of Regensburg, 93040 Regensburg, Germany}
\affiliation[b]{High Energy Theory Group, Physics Department, 
    Brookhaven National Laboratory, Upton, NY 11973, USA}
\affiliation[c]{Departamento de Física Teórica and Instituto de Física Teórica UAM/CSIC, 
    Universidad\\ Autónoma de Madrid, Cantoblanco, 28049, Madrid, Spain}

\abstract{We systematically study potential effects of Beyond the Standard Model 
physics in the \texorpdfstring{$e^+ e^- \rightarrow Z H$}{e+ e- -> Z H} process. 
To this end, we include all relevant dimension-6 Standard Model Effective Field 
Theory operators and work to next-to-leading order (NLO) accuracy in the 
electroweak coupling. We consider both polarized and unpolarized electron and 
positron beams and present results for $\sqrt{s}=240, ~365$ and $500~{\rm GeV}$, 
emphasizing contributions where the NLO predictions differ significantly from 
the leading order results. At NLO, a sensitivity arises to operators that 
do not contribute at tree level, such as the Higgs tri-linear coupling, CP-violating operators, and  dimension-6 operators involving the top quark,  among 
many others. We compare the prospects of future $e^+e^-$ colliders to explore 
these new physics effects with existing measurements from the LHC, electron EDMs 
(for CP violating operators), and Z pole measurements.}

\begin{document}

\preprint{IFT-UAM/CSIC-24-131}

\maketitle

%%%%%%%%%%%%%%%%%%%%%%%%%%%%%%%%%%%%%%%%%%%%%%%%%%

\section{Introduction} 
One of the primary goals of the high luminosity run of the LHC and future $e^+e^-$ colliders is to observe Beyond the Standard Model (BSM)  physics through precision measurements. A helpful approach is to parameterize deviations from the Standard Model (SM) using an effective field theory (SMEFT). The SMEFT approach requires calculations beyond the leading order (LO) both in the effective field theory expansion and in the SM limit in order to probe BSM physics accurately. Using the SMEFT allows experimental data from many sources to be included in a systematic framework to search for the presence of new physics effectively. The SMEFT is particularly effective in scenarios where BSM physics involves new particles too heavy to be directly observed at the LHC.

The SMEFT maintains the $\textrm{SU}(3)_C \times \textrm{SU}(2)_L\times \textrm{U}(1)_Y$ symmetry of the unbroken SM. Deviations from the SM are written as a power series in some heavy scale, $\Lambda$,
\begin{align}
    \mathcal{L}=\mathcal{L}_{\rm SM}+\sum_{\substack{\alpha\\d>4}} \frac{C^{(d)}_\alpha O_\alpha^{(d)}}{ \Lambda^{d-4}}\, ,
    \label{eq:lagr}
\end{align}
where all new physics is contained in the coefficients, $C_\alpha^{(d)}$, and the operators of canonical dimension $d$ contain only SM fields.

An $e^+e^-$ collider running at $\sqrt{s}=240~{\rm GeV}$ is ideally suited to probe new interactions in the electroweak sector through the process
$e^+e^-\rightarrow ZH$.   A complete SM NLO electroweak calculation of this process exists~\cite{Fleischer:1980ub,Kniehl:1991hk,Denner:1992bc}, along with partial NNLO results for polarized initial states~\cite{Freitas:2023iyx,Freitas:2022hyp,Chen:2022mre,Gong:2016jys,Sun:2016bel}. 
 Two loop initial state QED effects have been studied in~\cite{Blumlein:2019pqb}.
In this work, we present the complete SMEFT NLO result, including the effects of all dimension-6 operators and retaining the polarization dependence of the initial state electrons. These results can be used in global fits to SMEFT coefficients to understand the sensitivity of $e^+e^-\rightarrow ZH$ (Higgsstrahlung) to new physics and the correlations between different contributions.  An initial study was presented in \cite{Asteriadis:2024qim}.

In Sec.~\ref{sec:onel}, we review the tree level dimension-6 SMEFT Lagrangian along with the tree level amplitudes for the polarized $e^+e^-\rightarrow ZH$ SMEFT  scattering process to set the notation. Next, we describe the NLO calculation, including the virtual contributions, the renormalization procedure, and the treatment of the real contribution in the SMEFT. A detailed discussion of the infra-red (IR) subtraction procedure and the treatment of large corrections of the form $\log({s}/{ m_e^2})$ is included. Phenomenological results are in Sec.~\ref{sec:res}, with the most detailed results relegated to a series of tables and plots in an appendix. Sec.~\ref{sec:con} contains some conclusions and an outlook for future directions.

\section{SMEFT calculation of Higgsstrahlung}
\label{sec:onel}

The SMEFT describes heavy new physics as an expansion around the SM through the Lagrangian of Eq.~\eqref{eq:lagr}, 
where all new physics effects are contained in the coefficients $C_\alpha^{(d)}$.  Since we consistently truncate the expansion at dimension 6, we drop the superscript $d$ for brevity. The coefficients can be complex, as we will consider the effects of CP-violating operators that can contribute at NLO.   For our calculations, we adopt the Warsaw basis~\cite{Grzadkowski:2010es} and the Feynman rules of Ref.~\cite{Dedes:2017zog}.  
Furthermore, the 2-fermion and 4-fermion operators, many of which first become relevant at the loop level, are allowed to have arbitrary flavor structures, although we assume the CKM matrix to be the identity matrix, which restricts the flavor structures that can appear. This flexibility in flavor structure is essential for capturing the full range of possible new physics effects that might arise in the SMEFT framework.

The input parameters  are boson and fermion masses and the Fermi constant, which is determined via muon decay:
\begin{align}
    \label{eq:reps}
    M_W,~M_Z,~M_H,~m_f,~G_\mu \, .
\end{align}
The above inputs replace the original parameters of the SM  dimension $\leq 4$ Lagrangian.  The $SU(2)_L$ and $U(1)_Y$ coupling constants, Higgs vacuum expectation value and self-coupling, and the Yukawa couplings, 
\begin{align}
    \label{eq:lagpar}
    {\overline{g}},~{\overline{g}}',~v_T,~\lambda, y_f \, ,
\end{align}
are all expressed in terms of the parameters of Eq.~\eqref{eq:reps}.\footnote{We neglect here the fermion mixings and neutrino masses.}
For completeness and to establish notation, the SM Lagrangian in Eq.~\eqref{eq:lagr} is 
\begin{align}
  \mathcal{L}_{\text{SM}} = \mathcal{L}_{\text{gauge}}+\mathcal{L}_{\text{fermion}} +\mathcal{L}_{\text{Higgs}}+\mathcal{L}_{\text{Yukawa}} \, ,
\end{align}
with the Lagrangian for each sector given by:
\begin{itemize}
\item Gauge sector 
\begin{align}
    \mathcal{L}_{\text{gauge}} = -\frac{1}{4} G^a_{\mu \nu} G^{a \mu \nu} - \frac{1}{4} W^i_{\mu \nu} W^{i \mu \nu} - \frac{1}{4} B_{\mu \nu} B^{\mu \nu} \, ,
\end{align}
where
\begin{align}
    \begin{split}
        G^a_{\mu \nu} &= \partial_\mu G^a_\nu - \partial_\nu G^a_\mu - g_s f^{abc} G^b_\mu G^c_\nu \, , \\
        W^i_{\mu \nu} &= \partial_\mu W^i_\nu - \partial_\nu W^i_\mu - {\overline{g} }\epsilon^{ijk} W^j_\mu W^k_\nu \, , \\
        B_{\mu \nu} &= \partial_\mu B_\nu - \partial_\nu B_\mu \, ,
    \end{split}
\end{align}
and $f^{abc}$ ($\epsilon^{ijk}$) are $SU(3)$ ($SU(2)$) totally antisymmetric structure constants. 
\item Fermion sector
\begin{align}
    \mathcal{L}_{\text{fermion}} = \sum_{\text{generations}} \left( \bar{Q}_L i \slashed{D} Q_L + \bar{u}_R i \slashed{D} u_R + \bar{d}_R i \slashed{D} d_R + \bar{L}_L i \slashed{D} L_L + \bar{e}_R i \slashed{D} e_R \right) \, ,
\end{align}
where the covariant derivative is:
\begin{align}
    D_\mu = \partial_\mu + i g_s T^a G^a_\mu + i \bar{g} \tau^i W^i_\mu + i \bar{g}^\prime Y B_\mu \, .
\end{align}
Here, \( g_s \), \( \bar g \), and \( \bar{g}^\prime \) are the coupling constants for the strong, weak, and hypercharge interactions, respectively. \( T^a \), \( \tau^i \), and \( Y \) are the generators for the gauge groups \( SU(3)_C \), \( SU(2)_L \), and \( U(1)_Y \).
The left-handed fermion fields are grouped into $SU(2)_L$ doublets 
\begin{align*}
    Q_L = \begin{pmatrix} u_L \\ d_L \end{pmatrix} \, , \quad  Q_L \sim \left(3, 2, \frac{1}{6}\right)\, , \quad
    L_L = \begin{pmatrix} \nu_L \\ e_L \end{pmatrix} \, , \quad  L_L \sim \left(1, 2, -\frac{1}{2}\right) \, , 
\end{align*}
where \( Q_L \) is the left-handed quark doublet composed of the up, $u_L$, and down, $d_L$, quark fields,  and \( L_L \) is the left-handed lepton doublet consisting of neutrinos $\nu_L$, and left-handed $e_L$ charged leptons. The numbers in parentheses denote the representation under \( (SU(3)_C, SU(2)_L, U(1)_Y) \).
The right-handed fermions are $SU(2)_L$ singlets,
\begin{align}
    u_R &\sim \left(3, 1, \frac{2}{3}\right), & d_R &\sim \left(3, 1, -\frac{1}{3}\right), &  e_R &\sim \left(1, 1, -1\right) \, ,
\end{align}
with \( u_R \) and \( d_R \) representing the right-handed up and down quarks, and \( e_R \) are the right-handed charged leptons. All the fermionic fields defined above are understood as vectors in generation space.

\item Higgs sector
\begin{align}
    \mathcal{L}_{\text{Higgs}} = (D^\mu \Phi)^\dagger (D_\mu \Phi) - V(\Phi)\,.
\end{align}
The Higgs potential $V(\Phi)$ is:
\begin{align}
    V(\Phi) = -  \mu^2 \Phi^\dagger \Phi + \frac{\lambda}{2} (\Phi^\dagger \Phi)^2 \, .
\end{align}
\item Yukawa sector
\begin{align}
    \mathcal{L}_{\text{Yukawa}} =  \sum_{\text{generations}}  \left( - y_u \bar{Q}_L \tilde{\Phi} u_R - y_d \bar{Q}_L \Phi d_R - y_e \bar{L}_L \Phi e_R + \text{h.c.} \right) \, ,
\end{align}
where \( \tilde{\Phi} = 2i\tau^2 \Phi^* \) is the conjugate Higgs doublet, and \( y_u \), \( y_d \), and \( y_e \) are the Yukawa coupling matrices, which here are taken to be diagonal for simplicity, and thus we omit the summation indices. In practice, we take only the top quark Yukawa $y_t$ coupling to be non-zero. 
\end{itemize}

The Lagrangian parameters of Eq.~\eqref{eq:lagpar} are derived quantities. 
Up to the corrections due to dimension-8 SMEFT operators, 
\begin{align}
\label{eq:parshift}
\begin{split}
    \lambda &= \sqrt{2}G_\mu{M_H^2}\biggl\{1+X_H+\frac{1}{2\sqrt{2}G_\mu\Lambda^2}(C_{HD}-4C_{H\square})\biggr\}+\frac{3 C_H}{ \sqrt{2}G_\mu\Lambda^2} \, , \\
    y_t &= \sqrt{2}m_t(\sqrt{2}G_\mu)^{1/2}\biggl\{1+\frac 12 X_H+(\sqrt{2}G_\mu)^{-3/2}\frac{C_{u\phi}[3,3]}{2 \sqrt{2}m_t\Lambda^2}\biggr\} \, , \\
    \overline{g} &= 2M_W (\sqrt{2}G_\mu)^{1/2}\biggl\{1+ \frac 12 X_H\biggr\} \, , \\
    \overline{g}' &= 2(\sqrt{2}G_\mu)^{1/2} \sqrt{M_Z^2-M_W^2}\biggl\{1+\frac 12 X_H\biggr\} \\
    &\hspace{10pt}-\frac{1}{2 (\sqrt{2}G_\mu )^{1/2}\Lambda^2}\biggl\{4M_W C_{\phi WB}+\frac{M_Z^2}{\sqrt{M_Z^2-M_W^2}}C_{\phi D}\biggr\} \, ,
\end{split}
\end{align}
where 
\begin{align}
    X_H\equiv \frac{1}{2\sqrt{2}G_\mu\Lambda^2}\biggl\{
    C_{ll}[1,2,2,1]+C_{ll}[2,1,1,2]-2 (C_{\phi l}^{(3)}[1,1]+C_{\phi l}^{(3)}[2,2])\biggr\} \, ,
    \label{eq:xdef}
\end{align}
and the numbers in the square brackets denote the fermion generation. 
In addition, it is convenient to express the electric charge in terms of our input parameters as
\begin{align}
\label{eq:charge}
\begin{split}
     e^2 &= {G_\mu}\sqrt {2}
     \left(1+ X_H  \right) 4 M_W^2 \left( 1- \frac{M_W^2}{M_Z^2}\right)\\
     &-\frac{2M_W^3} {M_Z^2\Lambda^2}\biggl\{M_W C_{\phi D} +{4}\sqrt{M_Z^2-M_W^2}C_{\phi W B}\biggr\} +{\cal{O}}\biggl(\frac{1}{\Lambda^4}\biggr) \, .
\end{split}
\end{align}
We note that $e$ is not an independent parameter of the model ($e$ is defined as the coupling of the electron to the photon in SMEFT).

\begin{figure}
    \centering
    \includegraphics[height=69pt]{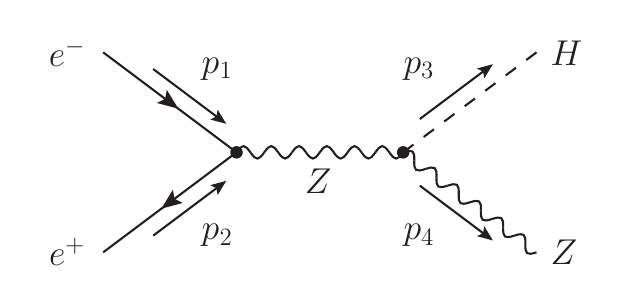}
    \includegraphics[height=69pt]{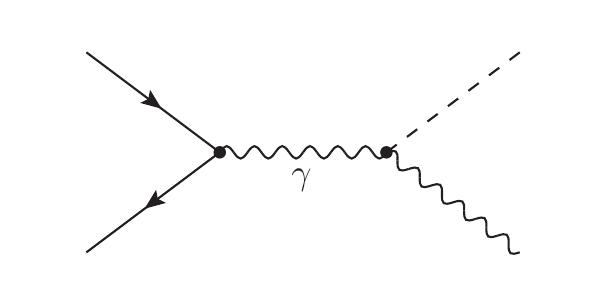}
    \includegraphics[height=69pt]{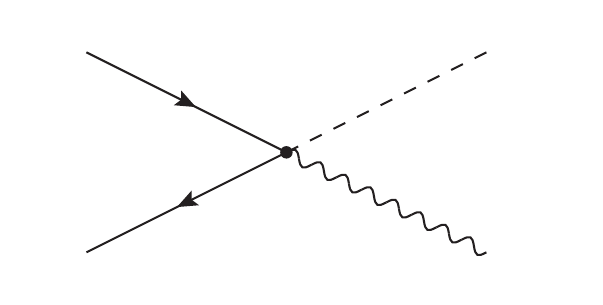}
    \caption{Tree level Feynman diagrams in the dimension-6 SMEFT. The first diagram is the only one present in the SM. It contains momentum conventions that are used throughout the paper. See text for details.} 
    \label{fig:treefd}
\end{figure}

The LO Feynman diagrams contributing to $e^-(p_1)+e^+(p_2)\rightarrow H(p_3)+Z(p_4)$ in the SMEFT are shown in Fig.~\ref{fig:treefd}.  We omit diagrams that lead to amplitudes proportional to the electron Yukawa coupling and only the left-most diagram is present in the SM.  

Tree-level dipole interactions in the SMEFT do not interfere with SM amplitudes at either LO or NLO, as the helicity is conserved in the limit of a massless electron. 
In particular, this means that there is no dependence in our results on $C_{eB}$ or $C_{eW}$. 

Thus, we will only compute vector-like amplitudes, which do not flip helicity on the lepton line. Remaining helicity amplitudes will necessarily lead to contributions to the cross-section that are suppressed by the electron mass. 
We will write the amplitude as
\begin{align}
\begin{split}
    \mathcal{A} &= M_L(s,t) \; {\overline {v}}(p_2)\gamma_\mu \frac{1-\gamma_5}{2}
    u(p_1) \epsilon^\mu(p_4) \\
    &+ M_R(s,t)\; {\overline {v}}(p_2)\gamma_\mu \frac{1+\gamma_5}{2}
    u(p_1) \epsilon^\mu(p_4) + \ldots \, ,
\end{split}
\end{align}
where the dots refer to helicity flipping terms whose contributions to the total cross-section are suppressed by factors of the electron mass.
We use the conventional definition of Mandelstam variables $s=(p_1+p_2)^2$, $t=(p_3-p_1)^2$.
We introduce the double expansion
\begin{align}
    \label{eq:adef}
    \mathcal{A} =\sum_{k}\mathcal{A}^{[k]} = \sum_{k,n}\mathcal{A}^{(k,n)} \, ,
\end{align}
where the first sum over $k$ enumerates the loop expansion, and the second expansion, enumerated by $n$, is performed in the inverse powers of $\Lambda$: $\Lambda^{-n}$. For example, the LO amplitude in SMEFT will be denoted by $\mathcal{A}^{[0]}$. At the same time, the tree-level dimension- 6 SMEFT contribution is $\mathcal{A}^{(0,2)}$.  We use this notation for both amplitudes and cross-sections. 

The LO SM amplitudes are
\begin{align}
    \begin{split}
        M_L^{(0,0)} &= \frac{2\sqrt{2}G_F M_Z}{(s-M_Z^2)}(2M_W^2 - M_Z^2) \, , \\
        M_R^{(0,0)} &= -\frac{4 \sqrt{2} G_F M_Z (M_Z^2-M_W^2)}{(s-M_Z^2)}\, .
    \end{split}
\end{align}
The tree level SMEFT result for Higgsstrahlung is computed using the FeynRules~\cite{Alloul:2013bka} $\rightarrow$ FeynCalc~\cite{Shtabovenko:2020gxv} tool-chain. 
We neglect all Yukawa couplings except for that of the top quark.

The NLO calculation of the $e^+e^-\rightarrow ZH$ inclusive cross-section has three ingredients: one-loop virtual contributions, the counter-terms needed for renormalization of the UV divergences, and the real photon correction, which is required to obtain an  IR finite result according to the celebrated Kinoshita-Lee-Nauenberg (KLM) theorem \cite{Kinoshita:1962ur, Lee:1964is}. 
Our complete SMEFT result reduces to the NLO SM result in the limit $\Lambda\rightarrow \infty$
and we find perfect agreement with the well-known SM one-loop result \cite{Fleischer:1982af, Kniehl:1991hk, ATLAS:2023mqy, Bondarenko:2018sgg, Greco:2017fkb}.

\subsection{Virtual Contributions}

\begin{figure}
    \centering
    \includegraphics[height=69pt]{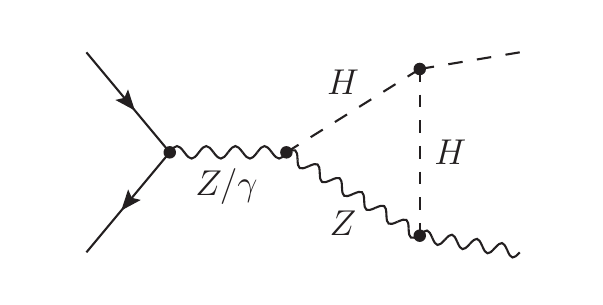}
    \includegraphics[height=69pt]{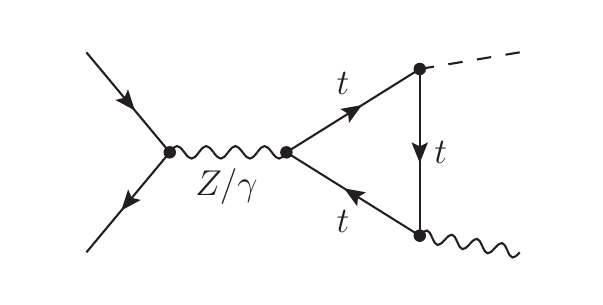}
    \includegraphics[height=69pt]{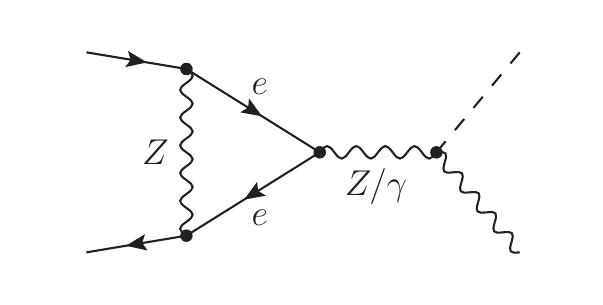}\\[8pt]
    \includegraphics[height=69pt]{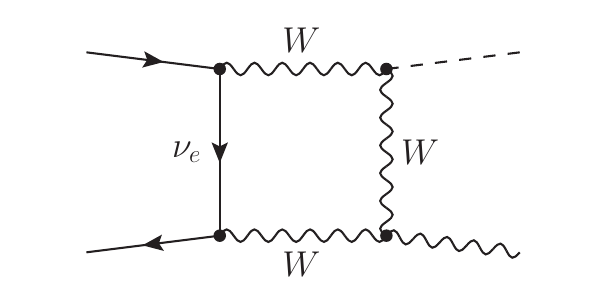}
    \includegraphics[height=69pt]{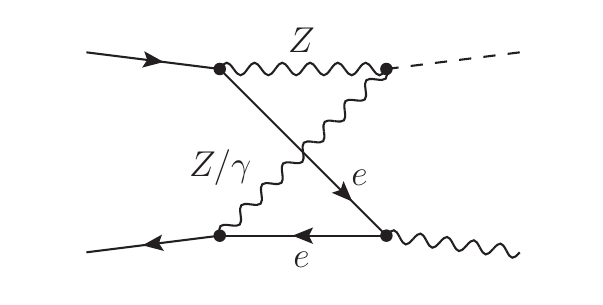}
    \includegraphics[height=69pt]{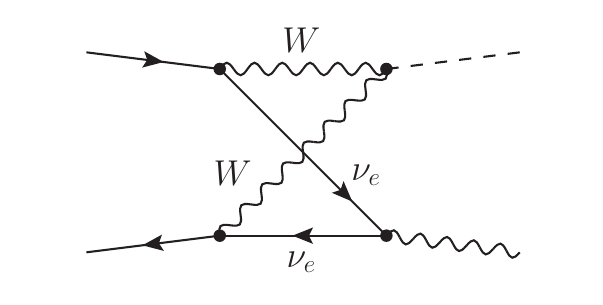}\\[8pt]
    \includegraphics[height=69pt]{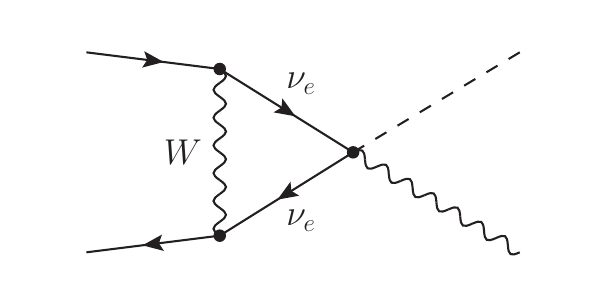}
    \includegraphics[height=69pt]{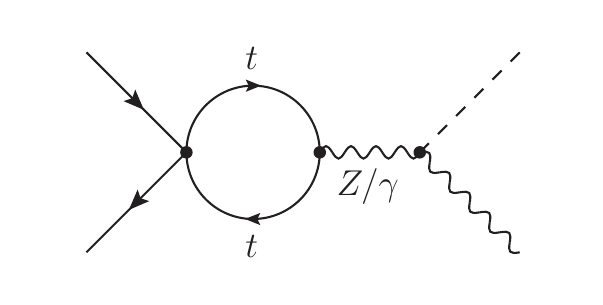}
    \includegraphics[height=69pt]{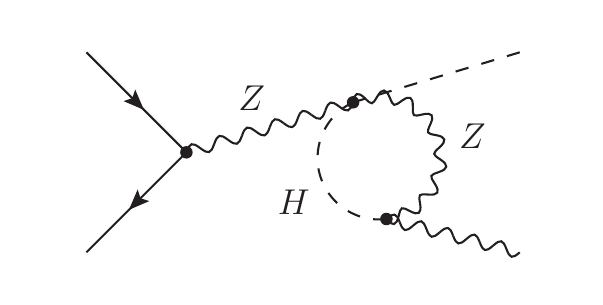}
    \caption{Sample virtual diagrams contributing to $e^+e^-\rightarrow HZ$ at NLO in the dimension-6 SMEFT. We show contributions with one-loop vertex corrections (first row), planar and non-planar box diagrams (second row), and contributions with 4-point interactions (third row).}
    \label{fig:1loop}
\end{figure}

One-loop virtual amplitudes are computed in dimensional regularization using the FeynRules~\cite{Alloul:2013bka} $\rightarrow$ FeynCalc~\cite{Shtabovenko:2020gxv} $\rightarrow$ Package-X \cite{Patel:2016fam} and LoopTools~\cite{Hahn:1998yk} tool-chain.
We neglect all fermion masses except that of the top quark in the virtual diagrams and UV renormalization terms. The electron mass will enter into our collinear subtraction procedure where we must account for the $\log(s/m_e^2)$ enhanced contributions in Sec.~\ref{sec:merestoration}.  Some sample virtual diagrams are shown in Fig.~\ref{fig:1loop}. We include all dimension-6 SMEFT operators that contribute, along with arbitrary flavor structures.  We work to linear order in ${1/\Lambda^2}$, which corresponds to keeping terms up $\sigma^{(1,2)}$ in the notation of Eq.~\eqref{eq:adef} .
Since the SMEFT is renormalizable order by order in the ${1/ \Lambda^2}$ expansion, we drop all higher-order terms of ${\cal{O}}({1/ \Lambda^4})$, {\it i.e.}, we only allow for a single insertion of a dimension-6 operator in the one-loop diagrams and perform a calculation including all terms of ${\cal{O}}({1/(16 \pi^2 \Lambda^2}))$. Double insertions of dimension-6 terms would require counter-terms proportional to dimension-8 operators, which we do not include.

Because the SMEFT vertices have up to 3 gamma matrices, treating $\gamma_5$ as anti-commuting is insufficient.  
As is well known, $\gamma_5$ is a  4-dimensional object and is not
well defined in $D=4-2\epsilon$ dimensions, where it is generally necessary to introduce a scheme to
perform traces involving $\gamma_5$ matrices consistently.
We implement the Kreimer scheme~\cite{Kreimer:1989ke,Korner:1991sx,Kreimer:1993bh}  for dealing with the complications that arise from the axial anomaly and the treatment of $\gamma_5$ in dimensional regularization.
The Kreimer scheme modifies the standard trace operations by introducing a specific reading point prescription for the non-cyclic trace to avoid ambiguities when dealing with $\gamma_5$. 

\subsubsection{UV Renormalization}
\label{sec:UVren}

We use a hybrid variation of the on-shell (OS) scheme for the UV renormalization of SMEFT. The SM couplings, masses, and Higgs vacuum expectation value are defined according to the %version of 
OS scheme~\cite{Jegerlehner:1990uiq},  where the input parameters are boson and fermion masses, and the Fermi constant is determined via muon decay. 
 
The tadpole renormalization is chosen to cancel the tadpole graphs completely~\cite{PhysRevD.23.2001}. Consequently, the renormalized vacuum expectation value of the Higgs field is the minimum of the renormalized scalar potential at each order in perturbation theory. This scheme  gives the one-loop relationship between the Higgs vacuum expectation value, $v_T$, and $G_\mu$ at dimension- 6,
\begin{align}
\label{eq:gmu:smeft}
\begin{split}
    \sqrt{2}G_\mu (1+X_H)&=\frac{1}{v_T^2}(1+\Delta r)\, ,
\end{split}
\end{align}
with $v_T$ representing the minimum of the potential. The quantity $\Delta r$~\cite{Sirlin:1980nh, Marciano:1980pb} is obtained from the one-loop corrections to muon decay, and an analytic form for both the SM and the SMEFT dimension-6 contributions can be found in Appendix C of~\cite{Dawson:2018pyl}.

In our computations, we use the complex mass scheme \cite{Denner:1999gp,Denner:2005fg}.
The renormalization condition, valid up to one-loop,\footnote{Beyond one loop, the $Z$ boson renormalization conditions differ from those of the $W$, due to $\gamma-Z$ mixing.} for the vector boson masses, for $V=W,Z$, relate the bare mass $M_{0,V}^2$ to the physical mass $M_V^2$:
\begin{align}
    M_V^2=M_{0,V}^2-\Pi_{VV}(M_V^2)\, ,
\end{align}
where $\Pi_{VV}(M_V^2)$ are the transverse parts of the one-loop corrections to the 2-point function evaluated at the physical mass $M_V^2$.
We can also introduce mass counter-terms $\delta M_V^2 \equiv M_{0,V}^2-M_{V}^2$, and 
\begin{align}
  \delta M_V^2   =\Pi_{VV}(M_V^2).
\end{align}
Similarly, for the Higgs boson, we have 
\begin{align}
    M_H^2=M_{0,H}^2-\Pi_{HH}(M_H^2)\,, \quad   \delta M_H^2   =\Pi_{HH}(M_H^2) \, ,
\end{align}
where $\Pi_{HH}(M_H^2)$ is the one-loop correction to the Higgs 2-point function evaluated at the physical Higgs mass. 

The wave-function renormalization NLO counter-terms are 
\begin{align}
    \delta  Z_{AA} =  - \left.  \frac{d \Pi_{AA}}{d k^2} \right|_{k^2 = 0} \, , \quad 
    \delta  Z_{VV} = - \left.  \frac{d \Pi_{VV}}{d k^2} \right|_{k^2 = M_V^2} \, , \quad 
    \delta  Z_H = - \left. \frac{d \Pi_{HH}}{d k^2} \right|_{k^2 = M_H^2} \, . 
\end{align}
The off-diagonal $\gamma-Z$ mixing term must vanish at both the  photon and Z boson poles
\begin{align}
   \Pi_{A Z}(0) =  \Pi_{A Z}(M_Z^2) = 0 \, ,
\end{align}
which introduces additional off-diagonal wave-function renormalization constants
\begin{align}
    \delta Z_{AZ} = -2 \frac{\Pi_{AZ}(M_Z^2)}{M_Z^2} \, , \quad  
    \delta Z_{ZA} = 2  \frac{\Pi_{AZ}(0)}{M_Z^2} \, .
\end{align}

The bare dimension-6 SMEFT coefficients, $C_{0,i}$ are renormalized at one-loop level in the ${\overline{\rm MS}}$ scheme, 
\begin{align}
    C_i(\mu) = C_{0,i}-\frac{1}{2}\left[\frac{1}{\epsilon}\bar \mu^{2\epsilon} \right]  \frac{1}{16\pi^2}\gamma_{ij}C_j(\mu) \, ,
\end{align}
where we introduce the UV renormalization scale $\bar \mu^2  =  \mu^2 \, { e^{\gamma_E}}/{(4 \pi)}$ in the  $\overline{\rm MS}$ scheme, and $\gamma_{ij}$
are the elements of the anomalous dimension matrix~\cite{Jenkins:2013zja,Jenkins:2013wua,Alonso:2013hga,Alonso:2014zka}. Consequently, the renormalized matching coefficients obey the renormalization group equation (RGE), 
\begin{align}
    16\pi^2 \, \frac{\textrm{d}C_i (\mu)}{\textrm{d}\ln\mu} = \gamma_{ij}C_j(\mu) \, ,
\end{align}
with the initial conditions determined through the matching at the new physics scale $ \mu\sim \Lambda$.  
The RGE allows us to resum the leading logarithmic corrections originating due to the large-scale separation of the new physics scale $\Lambda$ and the electroweak scale. 

\subsubsection{IR Subtraction}

The poles corresponding to UV divergences cancel after renormalization. However, the virtual amplitude $\mathcal{A}_{\rm V} $ still contains infrared (IR) poles, which have a universal form~\cite{Catani:1998bh, Gardi:2009qi, Becher:2009cu}. In QED, with massless electrons, the IR singularities of the UV finite virtual amplitude  are governed by   a multiplicative renormalization factor, where the form relevant to a process with electron-positron annihilation producing a neutral final state is
\begin{align}
    \label{eq:Z}
    A_{\rm V}(\mu) =  Z^{-1}(\epsilon,\mu) \mathcal{A}_{\rm V}(\epsilon,0) \, , 
\end{align}
where $ A_{V}(\mu)$ is the IR-subtracted virtual amplitude (usually referred to as the finite remainder), $\mathcal{A}_{\rm V}(\epsilon,0)$ is the dimensional regularized virtual amplitude which contains IR poles and  $Z^{-1}(\epsilon,\mu)$ is the universal renormalization factor. Up to one loop, we have
\begin{align}
    \label{eq:ZSCET}
    Z(\epsilon,\mu) = 1 + \frac{\alpha}{4\pi} \left(-\frac{\gamma_0^{\rm cusp}}{2\epsilon^2}-\frac{1}{2\epsilon}\gamma_0^{\rm cusp} \ln \frac{\bar \mu^2}{s}+ \frac{1}{\epsilon}\gamma_0^q  \right) + \mathcal{O}(\alpha^2) \, ,
\end{align}
with the cusp anomalous dimension $\gamma_0^{\rm cusp} =4$ and $\gamma_0^q = -3$. 
The fine structure constant is $\alpha = {e^2}/{4\pi}$,  with $e$ defined in Eq.~\eqref{eq:charge}.
Eq.~\eqref{eq:ZSCET} can be most easily proven in the framework of soft-collinear effective field theory (SCET) through matching of the SM amplitudes onto N-jet operators containing only massless degrees of freedom. Since the higher dimensional SMEFT operators trivially match onto power-suppressed SCET interactions,  Eq.~\eqref{eq:ZSCET} remains valid to all powers in the $\Lambda$ expansion. 
The explicit form of the one-loop finite remainder of the virtual amplitude can be obtained from Eq.~(\ref{eq:Z}) after expanding it in $\alpha$:
\begin{align}
    A^{[1]}_{\rm V}(\mu) =  \mathcal{A}^{[1]}_{\rm V}(\epsilon,0)- \delta Z(\epsilon, \mu)  \mathcal{A}^{[0]} \, , 
\end{align}
where as usual $\delta Z(\epsilon,\mu) \equiv Z(\epsilon,\mu) -1 $.

The total virtual contribution to the cross-section in SMEFT is,\footnote{We implicitly sum over final state polarizations and average over initial state polarizations using a normalized density matrix.}
\begin{align}
\label{eq:virtual}
    \begin{split}
        \int \textrm{d} \sigma^{(1,2)}_{\rm V}(s;\epsilon,0) &= \frac{1}{2 s} \int \textrm{d}{\rm PS}_2 \, {\rm Re} \biggl(2 \mathcal{A}^{(0,0)*} \mathcal{A}^{(1,2)}_{\rm V}(\epsilon,0) + 2 \mathcal{A}^{(0,2)*} \mathcal{A}^{(1,0)}_{\rm V}(\epsilon,0)\biggr)  \\
        &=\frac{1}{2 s} \int \textrm{d}{\rm PS}_2 \, {\rm Re} \biggl(2 \mathcal{A}^{(0,0)*} A_{\rm V}^{(1,2)}(\mu)+2 \mathcal{A}^{(0,2)*} A_{\rm V}^{(1,0)}(\mu)\biggr)\\
        &\hspace{10pt}+2 \delta Z(\epsilon,\mu) \int \textrm{d}\sigma^{(0,2)}(s) \, , 
    \end{split}
\end{align}
where $\textrm{d}{\rm PS}_n $ is the n-body phase space measure. 

\subsection{Real Contributions}

\begin{figure}
    \centering
    \includegraphics[height=69pt]{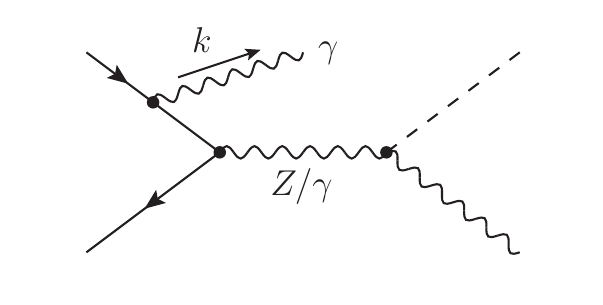}
    \includegraphics[height=69pt]{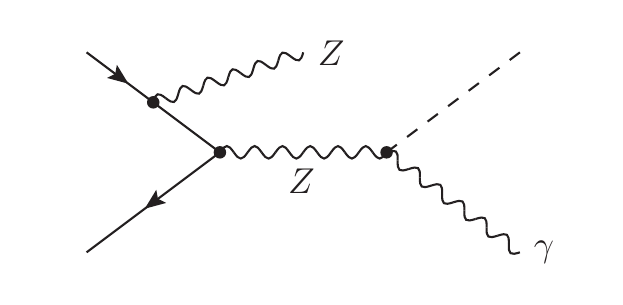}
    \includegraphics[height=69pt]{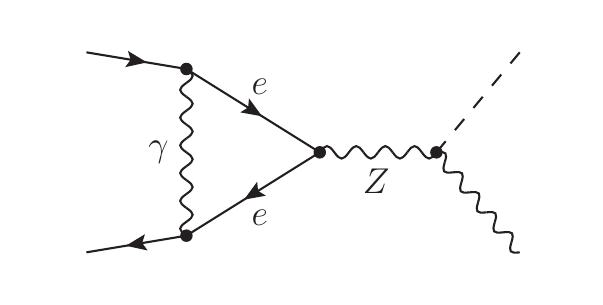}\\[-5pt]
    (a) \hspace{120pt} (b) \hspace{120pt} (c)
    \caption{NLO QED contributions to dimension-6 in the SMEFT. Diagrams~(a) and (b) are examples of the real corrections. In diagram~(a), additional momentum conventions are shown that are used throughout the paper, c.f. Fig.~\ref{fig:treefd} for more details. Diagram~(c) shows a QED vertex correction.}
    \label{fig:QED}
\end{figure}

Canceling IR divergences requires including the real photon emission contributions, $e^+e^-\rightarrow ZH\gamma$. We also need to restore the leading dependence on the electron mass, which is necessary to screen the collinear divergences in the initial state. In this section, we first describe the method we adopt to compute the real photon emission contribution, including the subtraction technique for IR divergences and then the restoration of the logarithmic dependence on the electron mass, which exploits the usual collinear factorization~\cite{Catani:2000ef,Penin:2005kf,Mitov:2006xs,Becher:2009kw,Ferroglia:2009ii}.

First, we compute the real contribution in dimensional regularization (DR) with a massless electron, which we label as $\textrm{d}\sigma_{\rm R}$.
The real contribution to the Higgsstrahlung cross-section can be written as,
\begin{align}
    \int \textrm{d}\sigma_{\rm R}(s;\epsilon,0)=\frac{1}{2 s}\int \textrm{d}{\rm PS}_3 \biggl( \mid A_{\rm R}\mid^2-\mid A_{\rm sub}\mid^2\biggr)+\int \textrm{d}\sigma_{\rm sub}(s;\epsilon,0) \, .
    \label{eq:realdr}
\end{align}
Here $A_{\rm R}$ is the tree-level amplitude for $e^-(p_1)+e^+(p_2)\rightarrow H(p_3)+Z(p_4)+\gamma(k)$ computed in SMEFT, see first and second diagrams in Fig.~\ref{fig:QED}. 
To isolate collinear and soft divergences, we use dipole subtraction~\cite{Catani:1996vz}, which is defined in terms of the square of the eikonal current,
\begin{align}
    | A_{\rm sub}|^2\ = 2 e^2 \frac{p_1 \cdot p_2}{p_1 \cdot k \; p_2 \cdot k} | \mathcal{A}^{[0]}|^2 \, ,
\end{align}
where the factor of two reflects that photons can be radiated from either of the initial legs, and we suppress the superscripts referring to the polarization of the LO amplitudes.  
The amplitudes of Eq.~\eqref{eq:realdr} are computed in dimensional regularization (DR) with massless electrons.

The integrated subtraction term contains
the $\epsilon$-poles from the collinear and soft emission
\begin{align} 
    \begin{split}
        \int \textrm{d}\sigma_{\rm sub}(s;\epsilon,0)&= \frac{1}{2 s}\int \textrm{d}{\rm PS}_3 \, | A_{\rm sub}|^2 \\
        &=\frac{\alpha}{2\pi}\int_0^1 \textrm{d}x\, \textrm{d}\sigma^{[0]}(xs) \biggl\{ G(s;\epsilon,0)\delta(1-x)+\left[{\cal{G}}(s,x; \epsilon,0)\right]_+\biggr\} \, ,
    \end{split}
\end{align}
and $\sigma^{[0]}(s)$ is the leading order SMEFT cross-section evaluated as a function of  $s=(p_1+p_2)^2$.
The integrated subtraction terms for the case of a massless electron are well known (see for example~\cite{Denner:2019vbn})
\begin{align}
\label{eq:gdefnew}
    \begin{split}
    G(s;\epsilon,0) 
    &=2\Gamma(1+\epsilon) \biggl(\frac{4\pi\mu^2}{s}\biggr)^\epsilon \biggl(\frac{1}{\epsilon^2}+\frac{3}{2\epsilon}\biggr)-\frac{2\pi^2}{3}+8 \\
    {\cal{G}}(s,x,\epsilon,0)
    &= 
    \biggl\{
    2{{P}}_{ee}(x)\biggl[ -\frac{1}{\epsilon}\frac{1}{\Gamma(1-\epsilon)}\left(\frac{4\pi\mu^2}{s}\right)^{\epsilon}  +2 \ln(1-x)\biggr] +2(1-x)\biggr\} \, ,
    \end{split}
\end{align}
where ${{P}}_{ee}(x) ={(1+x^2)}/{(1-x)}$ is the unregulated electron splitting function. 

Adding the virtual  (\ref{eq:virtual}) and real (\ref{eq:realdr}) contributions in DR (still with a massless electron),
\begin{align}
    \begin{split}
    \textrm{d}\sigma^{\rm NLO}(s;\epsilon,0)&=\textrm{d}\sigma^{[0]}(s)+\textrm{d}\sigma^{[1]}(s)\\
    &=\textrm{d}\sigma^{[0]}(s)+\textrm{d}\sigma_{\rm V}(s;\epsilon,0)+\textrm{d}\sigma_{\rm R}(s;\epsilon,0) \, , 
    \end{split}
    \label{eq:nlodr}
\end{align}
we obtain the complete NLO cross-section. 
Eq.~\eqref{eq:nlodr} is free of soft divergences. Still, it is not yet the physical answer, as it contains a singular term proportional to $\epsilon^{-1} P_{ee}$, that stems from the collinear divergence due to the massless electron. 

\subsubsection{Including the electron mass}
\label{sec:merestoration}

We now want to include the terms of ${\cal{O}}(\ln({s/m_e^2}))$, which will regulate the remaining collinear singularities.  
The DR  subtraction terms for a massive electron are given by\footnote{
 Note that the overall factors of 2 in Eqs.~(\ref{eq:gdefnew}) and~(\ref{eq:drsub}) reflect the fact that photons can be radiated from either of the two initial electron legs.} \cite{Dittmaier:1999mb},
\begin{align}
    \label{eq:drsub}
    G(s;\epsilon,m_e) &=  \mathcal{L}(s;\epsilon, m_e) -\frac{2\pi^2}{3}+4 \, , \nonumber \\
    \mathcal{L}(s;\epsilon,m_e) &= \frac{2\ln\left(\frac{m^2_e}{s}\right)+ 2}{\epsilon}
    + 2\left[\ln\left(\frac{m^2_e}{s}\right)+ 1\right] \ln \left(\frac{\bar{\mu}^2}{s} \right) - \ln^2 \left( \frac{m^2_e}{s}\right)+ \ln \left( \frac{m^2_e}{s}\right) \, , \nonumber\\
    \mathcal{G}(s,x;\epsilon,m_e) &= P_{ee}(x) \left[2\ln \left(\frac{s}{m_e^2}\right)-2\right] +2(1 -x) \, .
\end{align}
This subtraction scheme does not capture mass suppressed power corrections of ${\cal{O}}({m_e^2}/{s})$.

The differences between the subtraction schemes with and without a finite electron mass are,
\begin{align}
    \begin{split}
    \Delta \mathcal{G}(x;\epsilon,m_e) &\equiv \mathcal{G}(s,x;\epsilon,m_e)-\mathcal{G}(s,x;\epsilon,0) =2 P_{ee}(x) \left[\frac{1}{\epsilon}- \ln \left(\frac{m_e^2(1-x)^2}{\bar{\mu}^2}\right)-1\right] \, , \\ 
    \Delta G(\epsilon,m_e) &\equiv G(s;\epsilon,m_e)-G(s;\epsilon,0) \\
    &=-\frac{2}{\epsilon^2} - \frac{2 \ln \left(\frac{\bar{\mu}^2}{m_e^2}\right)+1}{\epsilon} - \ln^2 \left(\frac{\bar{\mu}^2}{m_e^2}\right)- \ln \left(\frac{\bar{\mu}^2}{m_e^2}\right)- 4 - \zeta_2 + \mathcal{O}(\epsilon) \,.
    \end{split}
    \label{eq:scheme_change}
\end{align}
Consequently, the real radiation contribution, including the leading power logarithmic corrections due to the electron mass, takes the following form 
\begin{align}
    \label{eq:realme}
    \begin{split}
        \int \textrm{d}\sigma_{\rm R}(s;\epsilon,m_e) &= \int \textrm{d}\sigma_{\rm R}(s;\epsilon,0) + \frac{\alpha}{2\pi} \int_0^1 \textrm{d}x\, \textrm{d}\sigma^{[0]}(xs) \biggl\{  \Delta G(\epsilon,m_e)\delta(1-x)\\
        &+ \left[{\Delta\cal{G}} (x;\epsilon,m_e)\right]_+\biggr\}+{\cal{O}}\left(\frac{m_e^2}{s}\right) \, .  
    \end{split}
\end{align}

The virtual contribution was calculated with a massless electron, and we can restore the logarithmically enhanced contributions from a massive electron using the massification \cite{Penin:2005kf, Penin:2005eh} jet function for the virtual amplitude $A_{\rm V}$~\cite{Mitov:2006xs}. We note that for a massive electron, Eq.~(\ref{eq:Z})  takes  the form 
\begin{align}
\label{eq:Zm}
 A_{\rm V}(\mu) 
 =  Z^{-1}(\epsilon,\mu,m_e) \mathcal{A}_{\rm V}(\epsilon,m_e) \, .
\end{align}
Thus, we can obtain the virtual amplitude with collinear singularities regularized by the electron mass as
\begin{align}
   \mathcal{A}_{\rm V}(\epsilon,m_e) =  Z_{q}(\epsilon,m_e) \mathcal{A}_{\rm V}(\epsilon,0) \, ,
\end{align}
where 
\begin{align}
   Z_{q}(\epsilon,m_e) \equiv Z(\epsilon,\mu,m_e)Z^{-1}(\epsilon,\mu) = 1 + \frac{\alpha}{4\pi }Z^{[1]}_{q}(\epsilon,m_e) + \mathcal{O}(\alpha^2) \, ,
\end{align}
and 
\begin{align}
Z_{q}^{\rm [1]}(\epsilon,m_e) = 
\frac{2}{\epsilon^2} + \frac{2 \ln \left(\frac{\bar{\mu}^2}{m_e^2}\right)+1}{\epsilon} + \ln^2 \left(\frac{\bar{\mu}^2}{m_e^2}\right)+ 
 \ln \left(\frac{\bar{\mu}^2}{m_e^2}\right)+ 4 + \zeta_2
 + \mathcal{O}(\epsilon) \, .
\end{align}

The leading corrections due to the electron mass at one loop are incorporated into the virtual amplitude by
\begin{align}
\mathcal{ A}^{[1]}_{\rm V}(\epsilon,m_e) = \mathcal{A}^{[1]}_{\rm V}(\epsilon,0) + \frac{\alpha}{4\pi } Z_{q}^{[1]}(\epsilon,m_e)\mathcal{A}^{[0]} \, ,
\end{align}
which implies
\begin{align}
    \label{eq:Vme}
    \textrm{d} \sigma_{\rm V} (s;\epsilon,m_e)  =   \textrm{d} \sigma_{\rm V} (s;\epsilon,0)  +  \frac{\alpha}{2\pi } Z_{q}^{[1]}(\epsilon,m_e) \textrm{d} \sigma^{[0]} (s) \, .
\end{align}
Combining virtual and real contributions, the dimensionally regularized massless result (\ref{eq:nlodr}) can be converted  to the  physical massive 
result using (\ref{eq:Vme}) and (\ref{eq:scheme_change}). The final result relating the computations with the massless and massive electrons is
\begin{align}
     \textrm{d}\sigma_{{\rm NLO}}(s;m_e) = \textrm{d}\sigma_{\rm NLO}(s;\epsilon,0)
    +\delta(s;\epsilon,m_e)\, ,
    \label{eq:deldef}
\end{align}
where the one-loop scheme conversion correction is 
\begin{align}
    \begin{split}
    \delta(s;\epsilon,m_e) &=
    {\frac{\alpha}{2\pi}} \int_0^1 \textrm{d}x\, \textrm{d}\sigma^{[0]}(xs) \biggl\{ \left( Z_{q}^{[1]}(\epsilon,m_e)+ \Delta G(\epsilon,m_e)\right)\delta(1-x)\\
    &\hspace{110pt}+\, \left[{\Delta\cal{G}} (x;\epsilon,m_e)\right]_+\biggr\}+{\cal{O}}\left(\frac{m_e^2}{s}\right) \\
    &= {\frac{\alpha}{2 \pi}}\int_0^1 \textrm{d}x\, \textrm{d}\sigma^{[0]}(xs)
    \left[{\Delta\cal{G}}(x;\epsilon,m_e)\right]_+\biggr\}+{\cal{O}}\left(\frac{m_e^2}{s}\right) \, .
    \end{split}
    \label{eq:anstot}
\end{align}

\subsubsection{Collinear factorization of \texorpdfstring{$\ln(s/m_e^2)$}{ln(s/me\textasciicircum 2)} terms}

We can interpret our results using an electron structure function (or electron parton distribution function - PDF) $\Gamma_e(x_1;\mu_F,m_e)$~\cite{Kuraev:1985hb, Cacciari:1992pz, Frixione:2019lga, Bertone:2019hks}. 
Using collinear factorization, we write the physical cross-section as a convolution of electron PDFs with an IR subtracted "partonic" cross-section\footnote{Partonic in this context refers to a massless electron, which can be treated as a parton inside the physical massive electron.} 
\begin{align}
    \textrm{d}\sigma_{\rm NLO}(s;m_e)=\int \textrm{d}x_1 \textrm{d}x_2 \Gamma_e(x_1;\mu_F,m_e)\Gamma_e(x_2;\mu_F,m_e) \textrm{d}\sigma_{\rm NLO}(x_1x_2 s;\mu_F) \, .
    \label{eq:dsigmaPDF}
\end{align}
Here, the IR subtracted partonic cross-section defined at the subtraction scale $\mu_F$ is 
\begin{align}
    \textrm{d}\sigma_{\rm NLO}(s, \mu_F)=\int \textrm{d}x_1 \textrm{d}x_2 \Gamma_e(x_1,\mu_F,\epsilon)
    \Gamma_e(x_2,\mu_F,\epsilon) \textrm{d}\sigma_{\rm NLO}(x_1x_2 s,\epsilon,0) \, , 
\end{align}
where $\textrm{d}\sigma_{\rm NLO}( s;\epsilon,0)$ was defined in~\eqref{eq:nlodr} and the $\overline{\rm MS}$ subtraction terms are
\begin{align}
    \Gamma_e(x;\mu_F,\epsilon)
    = \delta(1-x)
    +{\frac{\alpha}{4 \pi}}
    \left(\frac{s}{\overline{\mu}_F^2}\right)^\epsilon
    \frac{1}{\epsilon} P_{ee}(x)\, .
\end{align}
The $\Gamma_e(x;\mu,m_e)$ functions in~(\ref{eq:dsigmaPDF}) correspond to electron PDFs and can be calculated perturbatively in QED.   In the ${\overline{\rm MS}}$ scheme, they are given by~\cite{Frixione:2019lga}
\begin{align}
    \Gamma_e(x;\mu_F,m_e) = \delta(1-x)+{\frac{\alpha}{2 \pi}} \left[\frac{1+x^2}{1-x}\ln \frac{\overline{\mu}_F^2}{m_e^2 (1-x)^2}-1\right]_+ \, .
\end{align}
The physical cross-section $\textrm{d}\sigma_{\rm NLO}(s;m_e)$ is independent of the factorization scale. Apart from the simplicity of the computation, the approach employing factorization allows for straightforward resummation of logarithmically enhanced QED corrections. 
We note that expressing the result in terms of electron PDFs is merely a different interpretation of Eq.~\eqref{eq:deldef} to this order in $\alpha$, which may be more intuitive to some readers.

\subsubsection{QED corrections}
\label{sec:qeddef}

For convenience, we separate the total NLO correction into the purely QED part and the weak part
\begin{align}
    \textrm{d}\sigma_{\rm NLO}(s;m_e) =  \textrm{d}\sigma^{\rm QED}(s;m_e) + \textrm{d}\sigma^{\rm W}(s) \, ,
\end{align}
where the $ \textrm{d} \sigma^{\rm QED}(s;m_e)$ part contains all real emission contributions and the virtual diagrams which involve photon exchange and exhibit IR divergences, see Fig.~\ref{fig:QED}. In the SM, only electron wave-function renormalization and vertex corrections are included here. 
Virtual corrections which contain an internal photon exchange but are IR finite due to the additional suppression of SMEFT vertices,  as well as all the other virtual diagrams, contribute to $\textrm{d} \sigma^{\rm W}(s)$, which has strictly $2\to 2$ kinematics. 
We also note that the weak part does not exhibit the logarithmic dependence on the electron mass at leading power in the electron mass expansion.

\section{Results}
\label{sec:res}

This section delves into the possible implications of NLO SMEFT contributions on the measurements of the Higgsstrahlung process. To accomplish this, we utilize the experimental inputs previously employed in the SM calculation~\cite{Freitas:2023iyx}. We provide a summary of these parameters below:
\begin{align}
    G_\mu&=1.16638\times 10^{-5}~{\rm GeV}^{-2} \, , & M_H&=125.1~{\rm GeV} \, , \nonumber \\ 
    M_W^{\exp}&=80.379~{\rm GeV} \, , & m_t&= 172.76~{\rm GeV}\, , \\
    M_Z^{\exp}&= 91.1876~{\rm GeV} \, , & m_e&= 0.511~{\rm MeV}\, . \nonumber 
\end{align}
All other fermion masses are set to zero, and the electron mass is only retained in the logarithmic corrections. The vector boson masses used in the computation, $M_W$ and $M_Z$, are defined following Refs.~\cite{Freitas:2023iyx, Bardin:1988xt} to improve the treatment of finite width effects in calculations where the gauge boson decays are included,
\begin{align}
\label{eq:MZMZMZ}
    \begin{split}
    M_Z&=\frac{M_Z^{\exp}}{\sqrt{1+\left({\Gamma_Z^{\exp}}/{M_Z^{\exp}}\right)^2}} = 91.1535~{\rm GeV} \, , \\
    M_W&=\frac{M_W^{\exp}}{\sqrt{1+\left({\Gamma_W^{\exp}}/{M_W^{\exp}}\right)^2}} = 80.352~{\rm GeV}\, .
    \end{split}
\end{align}

At the tree level, the differential Higgsstrahlung cross-section depends on the 7 SMEFT coefficients,
\begin{align}
C_{\phi D}\, ,~C_{\phi \square}\, ,~
C_{\phi WB}\, ,~C_{\phi W}\, ,~
C_{\phi B}\, ,~C_{\phi e}[1,1]\, ,~C^+_{\phi l}[1,1] \equiv C_{\phi l}^{(1)}[1,1]+C_{\phi l}^{3)}[1,1]\, .
\end{align}
At the NLO level, significantly more operators play a role in the process. In particular, the 4-fermion top quark operators and the Higgs tri-linear self-coupling give essential contributions.
There are many blind directions, and the combinations of operators contributing to  NLO accuracy are listed in  Appendix~\ref{sec:combo}. It is worth repeating that we have considered the possibility of arbitrary flavor structures and included all operators relevant at the one-loop level.

At NLO, there are both weak and pure QED corrections. As the typical size of QED effects is significant due to the enhancement by large logarithms involving the electron mass, we separate these two corrections. 
We write the SM total cross-section as,
\begin{align}
    \sigma_{\rm SM,NLO}=\sigma_{\rm SM,NLO}^W(1+\delta_{\rm SM,QED}) \, , 
\end{align}
where $\sigma_{\rm SM, NLO}^W$ includes $\sigma_{\rm SM, LO}$ plus the one-loop virtual corrections that are not purely photonic as discussed in Sec. \ref{sec:qeddef}. The SM NLO result of \cite{Freitas:2022hyp} corresponds to $\sigma_{\rm SM, NLO}^W$. The purely photonic virtual and real photon emission corrections are included in $\delta_{\rm SM, QED}$; see the diagrams of Fig. \ref{fig:QED}.

At  NLO, the effects on the  unpolarized total cross-section to ${\cal{O}}({1 / \Lambda^2})$ are parameterized as
\begin{align}
    \label{eq:sigpar}
    \sigma_{\rm NLO} = 
    \sigma_{\rm SM,NLO}+\sigma_{\rm SM,NLO}^{W}\sum_i \frac{C_i(\mu)}{\Lambda^2} \bigg\{ \Delta_{i,\rm weak}^{\rm (NLO)} + \bar{\Delta}_i \log{\frac{\mu^2}{s}}
    +  \Delta_{i,\rm QED}
    \bigg\}
    \, ,
\end{align}
where
\begin{align}
    \begin{split}
    \Delta_{i,\rm weak}^{\rm (NLO)} &= \Delta_{i}^{\rm (LO)} + \Delta_{i,\rm weak}^{(\delta \rm  NLO)} \, , \\
    \Delta_{i}^{\rm (NLO)} &= \Delta_{i,\rm weak}^{\rm (NLO)}  + \Delta_{i,\rm QED} \, ,
    \end{split}
    \label{eq:nlodef}
\end{align}
and 
\begin{align}
    \label{eq:lodefs}
    \frac{\sigma_{\rm LO}}{\sigma_\textrm{SM,NLO}^{W}} &\equiv 1+ \sum_i \frac{C_i(\mu)}{\Lambda^2} \Delta_i^{(\rm LO)} \, .
\end{align}
Normalizing our results to the SM weak contribution, $\sigma_{\rm SM, NLO}^W$, is entirely arbitrary.
For polarized initial states, the notation is the straightforward generalization of Eqs.~\eqref{eq:nlodef} and~\eqref{eq:lodefs}.

The contributions $\Delta_i$ and ${\overline{\Delta}}_i$ originate only from virtual diagrams. They are separately IR finite and form a gauge invariant set. 
The contributions $\Delta_{i,\rm QED}$ combine the SMEFT contributions from the virtual vertex correction diagram shown on the left of Fig.~\ref{fig:QED} (which is IR divergent) and the complete set of SMEFT real radiation contributions (also shown in Fig.~\ref{fig:QED}). We note that the SMEFT real contribution contains diagrams with a different kinematic structure from that of the SM initial state radiation (as seen in the rightmost diagram of Fig.~\ref{fig:QED}), even though these diagrams do not contain IR divergences. 

Our results are presented in numerical form, expressing the total cross-sections using the parameterization given in Eq.~\eqref{eq:sigpar}. This parameterization will enable easy integration into global fitting programs. We report results for energies proposed for future $e^+e^-$ colliders: $\sqrt{s}=240,~365$, and $500~\rm{GeV}$. Additionally, we include the separate results for polarized beams at each of these energies.  

\begin{figure}
    \centering	
    \includegraphics[width=.32\textwidth]{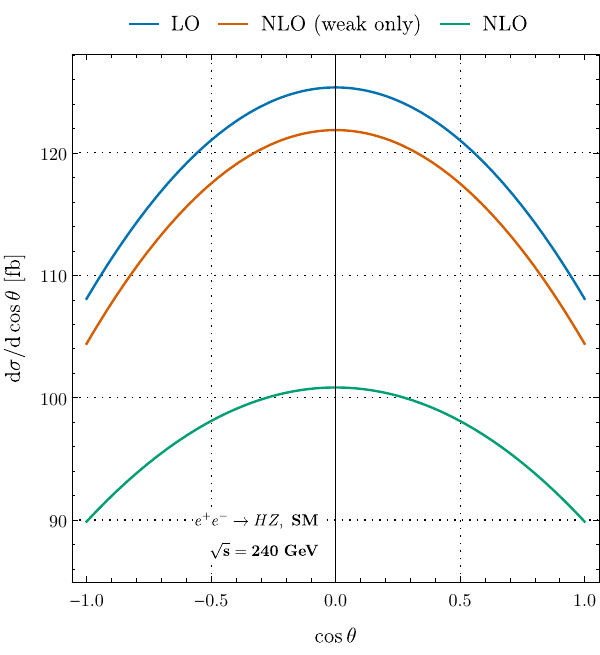}\hspace{5pt}
    \includegraphics[width=.32\textwidth]{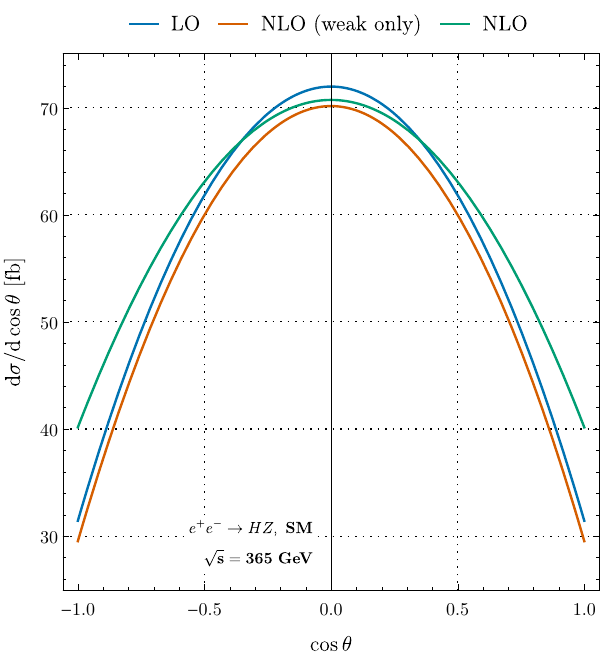}\hspace{5pt}
    \includegraphics[width=.32\textwidth]{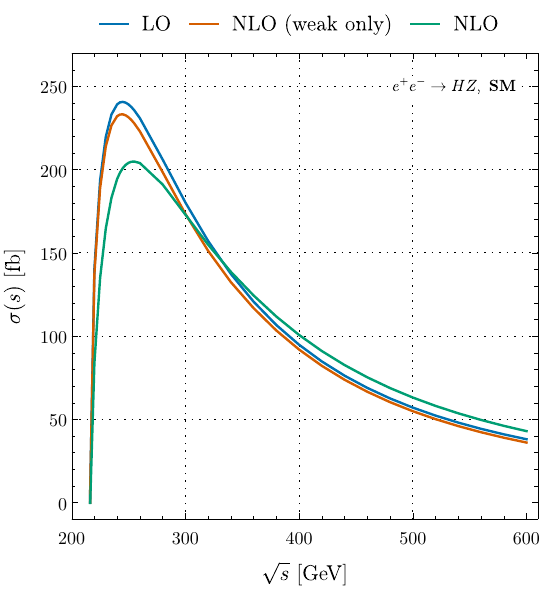} 
    \caption{SM cross-sections for $e^+e^-\rightarrow ZH$ and for unpolarized beams as a function of the angle between the $ e^-$ beam direction and the Higgs boson momentum and as a function of the center of mass energy. As the text describes, the purely weak contribution is separated from the total NLO result.}
    \label{fg:smdist}
\end{figure}

SM results are given in Table \ref{tab:results:sm}, with the weak and QED pieces separated.  We note that the sign of the QED corrections changes as the energy is increased from $\sqrt{s}=240~\rm{GeV}$ to $\sqrt{s}=365~\rm{GeV}$.  It is clear that at all energies, the largest contribution to the SM total rate is from the left-handed electron polarization.\footnote{Our SM results agree with Ref.~\cite{Freitas:2023iyx}.}  The SM angular distribution is shown in Fig.~\ref{fg:smdist}, where $\theta$ is defined as the angle between the incoming electron and outgoing $H$ in the center of the mass frame.

One-loop SMEFT results are given in tables in Appendix~\ref{sec:numbers}.  Table~\ref{tab:0fops} contains the purely weak contribution, $\Delta_{i, {\rm weak}}^{({\rm NLO})}$, and the contribution from the logarithmic running, ${\overline{\Delta}}_i$, from operators that contribute at tree -level for both polarized and unpolarized electrons.  These contributions must be combined with the QED contributions (both real and virtual) to obtain a physical result and are given in Table~\ref{tab:0fopsfin} (unpolarized beams), Table~\ref{tab:0fopsL} (left-handed electron polarization) and Table~\ref{tab:0fopsR} (right-handed electron polarization).  The quantity $\Delta_i^{({\rm NLO})}$ is the sum of the weak and QED corrections and is the total NLO result, including the LO contribution, Eq.~\eqref{eq:nlodef}. Table~\ref{tab:2fops} contains the NLO contributions from 2-fermion and scalar operators that first appear at one-loop, and Table~\ref{tab:4fops} has the one-loop contributions from 4-fermion interactions.  Since these operators do not contribute at the tree level, they have no contribution from real photon emission, which would be a part of the two-loop corrections. 

\begin{figure}
    \centering	
    \includegraphics[width=.32\textwidth]{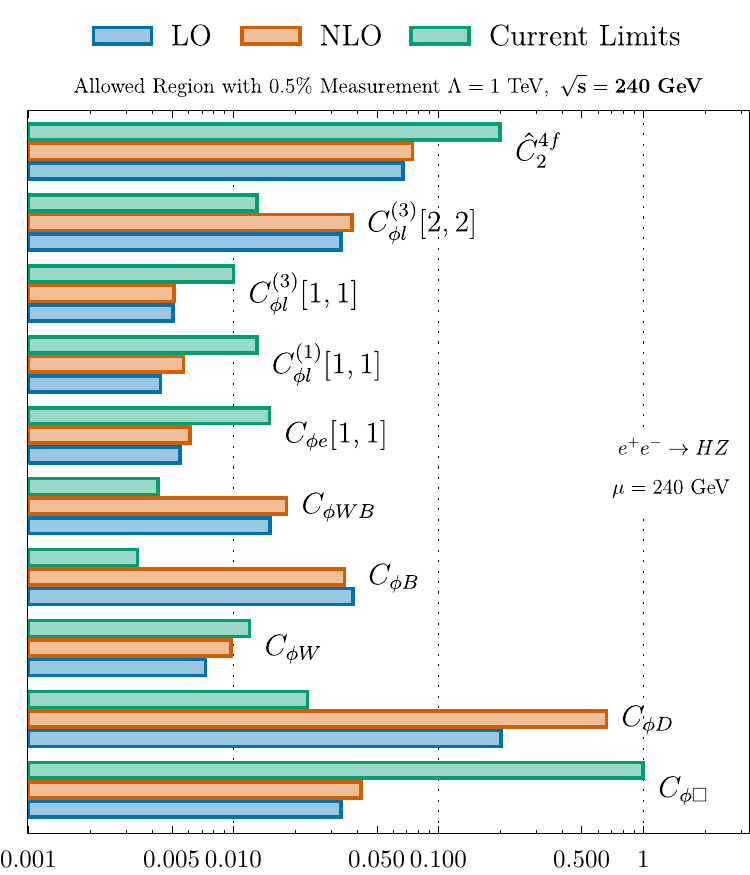}\hspace{5pt}
    \includegraphics[width=.32\textwidth]{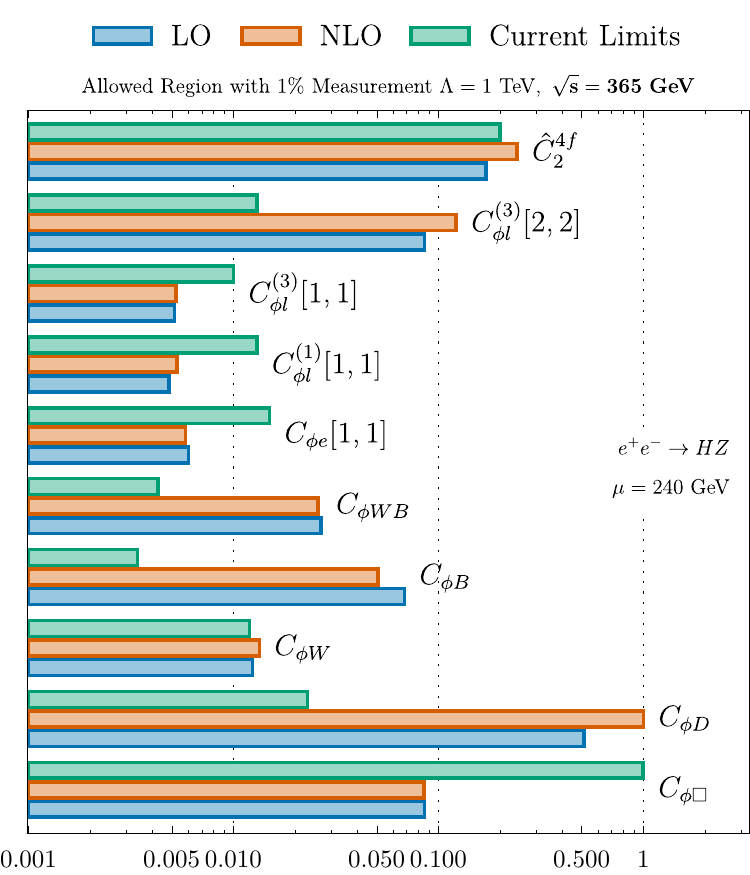}\hspace{5pt}
    \includegraphics[width=.32\textwidth]{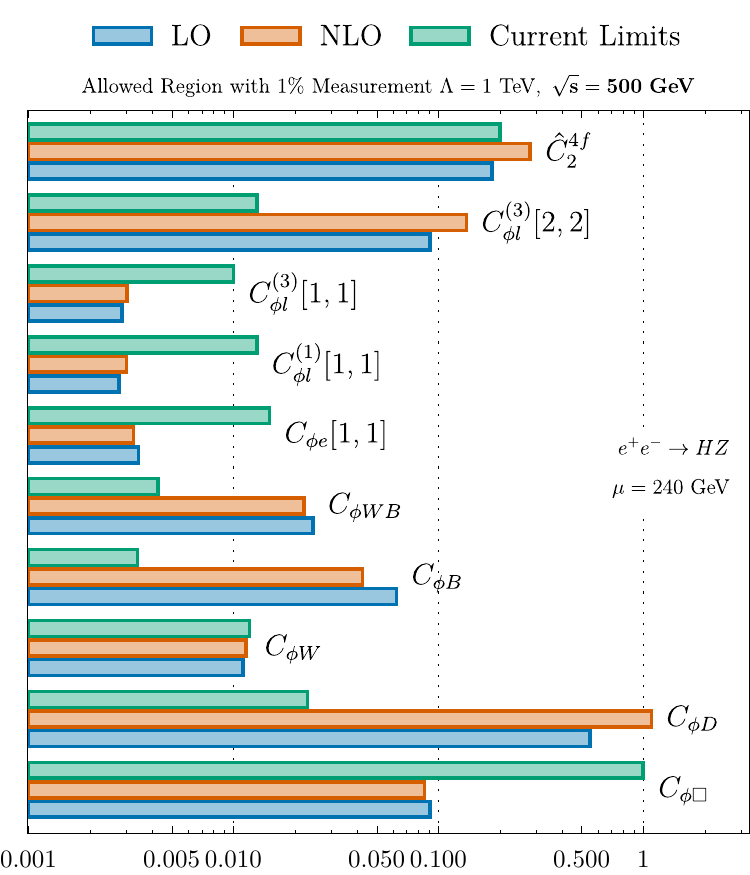}
    \caption{Single parameter range corresponding to a $0.5\%$ deviation of the Higgsstrahlung total cross-section from the SM NLO prediction at $\sqrt{s}=240~\textrm{GeV}$, including the LO and NLO SMEFT results and a comparison with the global fit of \cite{Ellis:2020unq}. Many of the global fit limits are asymmetric, and we have plotted the absolute value of the weakest limit. For $\sqrt{s}=365~\textrm{GeV}$ and $\sqrt{s}=500~\textrm{GeV}$, we assume a $1\%$ measurement.  Note that the scale is kept fixed at $\mu=240~\textrm{GeV}$ to aid in the comparison. The fits assume unpolarized beams.}
\label{fg:singlenlo}
\end{figure}

A comparison of the single parameter sensitivity to a $0.5\%$ measurement at $\sqrt{s}=240~{\rm GeV}$ from a measurement of the total Higgsstrahlung cross-section is shown in Fig.~\ref{fg:singlenlo} for those operators that contribute at tree- level. 
Unless otherwise specified, all of the coefficients in our plots are evaluated at $\mu=240~{\rm GeV}$ to enable direct comparisons between results. The single parameter results are obtained by looking at the deviations from the SM NLO predictions and shown as the allowed range of the coefficients. The global fit limits from a top-specific fit~\cite{Ellis:2020unq}  to LHC Higgs and top quark data and to precision electroweak data are shown for comparison.   Many of the limits in the global fit are asymmetric, and we have plotted the absolute value of the weakest limit.\footnote{There are several global fits available, which are performed with various assumptions, but the overall picture emerging from the different fits is qualitatively similar~\cite{Ethier:2021bye,Ellis:2020unq,Biekotter:2018ohn}. The global fits are typically performed using dimension-6 NLO QCD SMEFT predictions, while the NLO EW SMEFT predictions for the observables are not included.  Hence, the comparison of our results with existing global fits is for illustrative purposes only.} The most significant improvement arising from measurements of Higgsstrahlung relative to the global fit results is for $C_{\phi D}$.  For most tree-level operators, the sensitivity at NLO is similar to that at LO, with only minor shifts, and the measurement of the Higgsstrahlung total cross-section does not significantly improve our knowledge of these coefficients from the current global fits.
The exception is the result for $C_{\phi D}$. The LO sensitivity to  $O_{\phi D}$ is small due to the significant cancellation between the right and left-handed contributions, making the NLO corrections for $O_{\phi D}$  more important than for the other tree-level operators.  
For comparison, we show results at $\sqrt{s}=365~{\rm GeV} $ and $\sqrt{s}=500~{\rm GeV}$ assuming a $1\%$ measurement of the total cross-section at these energies~\cite{ILCInternationalDevelopmentTeam:2022izu,deBlas:2022ofj}.  The single parameter sensitivity results of Fig.~\ref{fg:singlenlo} do not show a significant energy dependence. 
The following sections, however, will emphasize the significant correlations between the different operator contributions that make the single parameter results extremely misleading in general. 

\begin{figure}
    \centering	
    \includegraphics[width=.32\textwidth]{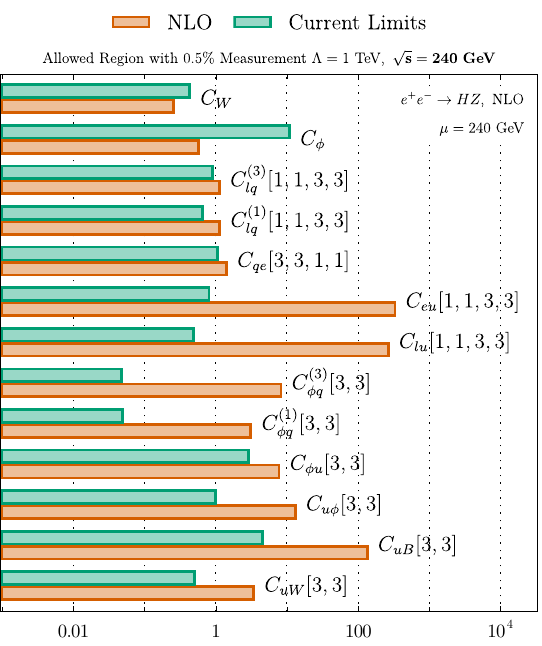}\hspace{5pt}
    \includegraphics[width=.32\textwidth]{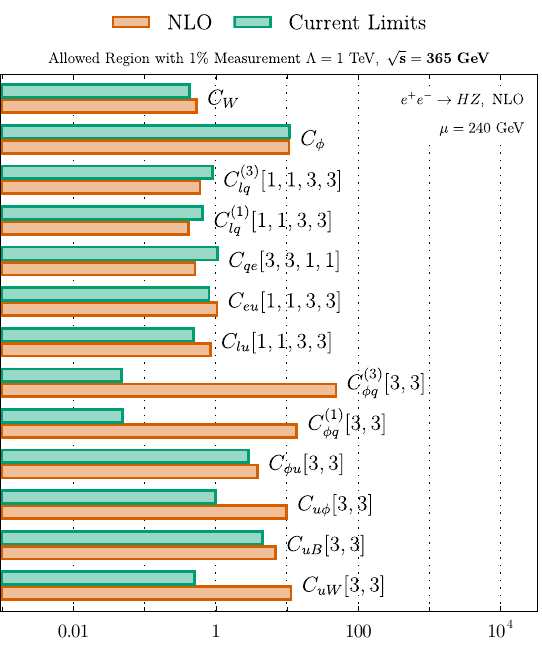}\hspace{5pt}
    \includegraphics[width=.32\textwidth]{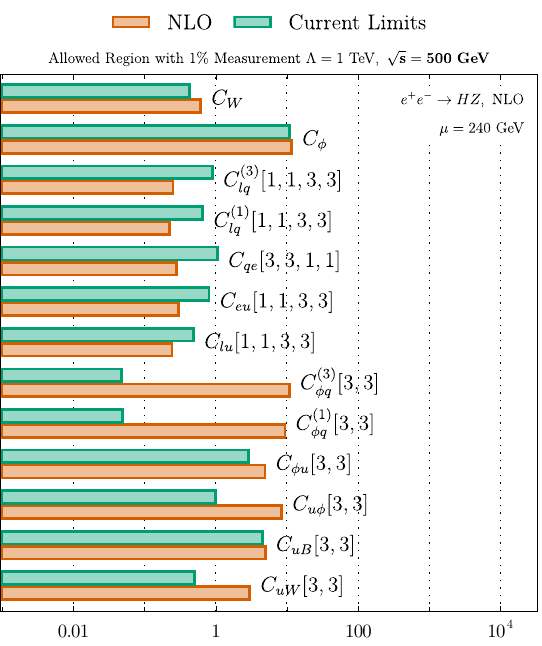} 
    \caption{  
    Single parameter range corresponding to a $0.5\%$ and $1\%$ deviation of the Higgsstrahlung total cross-section from the SM NLO prediction at $\sqrt{s}=240~\textrm{GeV}$, and $\sqrt{s}=365~\textrm{GeV}$ and $\sqrt{s}=500~\textrm{GeV}$, respectively. Results are compared with the global fit of \cite{Ellis:2020unq}. The current limits for the 4-fermion operators are from \cite{Bellafronte:2023amz}.  Many of the global fit limits are asymmetric, and we have plotted the absolute value of the weakest limit. Note that the scale is kept fixed at $\mu=240~\textrm{GeV}$ to aid in the comparison. The fits assume unpolarized beams.}
    \label{fg:single2f4f}
\end{figure}

At NLO, a sensitivity arises to 2-fermion and 4-fermion operators that do not contribute at tree level, along with sensitivity to anomalous Higgs tri-linear couplings ($O_\phi$) and anomalous 3-gauge boson interactions ($O_W$) and the single parameter sensitivities from Higgsstrahlung are shown in Fig.~\ref{fg:single2f4f}. 
The current limits on $C_W$ and the 2-fermion operators are from the global fit of~\cite{Ellis:2020unq}.
  The limits on the 4-fermion top quark operators are from NLO predictions for EWPOs in a flavor-agnostic scenario~\cite{Bellafronte:2023amz}. At $\sqrt{s}=240~\rm{GeV}$, the only single parameter limits improved by the NLO Higgsstrahlung total cross-section measurement are those for $C_W$ and $C_\phi$. The 4-fermion interactions contain contributions that are enhanced by ${s}/{\Lambda^2}$, and indeed, we see that the single parameter results for these operators improve with energy. 
Again, we emphasize that our results must be included in global fits for future colliders, including the crucial correlations between different operator contributions. 

\begin{figure}
    \centering	
    \includegraphics[width=.32\textwidth]{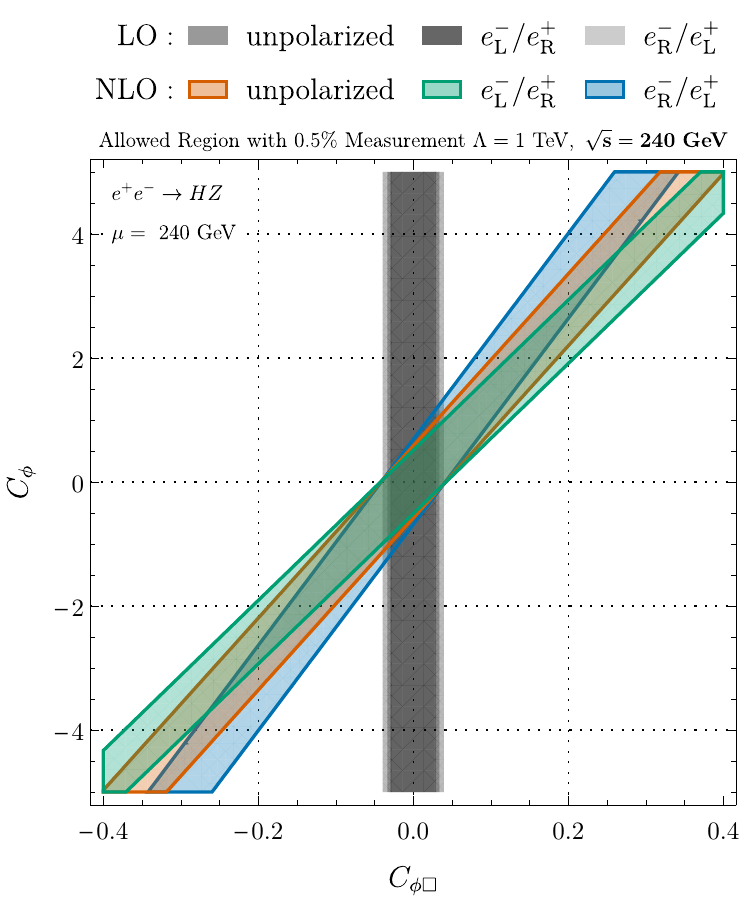}\hspace{5pt}
    \includegraphics[width=.32\textwidth]{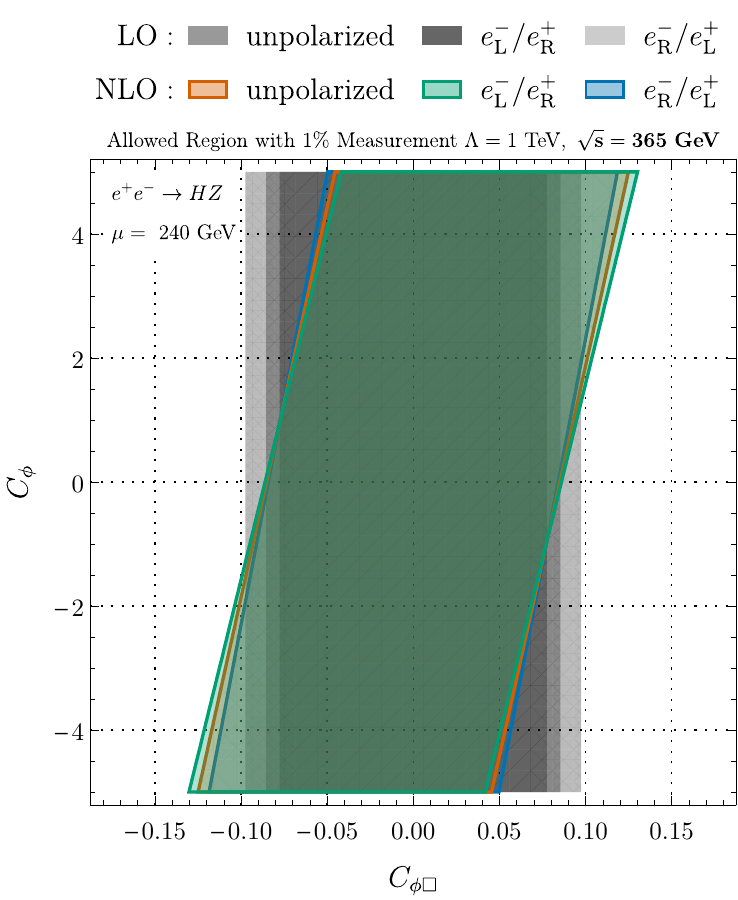}\hspace{5pt}
    \includegraphics[width=.32\textwidth]{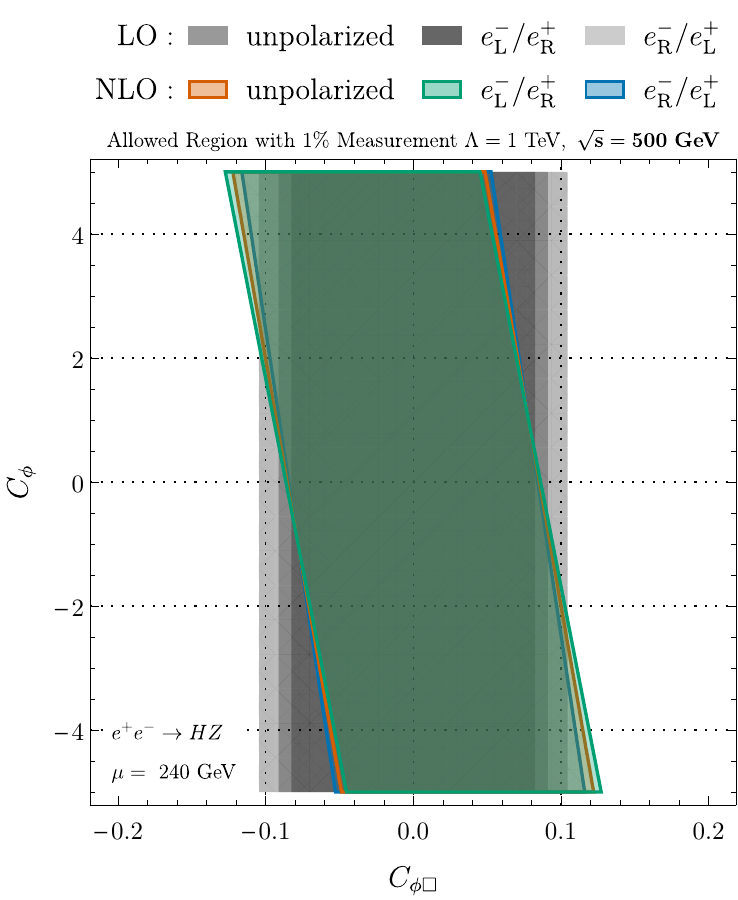}
    \caption{Contributions from modifications of the Higgs tri-linear coupling $C_\phi$ to the cross-section for $e^+e^-\rightarrow ZH$ correlated with those from $C_{\phi \square}$.   The sensitivity to a $0.5\%$, $1\%$ and $1\%$ measurement at $\sqrt{s}=240~{\rm GeV}$, $365~\textrm{GeV}$ and $500~\textrm{GeV}$, respectively, is shown. Note that there is no sensitivity to $C_\phi$ at tree level. }
    \label{fg:chchk}
\end{figure}

\begin{figure}
    \centering	
    \includegraphics[width=.32\textwidth]{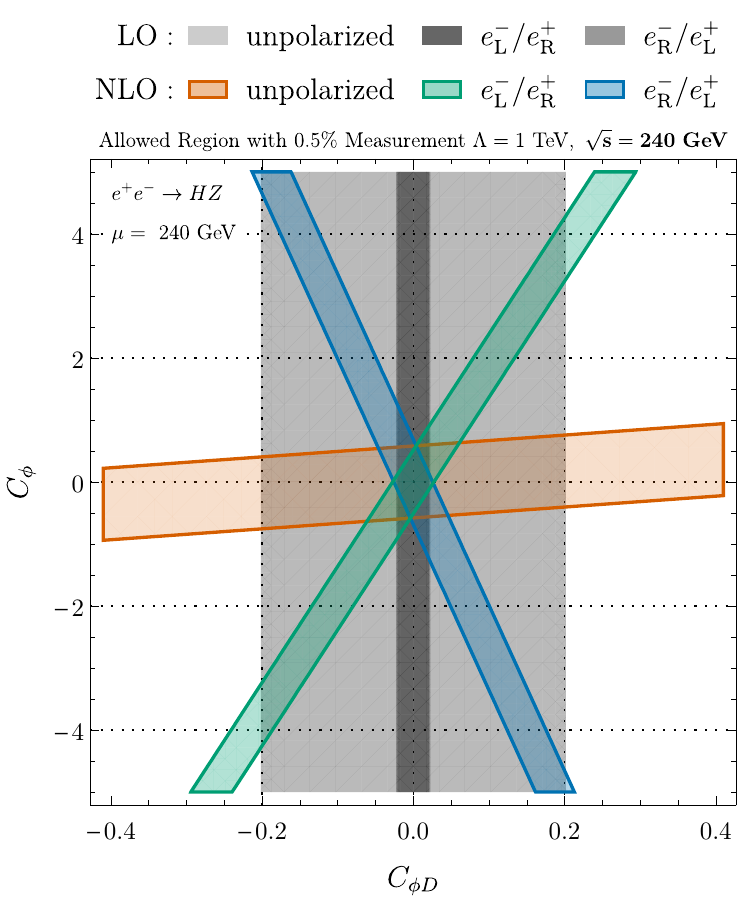}\hspace{5pt}
    \includegraphics[width=.32\textwidth]{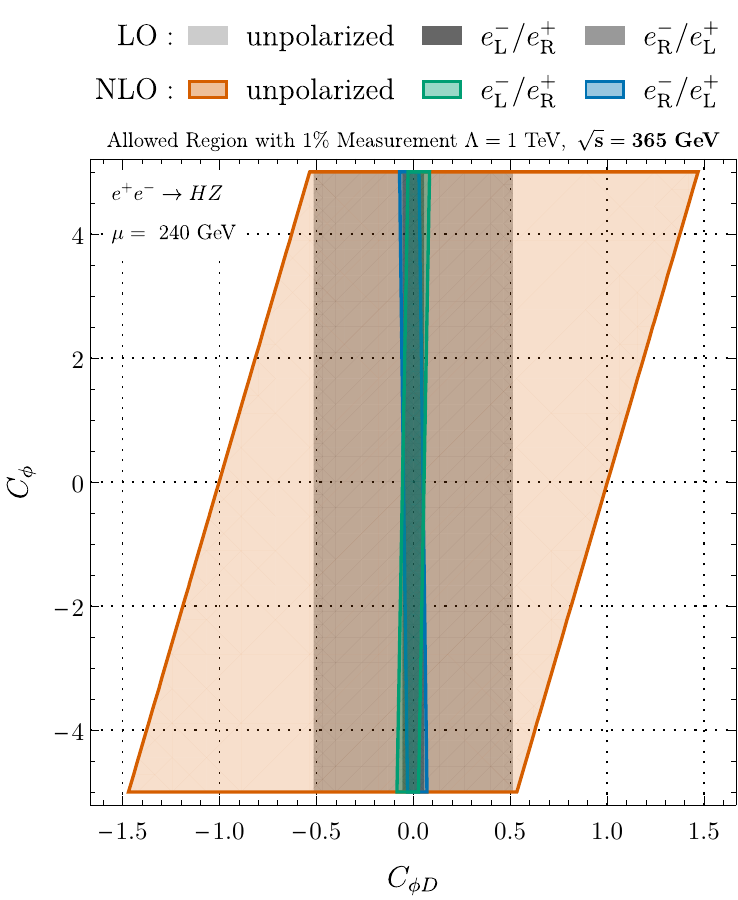}\hspace{5pt}
    \includegraphics[width=.32\textwidth]{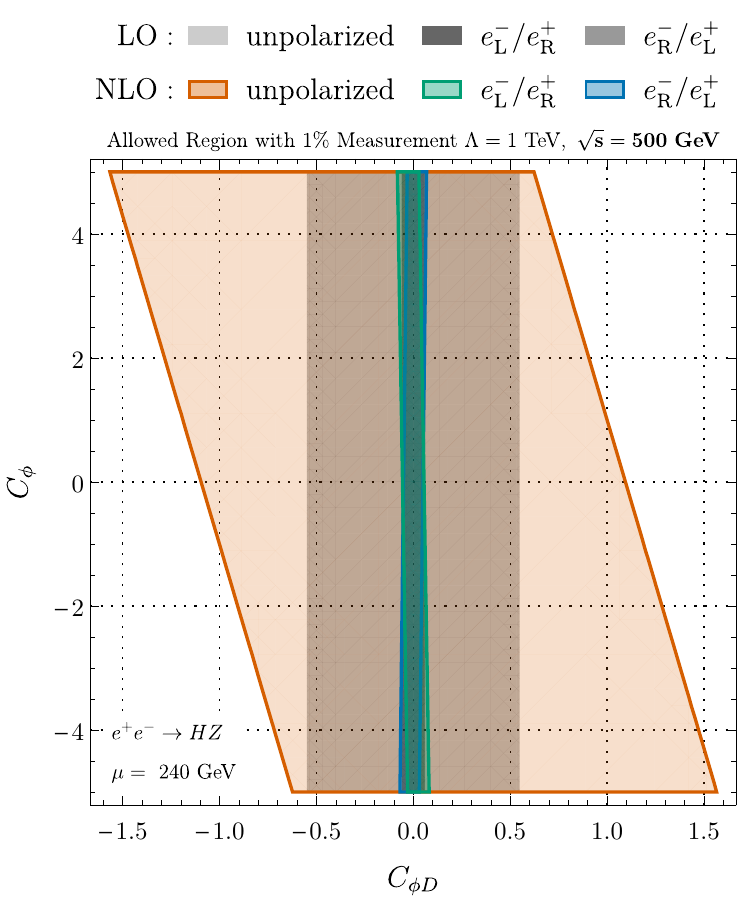}
    \caption{Contributions from modifications of the Higgs tri-linear coupling $C_\phi$ to the cross-section for $e^+e^-\rightarrow ZH$ correlated with those from $C_{\phi D}$.   The sensitivity to a $0.5\%$, $1\%$ and $1\%$ measurement at $\sqrt{s}=240$ ~GeV, $365~\textrm{GeV}$ and $500~\textrm{GeV}$, respectively, is shown. Note that there is no sensitivity to $C_\phi$ at tree level.}
    \label{fg:chchD}
\end{figure}

\begin{figure}
    \centering	 
    \includegraphics[width=.32\textwidth]{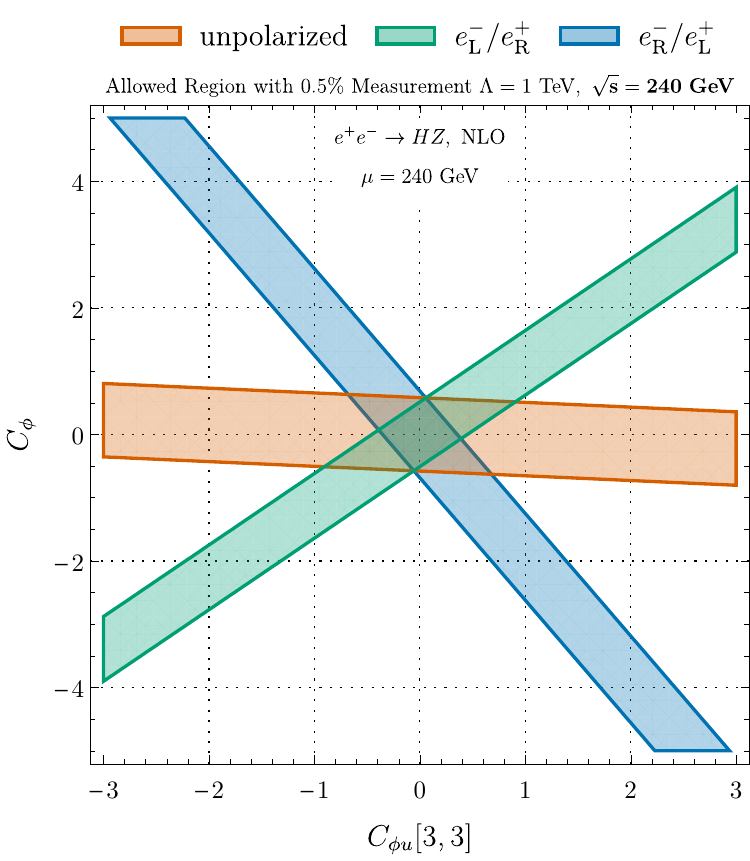}\hspace{5pt}
    \includegraphics[width=.32\textwidth]{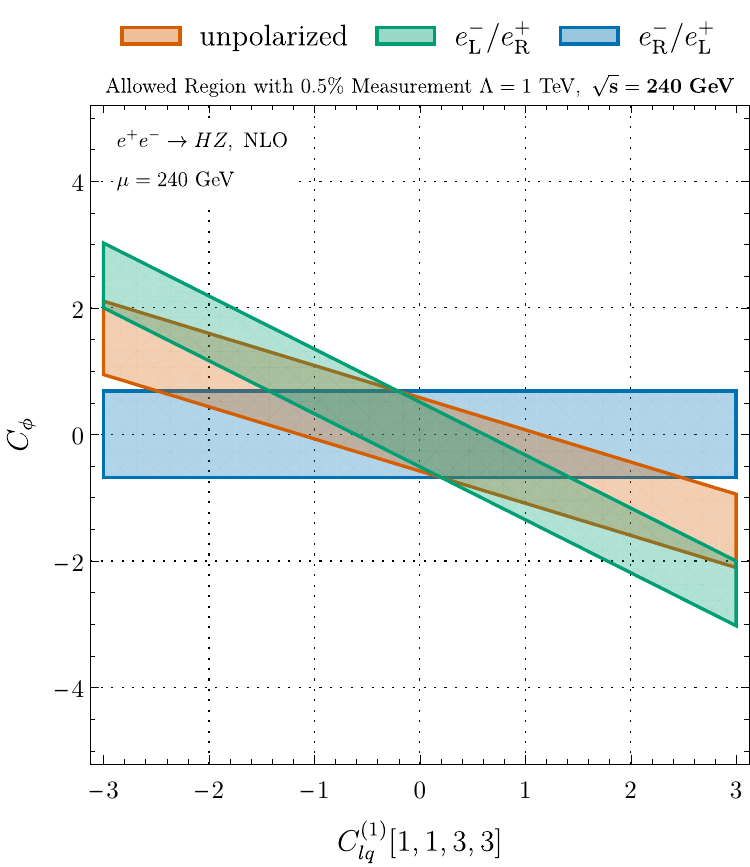}\hspace{5pt}
    \caption{Contributions from modifications of the Higgs tri-linear coupling $C_\phi$ on the cross-section for $e^+e^-\rightarrow ZH$ correlated with those from $C_{\phi u}[3,3]$, which modifies the $Z t{\overline{t}}$ vertex, and from $C_{lq}^{(1)}[1,1,3,3]$ vertex which modifies the $e^+e^-t{\overline{t}}$ interaction. The sensitivity to a $0.5\%$ measurement at $\sqrt{s}=240~{\rm GeV}$ is shown. Note that there is no sensitivity to $C_\phi$, $C_{\phi u}[3,3]$ or $C_{lq}^{(1)}[1,1,3,3]$ at tree level. 
    }
    \label{fg:chztt}
\end{figure}

\begin{figure}
    \centering	 
    \includegraphics[width=.32\textwidth]{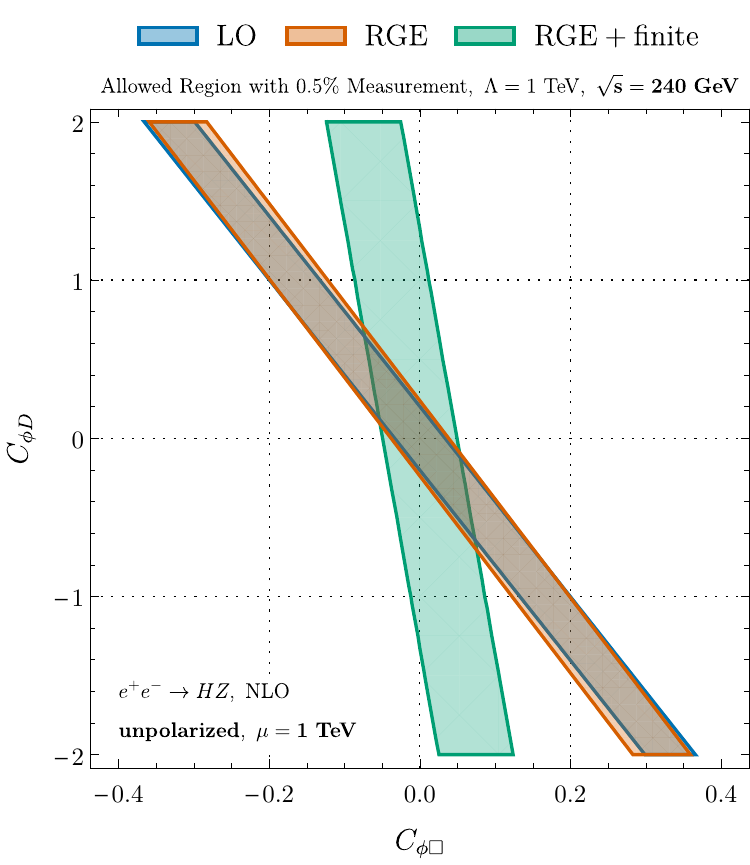}\hspace{5pt}
    \includegraphics[width=.32\textwidth]{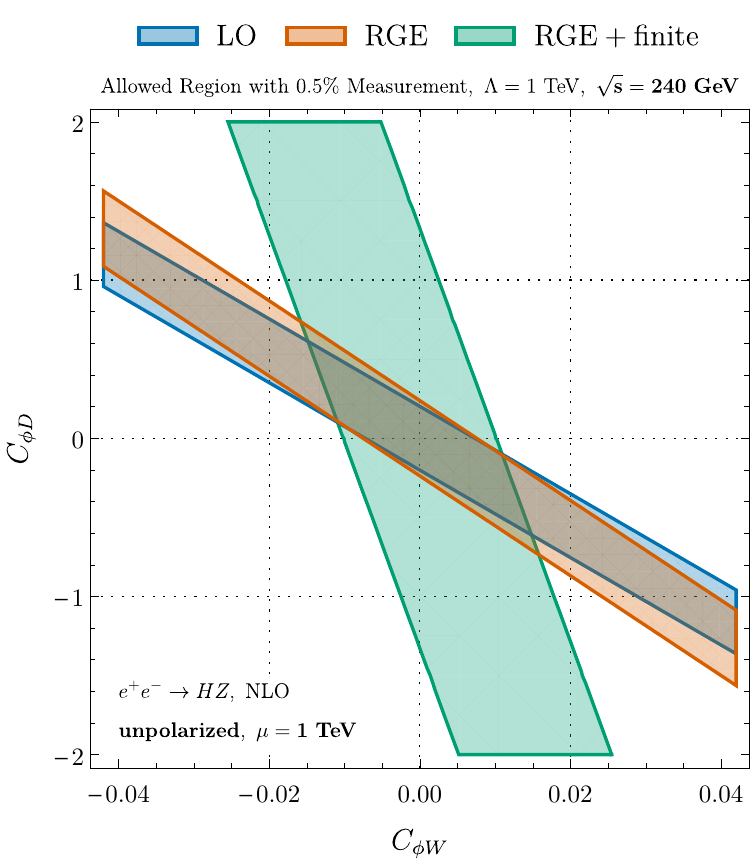}\hspace{5pt}
    \includegraphics[width=.32\textwidth]{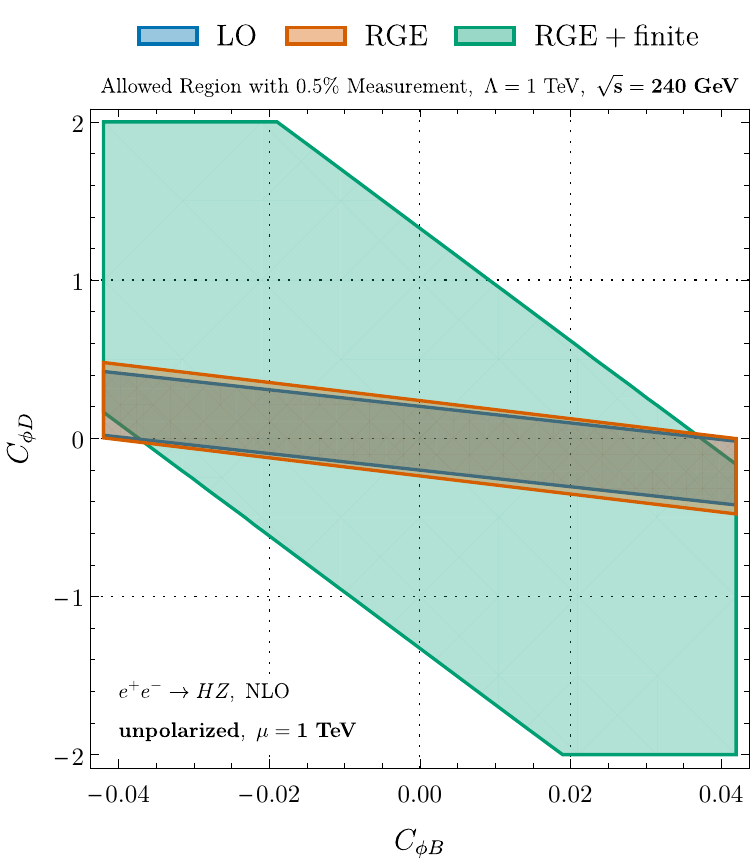}
    \caption{Sensitivity to $C_{\phi W}$ and $C_{\phi D}$ in $e^+e^-\rightarrow ZH$ at $\sqrt{s}=240~\textrm{GeV}$ when the scale $\mu=\Lambda=1~\textrm{GeV}$.  The curve labeled RGE contains only the RGE corrections to the LO result, while the RGE+finite curve includes the complete NLO result.}
    \label{fg:tmp}
\end{figure}

\subsection{Higgs Tri-linear Coupling}

Measuring the Higgs cubic self-coupling and constraining the shape of the Higgs potential $V$ is a major goal of both the HL-LHC and future colliders. In the SMEFT framework, modifications of the Higgs self-coupling  originate from the operator, $O_\phi$, along with terms arising from the redefinition of the parameters.  
We begin with the purely scalar terms in the Lagrangian,
\begin{eqnarray}
\mathcal{L}_\phi&=&(D_\mu \phi)^\dagger(D_\mu \phi)+\mu^2(\phi^\dagger\phi)-\lambda(\phi^\dagger\phi)^2\nonumber\\&+\frac1{\Lambda^2}&\biggl[\mathcal{C}_\phi(\phi^\dagger\phi)^3+\mathcal{C}_{\phi D}(\phi^\dagger D_\mu\phi)^*(\phi^\dagger D_\mu\phi)+\mathcal{C}
_{\phi\Box}(\phi^\dagger\phi)\Box(\phi^\dagger\phi)\biggr]\, .
\end{eqnarray}
In the physical gauge, $\phi^T\to(0,(v_T+H)/\sqrt{2})$, the kinetic part of the Lagrangian is
\begin{eqnarray}
    \mathcal{L}_H=\frac12 (\partial_\mu H)^2(1+2 \frac{v_T^2}{\Lambda^2}\mathcal{C}_{\rm Kin})\, ,
\end{eqnarray}
with $\mathcal{C}_{\rm Kin}= \mathcal{C}_{\phi D}/4-\mathcal{C}_{\phi\Box}$, so we apply the substitution $H\to h Z_H^{-1}$ with $Z_H=1+ (v_T^2/\Lambda^2) \mathcal{C}_{\rm Kin}$, so that the kinetic part of the Lagrangian for the field $h$ has the canonical form. Similarly to the equivalent SM calculation, we can extract the relation between $\mu^2$ and the rest of the Lagrangian parameters by calculating the tadpole term of the potential 
\begin{eqnarray}
    V_t(h)=v_T (v_T^2\lambda-\mu^2-\frac34 \frac{v_T^4}{\Lambda^2}\mathcal{C}_\phi)h Z_H^{-1}
\end{eqnarray}
and imposing $V_t=0$. This allows us to write the quadratic term of the scalar potential as \begin{eqnarray}
    V_2(h)= \biggl(v_T^2\lambda -\frac{3}{2}\frac{v_T^4}{\Lambda^2}\mathcal{C}_\phi-2\frac{v_T^4}{\Lambda^2}\lambda\mathcal{C}_{\rm Kin}\biggr)h^2\equiv\frac12 M_H^2 h^2\, ,
\end{eqnarray}
from which we extract the  physical scalar mass, $M_H$, and use it to eliminate the dependence on $\lambda$ from the Lagrangian. Contrary to what happens in the SM, we can distinguish two contributions to the Lagrangian involving 3 field insertions: a purely cubic contribution that does not involve derivatives of the fields
\begin{eqnarray}
    V_3(h)=\frac{M_H^2}{2 v_T}\biggl(1-\frac{v_T^2}{\Lambda^2}(\mathcal{C}_{\rm Kin}+2\frac{v_T^2}{M_H^2}\mathcal{C}_\phi)\biggr)h^3\, ,
\end{eqnarray}
and a mixed kinetic coupling, 
\begin{eqnarray}
    V_{3K}(h)=-\frac{M_H^2}{2 v_T}\biggl(\frac{v_T^2}{\Lambda^2}\frac{4}{M_H^2}\mathcal{C}_{\rm Kin}\biggr)h (\partial_\mu h)^2\, ,
\end{eqnarray} which involves the derivatives of the Higgs field, and 
thus implies momentum dependent contributions to the Feynman rule of the triple Higgs vertex, which is apparent in the Feynman rules \cite{Dedes:2017zog}.
Establishing the connection between the SMEFT notation and the commonly used phenomenological parametrization,
   \begin{eqnarray}
    V_{pheno}(h)\equiv \frac{M_H^2}{2 v_T}\kappa_\lambda h^3\, ,
\end{eqnarray}
we obtain 
\begin{equation}
\kappa_\lambda=\biggl(1+\frac{v^2}{\Lambda^2}\biggl[3\mathcal{C}_{\rm Kin}-2\frac{v^2}{M_H^2}\mathcal{C}_\phi\biggr]\biggr)\, ,
\label{eq:kdef}
\end{equation}
 where $\kappa_\lambda$ is expanded only to linear order in $1/\Lambda^2.$
This correspondence allows us to extract limits on $C_\phi$ from the existing LHC experimental results, which do not use the SMEFT framework directly.
Current experimental limits from ATLAS~\cite{ATLAS:2024fkg} and CMS~\cite{CMS:2024awa} from combining single and double Higgs production assuming SM couplings except for the Higgs tri-linear coupling give the $95 \%$ CL limits if we set ${\mathcal{C}}_{Kin}\rightarrow 0$,
\begin{eqnarray}
    {\text{ATLAS}}:& -0.4 < \kappa_\lambda < 6.3\quad &\longrightarrow \quad -11 < C_\phi \biggl(\frac{1~\textrm{TeV}}{\Lambda}\biggr)^2 <  3\, ,\\[5pt]
    {\text{CMS}}:& -1.2 < \kappa_\lambda < 7.5 \quad  &\longrightarrow \quad  -14 < C_\phi \biggl(\frac{1~\textrm{TeV}}{\Lambda}\biggr)^2 <  4.6 \, .
\end{eqnarray}
Currently, the LHC limits from single Higgs production are similar to those from double Higgs production. From Table \ref{tab:2fops}, we see that a measurement of the total Higgsstrahlung cross-section at $\sqrt{s}=240~{\rm GeV}$ to $0.5\%$ accuracy would restrict $|C_\phi| < 0.58$. It is essential to remember that this assumes that the only new physics exists in the Higgs self-coupling, which seems like a rather unlikely scenario.  

A pioneering study\footnote{ See also \cite{Craig:2014una,Beneke:2014sba,DiVita:2017vrr,Maltoni:2018ttu}.}~\cite{McCullough:2013rea} examined the sensitivity of Higgsstrahlung to anomalous Higgs self-couplings in the presence of momentum-independent modifications of the $ZZH$ vertex using an effective parameterization,
\begin{align}
    L_{HZZ}\equiv2M_Z^2\sqrt{\sqrt{2}G_\mu} (1+\delta_Z) HZ^\mu Z_\mu\, .
\end{align} 
In the context of the SMEFT, the parameter $\delta_Z$ can be expressed as a combination of several coefficients
\begin{align}
    \delta_Z\sim \frac{1}{4\sqrt{2}G_\mu \Lambda^2} \biggl(  4 C_{\phi\square}+C_{\phi D}\biggr)+\frac{X_H}{2} \, ,
    \label{eq:dzdef}
\end{align}
A fuller and theoretically consistent picture can be obtained by including all SMEFT operators contributing to Higgsstrahlung. We note that $O_{\phi WB}$
introduces momentum-dependent interactions, along with the constant shift in the $HZZ$ vertex. Further, the coefficients in Eq.~(\ref{eq:dzdef}) contribute to many interactions besides the $HZZ$ vertex.
Our results allow us to correlate the sensitivity to $C_\phi$ with all dimension-6 SMEFT operators contributing to Higgsstrahlung at NLO.

Figs.~\ref{fg:chchk} and~\ref{fg:chchD} and  show the two parameter sensitivity to $O_\phi$ correlated  with $O_{\phi\square}$ and $O_{\phi D}$. At  LO, there is no sensitivity to $O_\phi$.
The correlations  of $O_\phi$  with the other operators that contribute at tree level, $O_{\phi W}$, $O_{\phi B}$, $O_{\phi W B}$,  $O_{\phi e}[1,1]$,   $O_{\phi l}^{(1,3)}[1,1]$, $O_{\phi l}^{(3)}[2,2]$ and ${\hat{O}}^4_2$ are given in Figs.~\ref{fg:tree1} and~\ref{fg:figmore} in Appendix~\ref{sec:numbers}. Fig.~\ref{fg:chchk} shows only a slight sensitivity to polarization at $\sqrt{s}=240~\textrm{GeV}$, while at higher energies, the total cross-section is dominated by the left-handed polarization. The correlation between $C_\phi$ and $C_{\phi \square}$ is significant at $\sqrt{s}=240$ GeV but becomes small at higher energies. Fig.~\ref{fg:chchD} shows that the polarized and unpolarized sensitivity to the total cross-section are quite different at $\sqrt{s}=240$ GeV and show interesting correlations with $C_\phi$. At higher energies, the $C_\phi-C_{\phi D}$ correlation is smaller and shows less dependence on polarization. 
We have observed that including or excluding contributions from different operators significantly impacts the size of the constraints and the interpretation of the results. Therefore, it is crucial to incorporate a complete set of operators beyond LO accuracy when conducting a model-independent EFT study.

Experimental limits on $C_\phi$ have also been obtained assuming the top-Higgs Yukawa coupling modification. 
The effects of modifying the top quark Yukawa coupling have been considered in Fig.~3 of~\cite{Asteriadis:2024qim}.
Aside from the constant shift in the top quark Yukawa coupling of Eq.~\eqref{eq:parshift}, there are  2-quark and 4-quark operators involving the top quark that contribute to Higgsstrahlung beginning at NLO and that generate new momentum-dependent structures in the interactions (Tables~\ref{tab:2fops} and~\ref{tab:4fops}), and are often enhanced by factors of $m_t^2$. It is impossible to obtain these effects using the phenomenological $\kappa$ formalism, and the correlations between the SMEFT operators involving the top quark open a new window for discovery.
The left hand side of Fig.~\ref{fg:chztt} shows the sensitivity to $C_\phi$ 
and  $C_{\phi u}[3,3]$. 
The modifications to the $Z t {\overline{t}}$ vertex from $O_{\phi u}[3,3]$ are at present poorly constrained, and Fig.~\ref{fg:chztt} shows the sensitivity at $\sqrt{s}=240~\textrm{GeV}$. We also see that polarization plays an important role, and the sensitivity to left- and right-handed initial states is quite different.
The correlations with the remaining 2-top operators that contribute at NLO are in Fig.~\ref{fg:chtop2} of Appendix~\ref{sec:numbers}.
The right hand side of Fig.~\ref{fg:chztt} shows the sensitivity to $C_\phi$ 
and  $C_{lq}[1,1,3,3]$, a 4-fermion operator connecting left-handed electrons and left-handed top quarks.  Again, we see the significance of polarization.
The sensitivity to  the remaining 4-fermion operators involving $e^+e^-t{\overline{t}}$ 4-point interactions are shown in Fig.~\ref{fg:4fch}  of Appendix~\ref{sec:numbers}. These interactions are poorly constrained, with the most significant limits from NLO corrections to Z-pole observables~\cite{Bellafronte:2023amz, Dawson:2022bxd}. We note that the limits from Z-pole observables depend strongly on the assumed flavor structure.  

\begin{figure}
    \centering	
    \includegraphics[width=.32\textwidth]{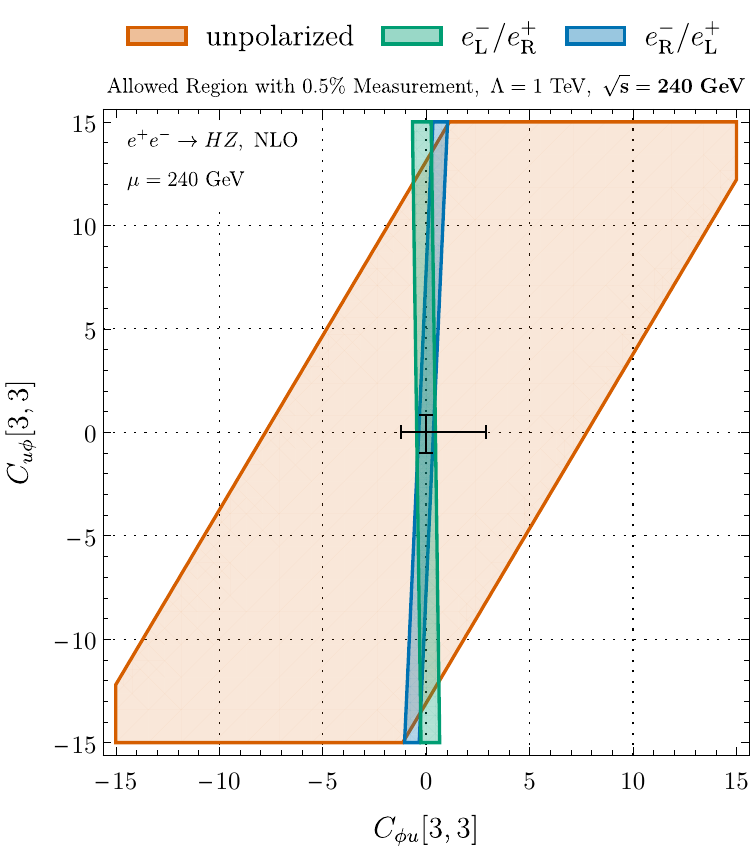}\hspace{5pt}
    \includegraphics[width=.32\textwidth]{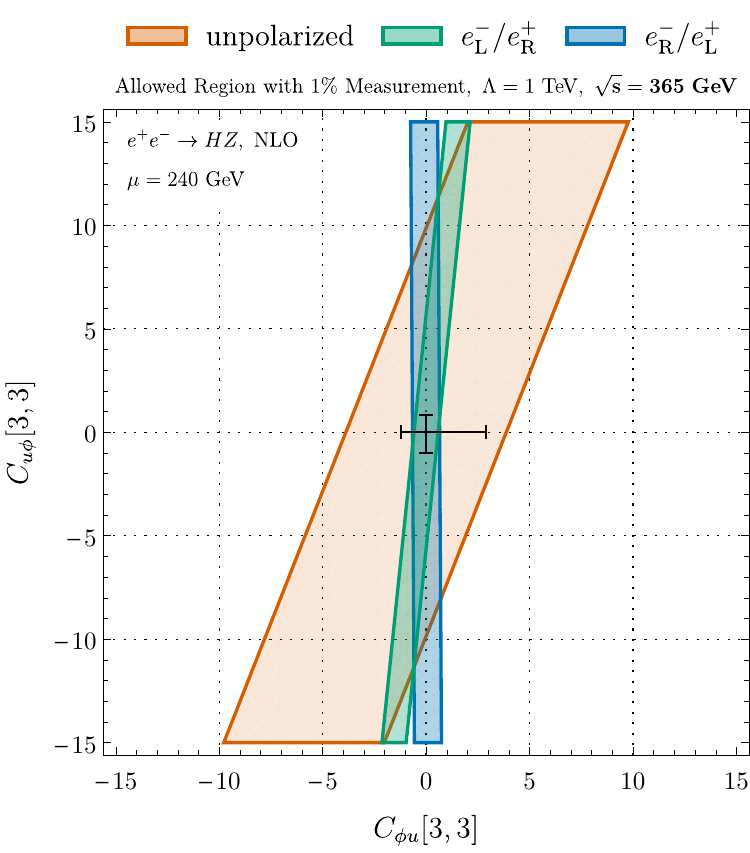}\hspace{5pt}
    \includegraphics[width=.32\textwidth]{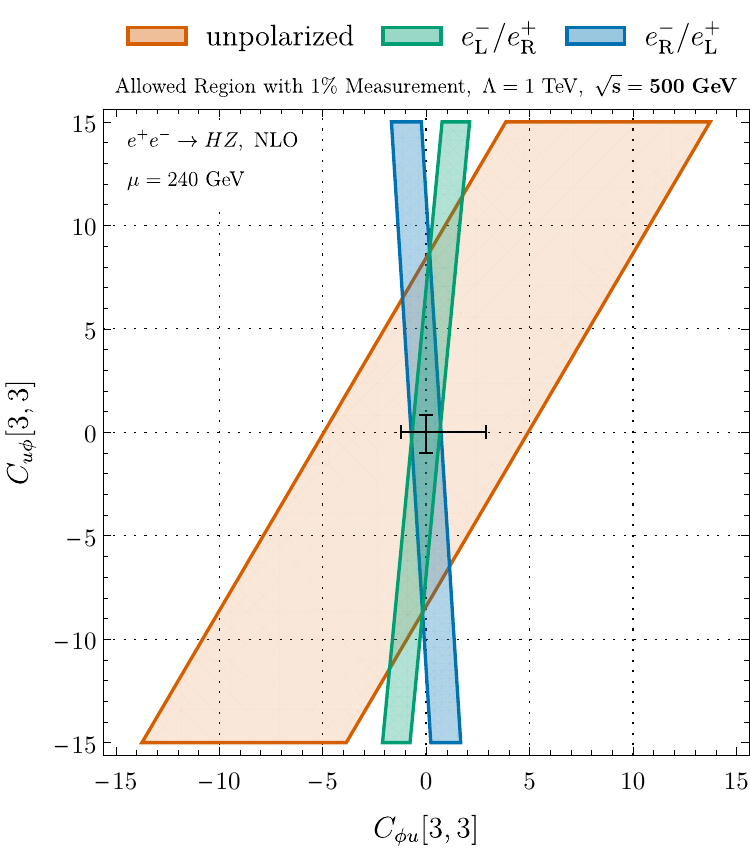}
    \caption{Sensitivity to $C_{\phi u}[3,3]$ and $C_{u\phi}[3,3]$ from measurements of the cross-section for $e^+e^-\rightarrow ZH$.  The sensitivities to  $0.5\%$, $1\%$, and $1\%$ measurements at $\sqrt{s}=240~{\textrm{GeV}}$, $\sqrt{s}=365~{\textrm{GeV}}$, and $\sqrt{s}=500~{\textrm{GeV}}$ are  shown.  Data points show single parameter limits and are taken from Ref.~\cite{Ellis:2020unq}.}
    \label{fg:chtopddd2}
\end{figure}

Finally, we consider the correlation between the 2-fermion operators modifying the top quark Yukawa, $O_{u\phi}[3,3]$, and the $Z t {\overline{t}}$ vertex, $O_{\phi u}[3,3]$. The correlations at $\sqrt{s}=240,~365$ and 
$500~\textrm {GeV}$ are shown in Fig.~\ref{fg:chtopddd2} and the enhancement near the $t {\overline{t}}$ threshold is apparent.  The correlation is the largest at all energies for unpolarized beams. We note that a more accurate calculation would need to include top quark width effects for $\sqrt{s}=365 ~\textrm{GeV}$.

\begin{figure}
    \centering	
    \includegraphics[width=.32\textwidth]{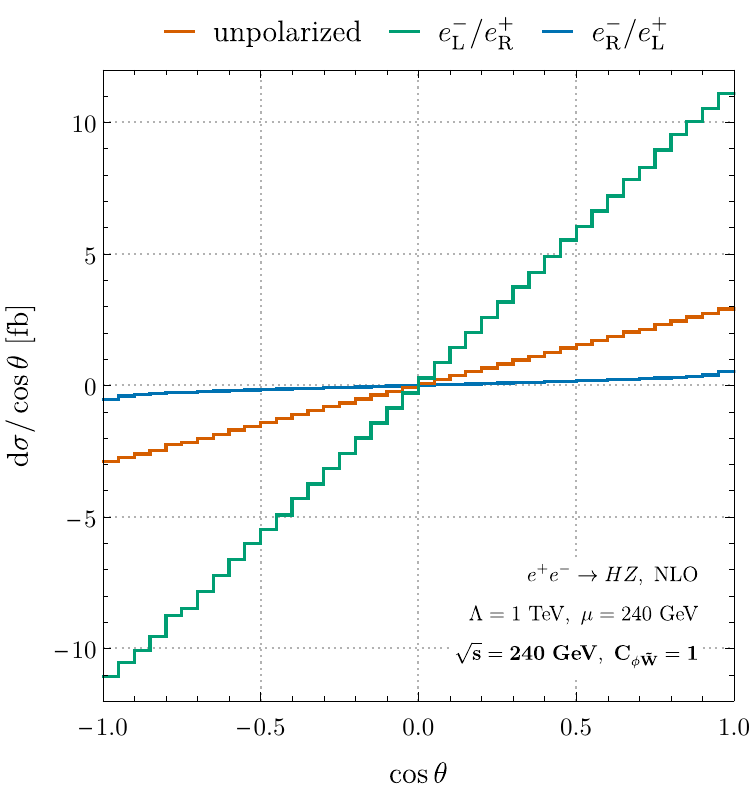}\hspace{5pt}   
    \includegraphics[width=.32\textwidth]{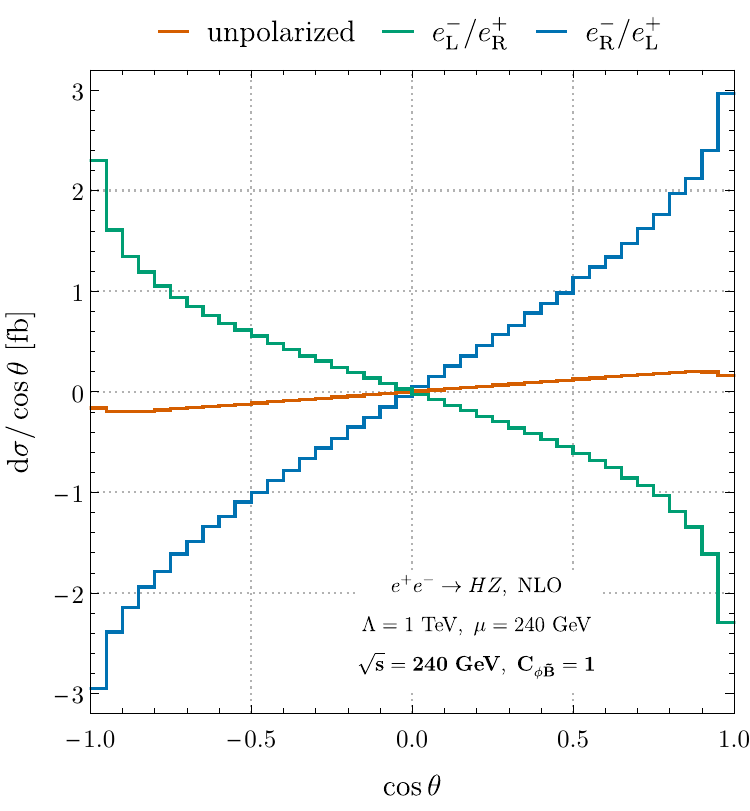}\hspace{5pt}   
    \includegraphics[width=.32\textwidth]{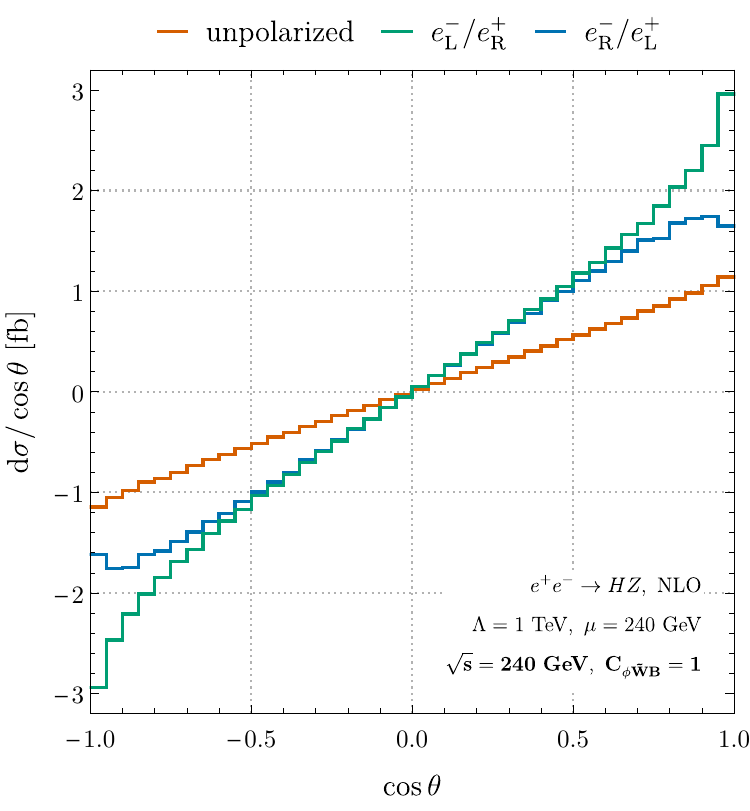}
    \caption{Angular distributions illustrating the effects of CP violating operators in $e^+e^-\rightarrow ZH$ at $\sqrt{s}=240~\textrm{GeV}$. These contributions first arise at NLO  and each plot only contains contributions from a single SMEFT coefficient.}
    \label{histo:th_h_cp_violation}
\end{figure}

At one loop, CP-violating operators can contribute to $e^+e^-\rightarrow ZH$ due to the imaginary contributions of the Passarino-Veltman integrals. (We take the CKM matrix to be diagonal and neglect the CP-violating phase.) 
The contributions from CP-violating operators cancel in inclusive observables~\cite{Beneke:2014sba}, as can be seen in Fig.~\ref{histo:th_h_cp_violation}, and we note that the effects can be ${\cal{O}}(2\%)$ in the distributions. Results for polarized initial states demonstrate that the most prominent effects arise from the $e^+_Le^-_R$ initial state. Results for $\sqrt{s}=350~{\textrm{GeV}}$ and $\sqrt{s}=500~\textrm{GeV}$ are shown in Fig.~\ref{histo:th_h_cp_violation_summary} of Appendix~\ref{sec:numbers} and we see that the relative size of the deviations are similar to those at $\sqrt{s}=240~{\rm GeV}$, although the overall rates are much smaller.  

We form a CP violating asymmetry to parameterize the sensitivity to each operator and summarize the numerical results in Table~\ref{tab:cp},
\begin{align}
    A_{{\rm CP},i} \equiv \frac{C_i(\mu)}{\Lambda^2} \, |\Delta_{i,\rm weak}^{\rm (NLO)}(\cos \theta < 0) - \Delta_{i,\rm weak}^{\rm (NLO)}(\cos \theta > 0)| \, .
    \label{eq:acp}
\end{align}
At NLO, no logarithmically enhanced contributions to $A_{{\rm CP},i}$ arise, and QED corrections start contributing at NNLO.

\begin{figure}
    \centering
    \includegraphics[width=.32\textwidth]{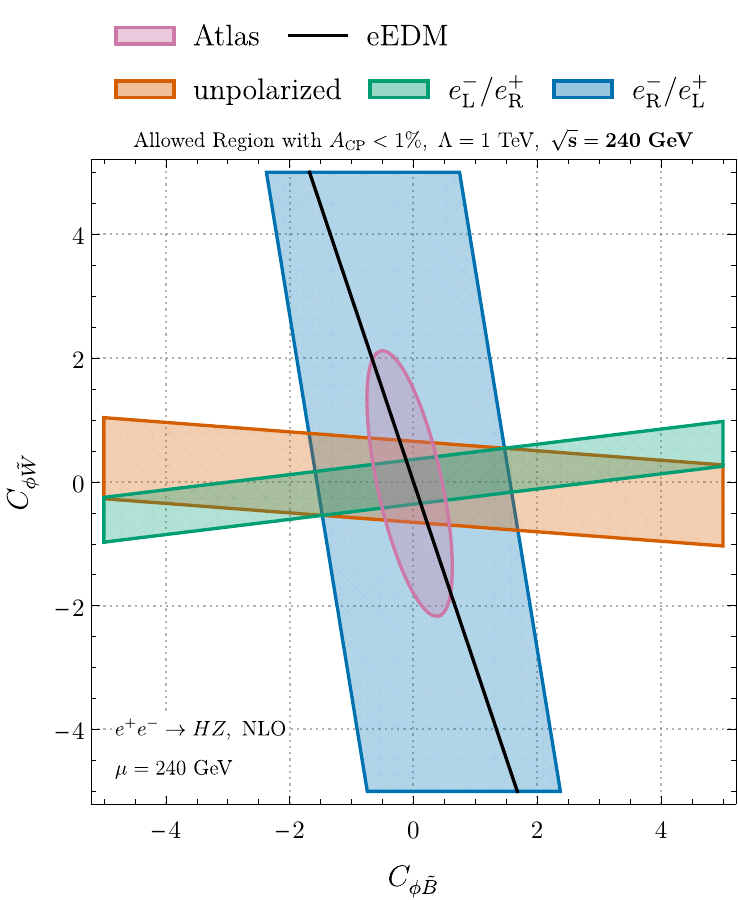}\hspace{5pt}
    \includegraphics[width=.32\textwidth]{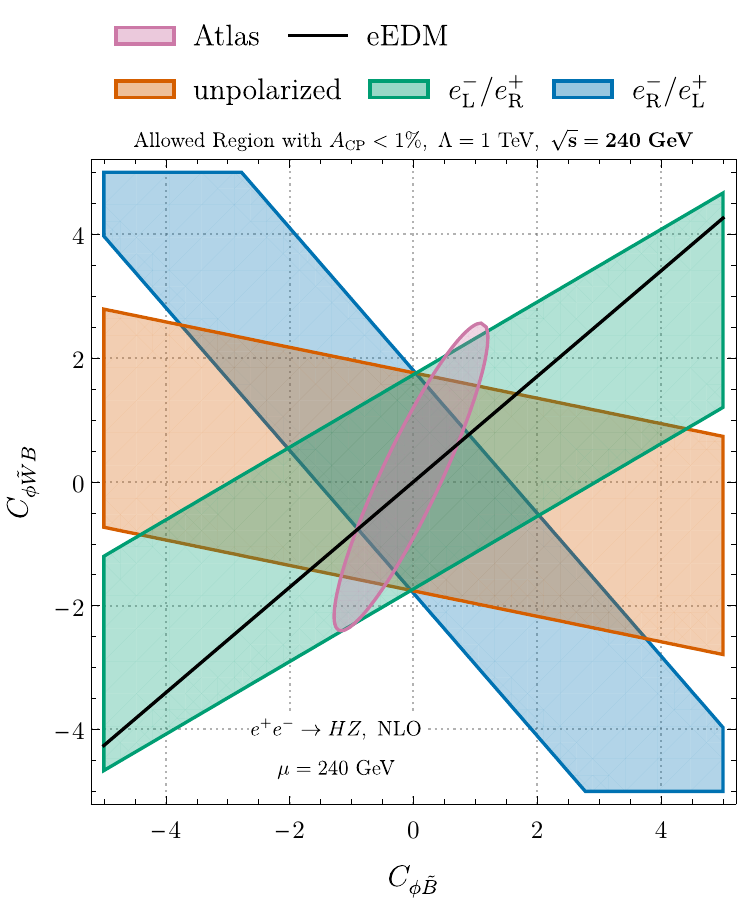}\hspace{5pt}
    \includegraphics[width=.32\textwidth]{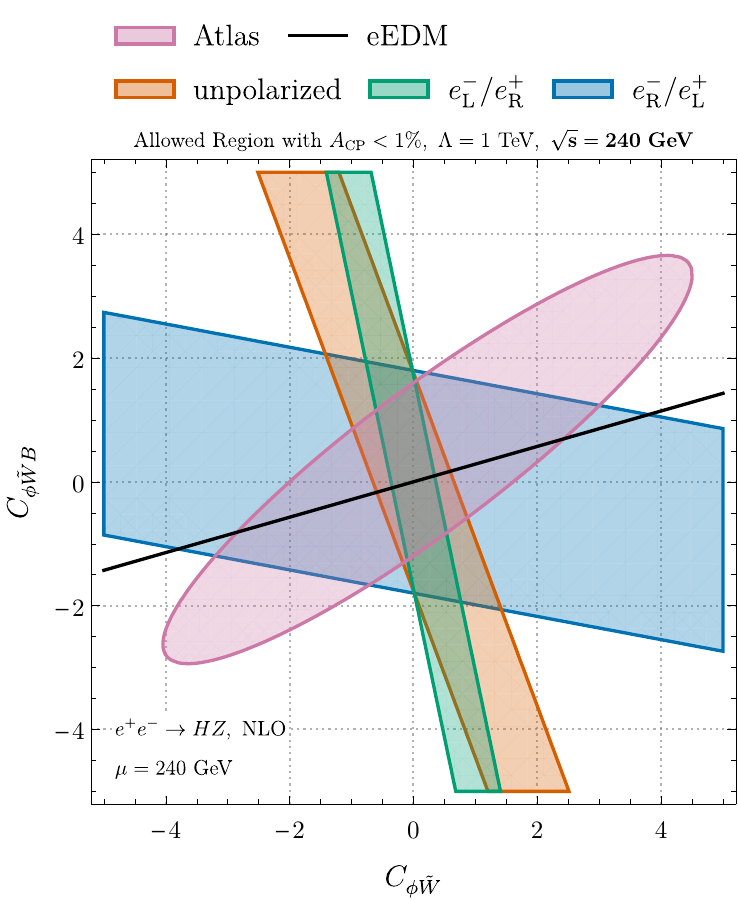}
    \caption{Sensitivity to CP violating dimension-6 operators in $e^+e^-\rightarrow ZH$ at $\sqrt{s}=240~\textrm{GeV}$ assuming a $1\%$ measurement of $A_\textrm{CP}$, c.f. Eq.~\eqref{eq:acp}. The current ATLAS limits from measurements of weak boson fusion with $H\rightarrow 4l$ decay~\cite{ATLAS:2023mqy} and ACME-II limits from measurements of the electron electric dipole moment are also shown. See text for details.}
    \label{histo:cp_violation_all}
\end{figure}

The best limits on the bosonic CP violating coefficient functions $C_{\phi{\tilde{W}}}$, $C_{\phi{\tilde{B}}}$ $C_{\phi{\tilde{W}}B}$  come from electric dipole moments (EDMs). The ACME limit \cite{ACME:2018yjb,ACME:2016sci,ACME:2013pal} on the electron EDM is $| {d_e}/{ e}|< 1.1\times 10^{-29}$.
At tree- level dimension-6 SMEFT~\cite{Panico:2018hal,Kley:2021yhn}, the electric dipole moment, $d_e$, is related to the Wilson coefficients of two operators
\begin{align}
    \label{eq:edm}
    d_e = \sqrt{2}v \ {\rm Im}\bigg\{ \sin\theta_{\rm W} \frac{C_{eW}}{\Lambda^2}-\cos\theta_{\rm W}\frac{C_{eB}}{\Lambda^2}\bigg\} \, ,
\end{align}
where $\cos\theta_{\rm W}=M_{\rm W}/M_{\rm Z}$ and the coefficients $C_{eW}$ and $C_{eB}$ are evaluated at the scale $\mu=M_{\rm W}$.  The CP-violating coefficients,
$C_{\phi{\widetilde W}}$, $C_{\phi{\widetilde B}}$, and $C_{\phi{\widetilde W}B}$ mix through 
 renormalization group running into $C_{eW}$ and $C_{eB}$, and following
\cite{Kley:2021yhn}, we obtain 
\begin{align}
    \frac{d_e}{e}=\biggl( 1.067 \, C_{\phi{\tilde{B}}} + 0.3583 \, C_{\phi {\tilde{W}}} - 1.2533 \, C_{\phi {\tilde{W}}B}\biggl)\biggl(\frac{1~{\rm TeV}}{\Lambda}\biggr)^2\times 10^{-24} \, .
\end{align}

A comparison of sensitivity to CP-violating operators is shown in Fig.~\ref{histo:cp_violation_all}.  Results for Higgsstrahlung, the LHC sensitivity from Higgs decays to 4 leptons in vector boson scattering, and from the electron eDM are shown for comparison.  The correlations in each of these processes are different, demonstrating the importance of Higgsstrahlung in determining the CP violating SMEFT coefficients.

\subsection{Triple Gauge Boson couplings}

\begin{figure}
    \centering
    \includegraphics[width=.32\textwidth]{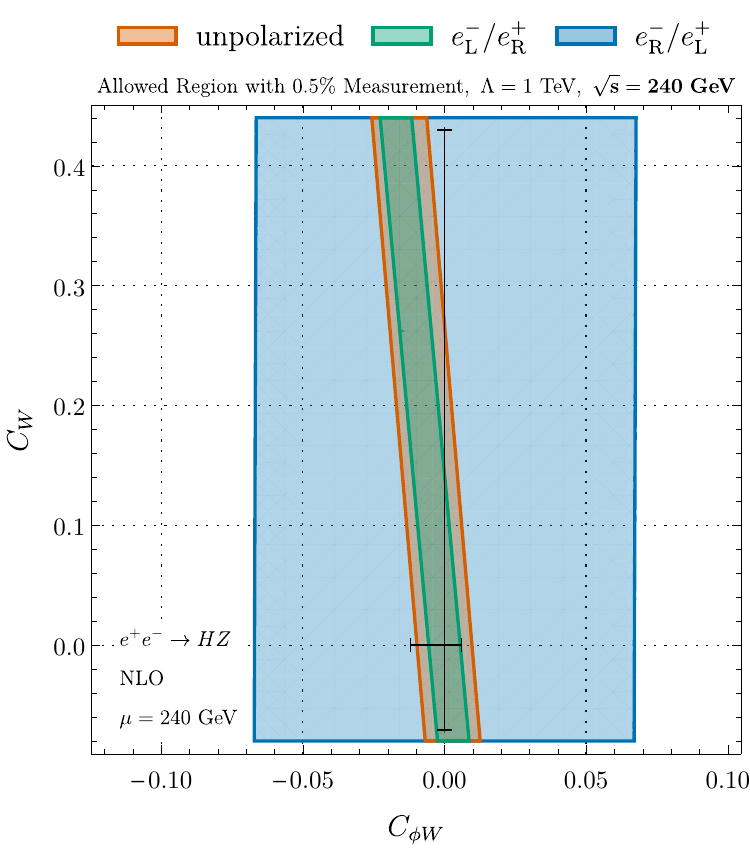}\hspace{5pt}
    \includegraphics[width=.32\textwidth]{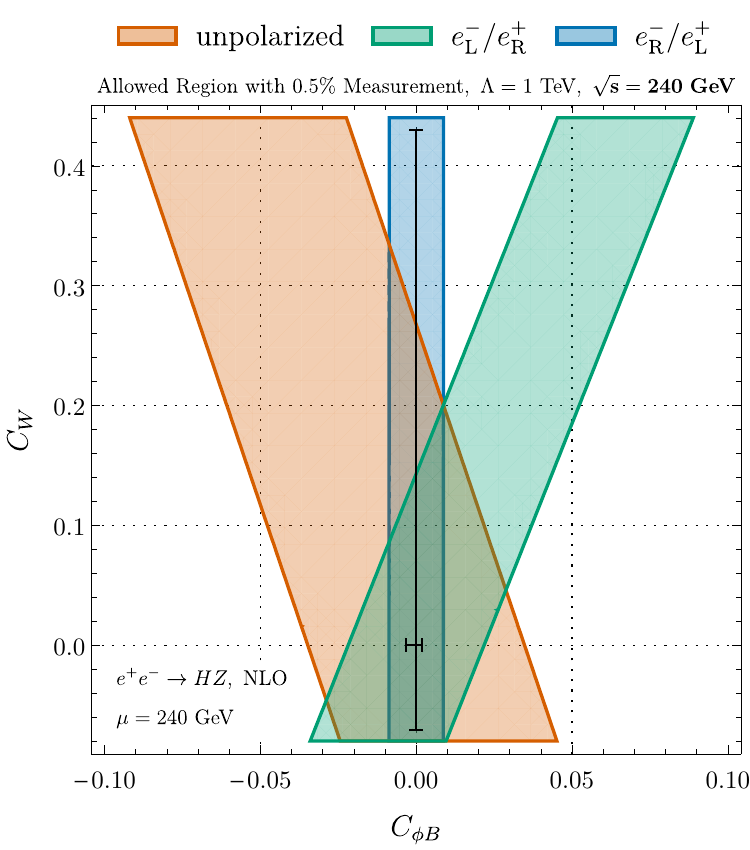}\hspace{5pt}
    \includegraphics[width=.32\textwidth]{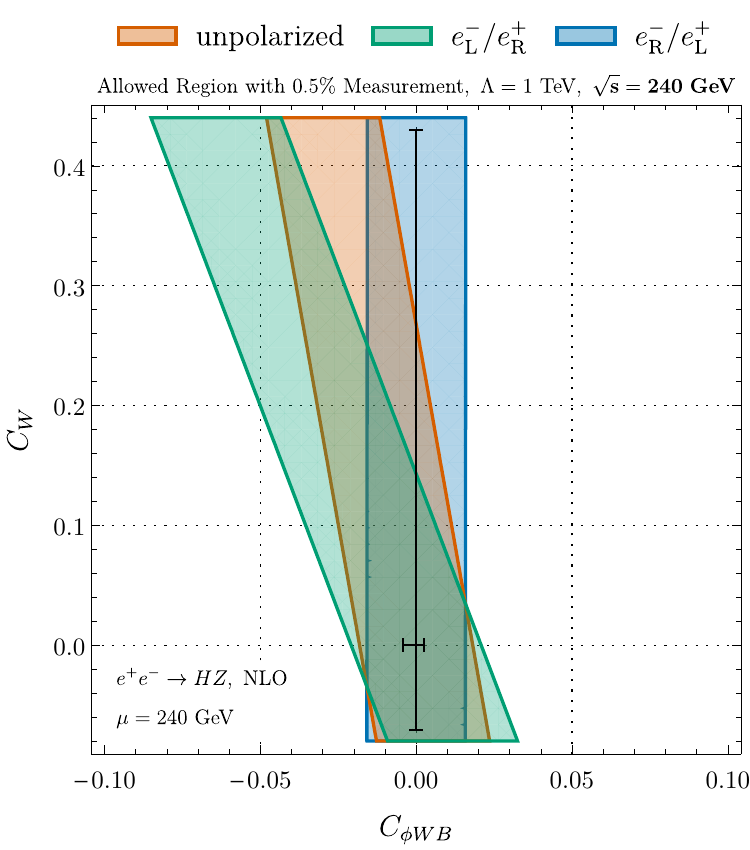}
    \caption{Sensitivity to purely bosonic operators contributing to $e^+e^-\rightarrow ZH$ at NLO. Data points show single parameter limits and are taken from Ref.~\cite{Ellis:2020unq}.}
    \label{fig:CW}
\end{figure}

Measurements of triple gauge boson couplings are a strong test of the structure of the gauge theory. If the couplings differ from their SM values, unitarity is violated in gauge boson scattering, and new physics must exist to restore unitarity. The operator $O_W$ contains non-SM $WWZ$ couplings and is limited from one-loop contributions to Z-pole observables and from Drell Yan production~\cite{Farina:2016rws,Dawson:2021ofa,Dawson:2018dxp}, along with $W^+W^-$ production at both the LHC and future colliders. NLO contributions to Higgsstrahlung can contribute further information about anomalous triple gauge boson couplings.  The CP conserving $WWZ$ vertex depends on the operators $O_W$ and $O_{\phi W B}$, plus the constant shifts generated by the rescaling of the input parameters, Eq.~\eqref{eq:parshift}. The coefficient $C_{\phi WB}$ is strongly constrained by precision electroweak measurements, so we will focus on the effects of $C_W$. Fig.~\ref{fig:CW} shows the sensitivity to the purely bosonic operator $C_W$ correlated with $C_{\phi B}$, $C_{\phi W}$ and $C_{\phi W B}$, along with the current global fit limit at $\sqrt{s}=240~{\textrm GeV}$. We note that for the $e_R^-e_L^+$ initial state, there is no dependence on $C_W$.  The sensitivity to $C_W$ is strongly correlated with $C_{\phi B}$, but there is less correlation with $C_{\phi W B}$. It is clear that Higgsstrahlung has the potential to contribute information about $C_W$ in a future NLO global fit. 

\subsection{Energy Dependence}
\begin{figure}
    \centering
    \includegraphics[width=.32\textwidth]{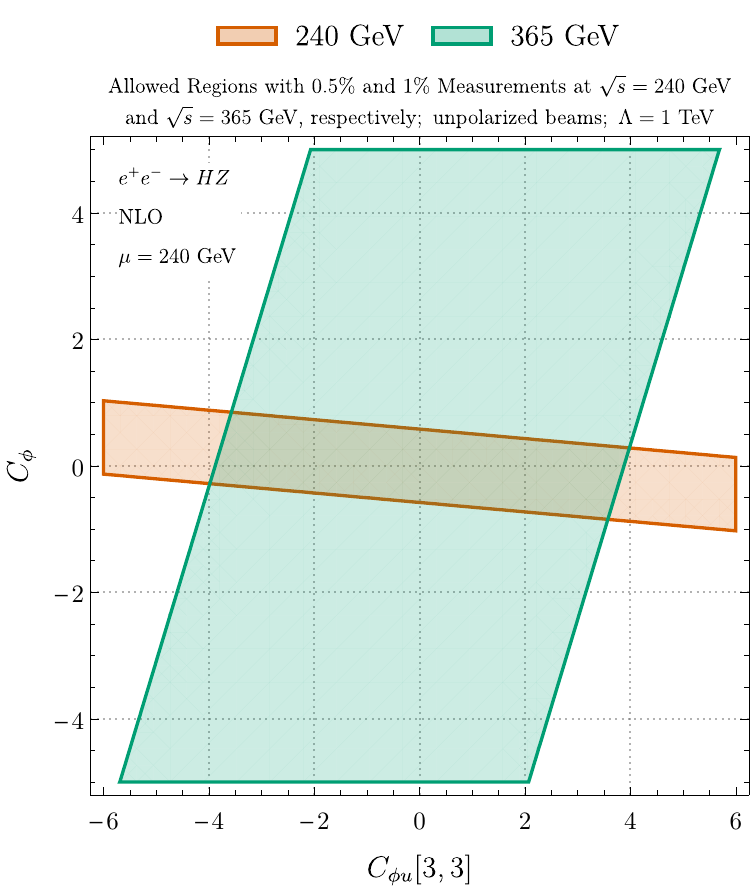}\hspace{5pt}
    \includegraphics[width=.32\textwidth]{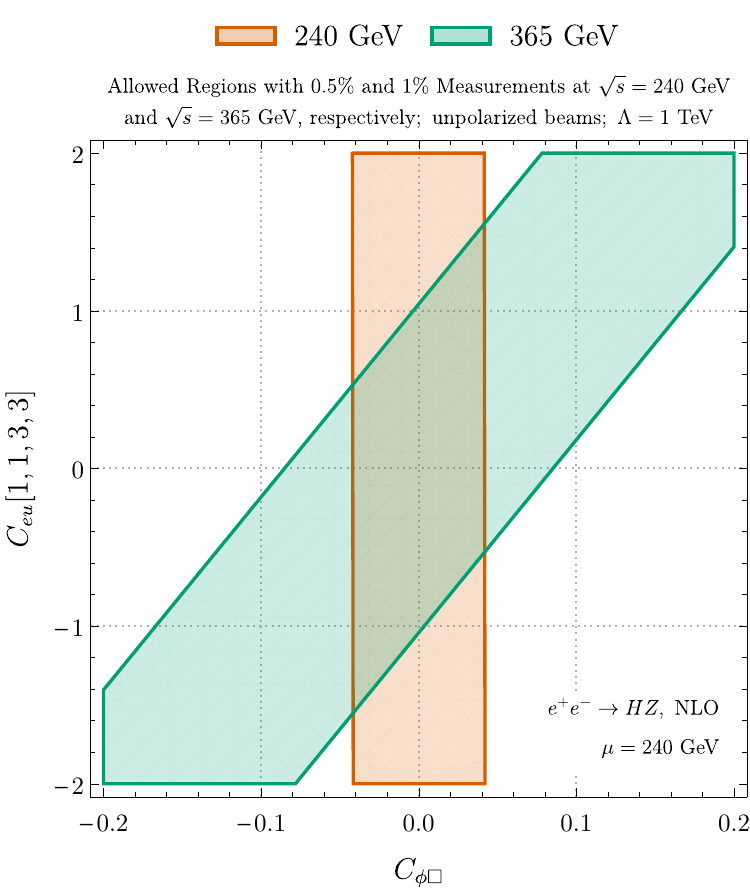}\hspace{5pt}
    \includegraphics[width=.32\textwidth]{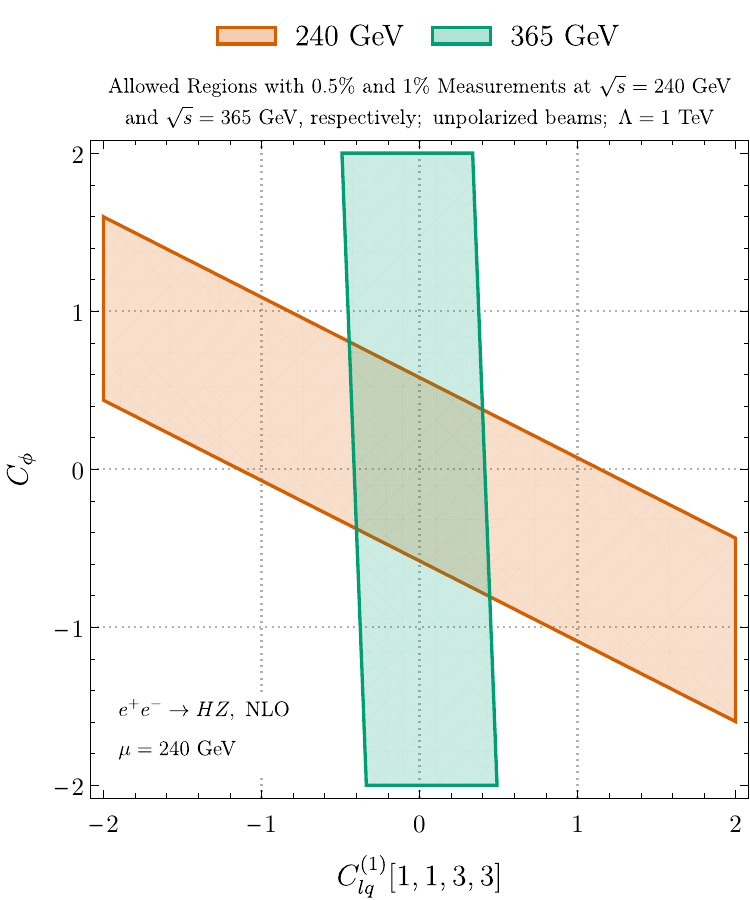}
    \caption{Sensitivity of limits on combinations coefficients to a variation in center of mass energy. All other coefficients are set to $0$. The overlapping green and orange regions can be probed by combining runs at $\sqrt{s}=240$ and $365$ {\textrm{GeV}}.}
    \label{fig:cenergy}
\end{figure}

The sensitivity of the Higgstrahlung process to the SMEFT operators is very dependent on the 
energy for some, (but not all), of the operators. In Fig.~\ref{fig:cenergy}, we demonstrate the impact of combining a run with unpolarized beams at $\sqrt{s}=240$ GeV and $365$ GeV with sensitivities of $0.5\%$ and $1\%$, respectively.  The plots correlate the sensitivity of different operators involving the top quark with $C_\phi$ (which first arises at NLO) and $C_{\phi D}$ (which contributes at tree level).  

\begin{figure}
    \centering
    \includegraphics[width=.32\textwidth]{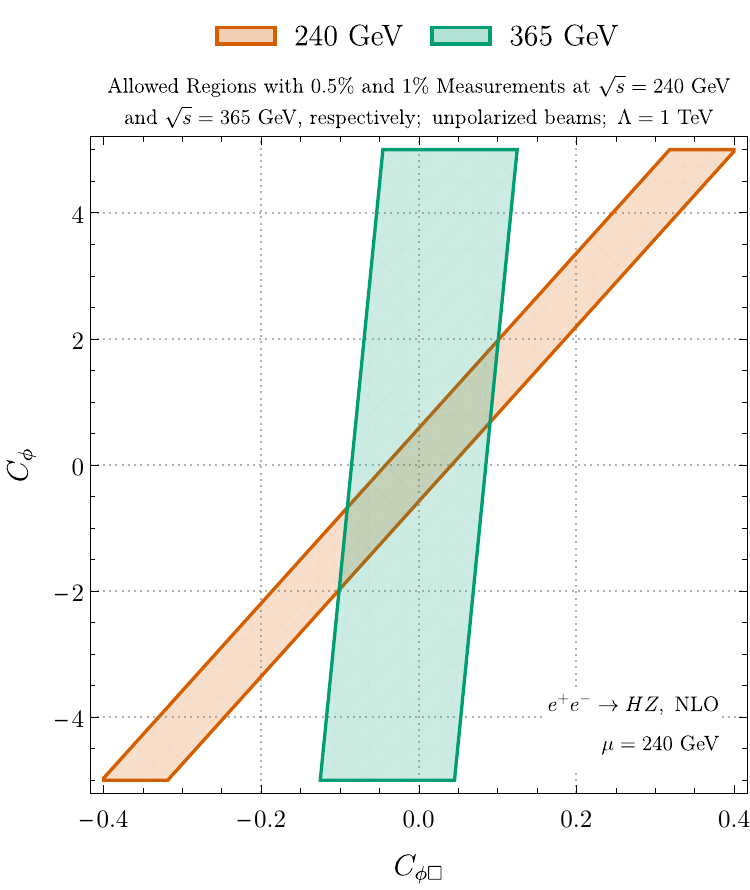}
    \caption{ 
    Sensitivity of limits on $C_{\phi\square}$ and $C_\phi$ to a variation in center of mass energy.  All other coefficients are set to $0$. The overlapping green and orange regions can be probed by combining runs at $\sqrt{s}=240$ and $365$ {\textrm{GeV}}. See Eq. \ref{eq:com} for a correspondence between the SMEFT operators shown here and the $\kappa$ framework.}
    \label{fig:kappaenergy}
\end{figure}

The impact of running at 2 different energies has been demonstrated in Refs.~\cite{Maltoni:2018ttu,DiVita:2017vrr} using the $\kappa$ formalism.  If we assume that only $C_{\phi\square}$ and $C_\phi$ are non-zero, then from Eqs.~(\ref{eq:kdef},~\ref{eq:dzdef}), 
\begin{align}
\begin{split}
    \frac{1}{\sqrt{2}G_\mu\Lambda^2}C_{\phi\square}&\sim  \delta_Z \, , \\
    \frac{1}{\sqrt{2}G_\mu\Lambda^2}C_\phi&\sim -{2\sqrt{2}G_\mu M_H^2}(\kappa_\lambda-1+3\delta_Z)\, ,
    \label{eq:com}
\end{split}
\end{align}
and a comparison with  the results of Refs.~\cite{Maltoni:2018ttu,DiVita:2017vrr} can be made.  In Fig.~\ref{fig:kappaenergy},
we show the limits on $C_{\phi\square}$ and $C_\phi$ that can be obtained by combining measurements at $\sqrt{s}=240$ GeV and $365$
GeV.  Using Eq.~\eqref{eq:com}, we see that the results are qualitatively similar to those of Refs.~\cite{Maltoni:2018ttu,DiVita:2017vrr}. We emphasize, however, that in general, a UV complete model would generate a plethora of SMEFT coefficients beyond $C_{\phi\square}$ and $C_\phi$.

\subsection{On the validity of the expansions}

SMEFT predictions involve two expansions:  The first expansion is in powers of $1/\Lambda$, and the second expansion is the standard perturbative expansion in the number of loops \cite{Buchalla:2022vjp}. As usual, the truncation error provides an estimate of the theoretical uncertainty and determines the effective convergence radius.  
We consistently calculate cross-sections to one-loop and to ${\cal{O}}({1}/{\Lambda^2})$: 
\begin{align}
    \sigma_{\rm NLO} = \sigma^{(0,0)}+\sigma^{(1,0)}+\sigma^{(0,2)}+\sigma^{(1,2)} \, ,
\end{align}
in the notation of Eq.~\eqref{eq:adef}.  
For the expansions to provide a good approximation to the complete results, the terms we have omitted, stemming from $\mathcal{A}^{(0,4)}\sim {1 /\Lambda^4}$,  $\mathcal{A}^{(1,4)}\sim {1 / (16\pi^2\Lambda^4)}$, and $\mathcal{A}^{(2,0)}\sim {1 / (16\pi^2)^2}$ need to be much smaller than the terms we have included. 

\begin{figure}
    \centering  
    \includegraphics[width=.32\textwidth]{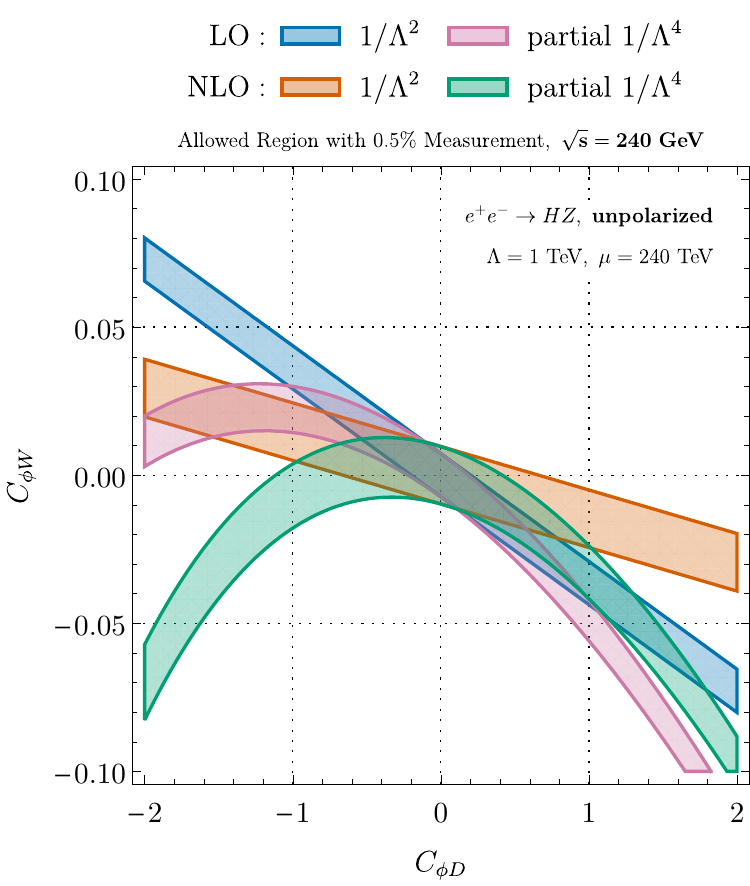}\hspace{5pt}
    \includegraphics[width=.32\textwidth]{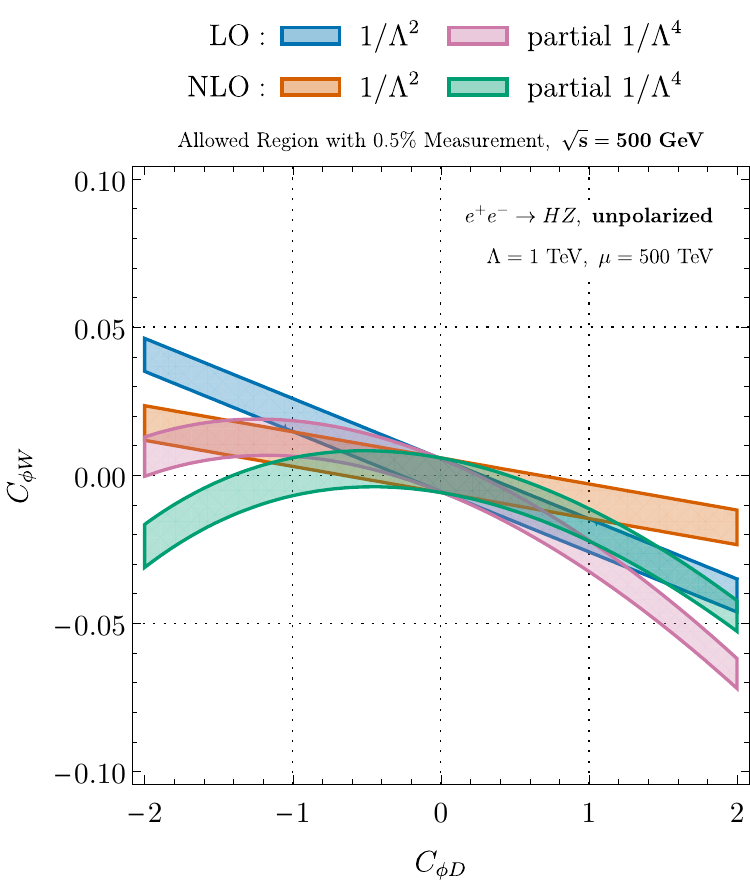}
    \caption{A comparison of the truncation effects for perturbative and SMEFT expansions  at $\sqrt{s}=240$ and $500$~{\textrm{GeV}}. Both figures show the effects of a $0.5\%$ deviation from the SM. See text for details.} 
    \label{fig:valid}
\end{figure}

In the SM, the 2-loop contribution to the cross-section for $e^+e^- \to HZ$ is significantly smaller than the NLO piece in the energy range we consider, thus justifying our neglect of the two loop effects,  $\mathcal{A}^{(2,0)}$. A calculation of $A^{(0,4)}$ requires the input parameter relations of Eq.~\eqref{eq:adef} evaluated to ${\cal{O}}({1}/{\Lambda^4})$, along with double insertions of dimension-6 operators and a single insertion of dimension-8 operators.  
To gauge the size of missing terms, we approximate $\sigma^{(0,4)}$ by including only the contribution of $|\mathcal{A}^{(0,2)}|^2$, 
labeled as "LO partial $1/\Lambda^4$" in Fig.~\ref{fig:valid}.  We emphasize that the effect of the omitted $\mathcal{O}(1/\Lambda^4)$ terms could be larger.

The region where the pink and the blue curves begin to diverge in this figure is the region where a full ${\cal{O}}({1/\Lambda^4})$ calculation is warranted.  In a similar vein, we have included the partial ${1}/{\Lambda^4}$ results with the NLO calculation (as a proxy for $\sigma^{(1,4)}$), which is depicted by the green curve in Fig.~\ref{fig:valid}. Again, the region where the green and orange curves diverge is the region where EFT expansion in ${1}/{\Lambda}$ may not be truncated at dimension-6, and further studies are required.  We show the effects of a $0.5\%$ variation at both $\sqrt{s}=240$ and $500$ {\textrm{GeV}}.  In this specific example, the effects of the quadratic terms does not show a large energy dependence.

\section{Conclusion}
\label{sec:con}
In this work, we have investigated the $e^+e^- \to HZ $ process within the framework of SMEFT at the next-to-leading order in the electroweak expansion and to ${\cal{O}}({1/ \Lambda^2})$ in the SMEFT expansion. Our NLO effort has been motivated by the need to explore, in a model-agnostic way, the impact of new physics in precision measurements at future electron-positron colliders, such as the FCC-ee, CEPC, ILC, and CLIC. These proposed Higgs factories will allow unprecedented precision in tests of the Higgs boson properties.

Through the NLO analysis presented here, we have extended previous partial results that included only a small subset of operators by incorporating the complete set of dimension-six operators contributing to the next-to-leading order corrections. 
Except for the contribution from the Higgs tri-linear coupling, $O_\phi$, current sensitivity studies for future $e^+e^-$ colliders utilize LO SMEFT predictions \cite{deBlas:2019rxi,deBlas:2022ofj,Celada:2024mcf}.
In particular, our results demonstrate that limits on the Higgs tri-linear coupling are highly correlated with the contributions of other operators, including those that first occur at NLO.  

Our results demonstrate that including dimension-six operators beyond LO can significantly alter the total cross-section for the 
$e^+e^- \to HZ $ process and provide sensitivity to many new operators. 
Bounds, or sensitivity studies, including only a single parameter, are typically highly misleading at NLO and oversimplify the complex structure of the complete NLO result. 
We note significant correlations exist between the contributions of different operators at both LO and NLO for most operator combinations. Nonetheless, even at NLO, there are still many flat directions in the parameter space, which can be resolved only by a global analysis that goes beyond the study of the $e^+e^- \to HZ $ process.
The dependence of the Higgsstrahlung cross-section on SMEFT operators at NLO has, in general, a strong energy dependence, and our results demonstrate that measurements at multiple energies are crucial for disentangling the effects of correlated operators and for probing operator contributions along flat directions. 
The NLO corrections to the total cross-section are not small and differ from the RGE-improved LO SMEFT results, indicating that higher-order electroweak SMEFT corrections should not be neglected in sensitivity studies for future $e^+e^-$ colliders. Including NLO corrections in SMEFT predictions will be essential for an accurate interpretation of experimental data at future Higgs factories and could provide further sensitivity to deviations from the SM, potentially signaling the presence of new physics.

We have separated the purely weak contributions from the QED corrections to aid in the future inclusion of our results in Monte Carlo studies. Our numerical results can be easily implemented in global fit programs with arbitrary scale dependence. These are available for download~\cite{GITLAB:eehz}.

A more comprehensive study across a broader range of processes beyond $e^+e^- \to HZ $ will be essential to gain a full understanding of the reach that future collider observables can have in constraining  SMEFT operators. Theoretical results now exist at NLO SMEFT for both Higgsstrahlung, as presented in this work, and for $Z$ pole observables. A concerted theory effort is needed to obtain results at NLO SMEFT for the remaining processes, contributing to sensitivity studies for future colliders. The SMEFT framework is essential for this purpose as it can be easily interfaced in a theoretically consistent manner with SM predictions to arbitrary order.

The ongoing development of future precision programs at colliders, combined with accurate theoretical predictions, will enable us to explore new territories in particle physics. Continued efforts in this direction will be vital for uncovering possible signs of physics beyond the Standard Model, contributing to a more complete understanding of the fundamental forces in nature.

\section*{Acknowledgements}

We are grateful to A. Freitas for helpful discussion.
K.A. thanks Brookhaven National Laboratory, where a significant portion of this research was conducted. 
S. D. and R.S.  are supported by the U.S. Department of Energy under Grant Contract~DE-SC0012704. Digital data is provided in the ancillary file.
P.P.G. is supported by the Ramón y Cajal grant~RYC2022-038517-I funded by MCIN/AEI/10.13039/501100011033 and by FSE+, and by the Spanish Research Agency (Agencia Estatal de Investigación) through the grant IFT Centro de Excelencia Severo Ochoa~No~CEX2020-001007-S. 

\bibliography{eezh.bib}

\appendix
\label{sec:appen}

\section{One-Loop Operator Combinations}
\label{sec:combo}

At NLO, new operators enter the virtual calculation that do not contribute at tree level.
At one-loop, the virtual amplitude depends on the following coefficients that contribute independently,
\begin{align}
\begin{split}
    &C_{\phi e}[1,1],~C_{\phi l}^{(1)}[1,1],~C_{\phi l}^{(3)}[1,1],~C_{\phi l}^{(1)}[2,2],~C_{\phi l}^{(3)}[2,2],~C_{\phi u}[3,3],  \\ 
    &C_{\phi q}^{(1)}[3,3],~C_{\phi q}^{(3)}[3,3],~C_{uW}[3,3],~C_{uB}[3,3],~C_{u\phi}[3,3] \, .
\end{split}
\end{align}
In addition, the following 
combinations of  2-fermion coefficients contribute:
\begin{align}
\begin{split}
    {\hat{C}}^{2f}_1&= C_{\phi e }[2,2]+C_{\phi e }[3,3]+
    C_{\phi d }[1,1]+C_{\phi d}[2,2]+C_{\phi d }[3,3]\nonumber \\ 
    &\hspace{10pt} -2 C_{\phi u}[1,1] -2 C_{\phi u}[2,2]
    -C_{\phi q}^{(1)}[1,1] -C_{\phi q}^{(1)}[2,2] +C_{\phi l}^{(1)}[3,3]
    \nonumber \\
    {\hat{C}}^{2f}_2&= C_{\phi l}^{(3)}[3,3]+3C_{\phi q}^{(3)}[1,1] +3C_{\phi q}^{(3)}[2,2]
    \, . 
\end{split}
\end{align}

For the 4-fermion operators,
the following coefficients contribute independently beginning at NLO,
\begin{align}
\begin{split}
    &C_{eu}[1,1,3,3],~C_{qe}[3,3,1,1],~C_{lu}[1,1,3,3],~C_{lq}^{(1)}[1,1,3,3],~C_{lq}^{(3)}[1,1,3,3], \\
    &C_{lq}^{(3)}[2,2,3,3],~C_{le}[1,1,1,1],~C_{ll}[1,1,1,1],~C_{ee}[1,1,1,1] \, . 
\end{split}
\end{align}
 There are an additional 6 combinations of coefficients that  first contribute at NLO:
\begin{align}
\begin{split}
    \hat{C}^{4f}_1 &= C_{ld}[1,1,1,1]+C_{ld}[1,1,2,2]+C_{ld}[1,1,3,3]-2\left(C_{lu}[1,1,1,1]+C_{lu}[1,1,2,2]\right)  \\ 
    &\hspace{10pt}-C_{lq}^{(1)}[1,1,1,1]-C_{lq}^{(1)}[1,1,2,2]+C_{ll}[1,1,3,3]+C_{ll}[3,3,1,1]  \\ 
    &\hspace{10pt}+C_{le}[1,1,2,2]+C_{le}[1,1,3,3] -\frac{3M_W^2}{(M_Z^2-M_W^2)}\left(C_{lq}^{(3)}[1,1,1,1]+C_{lq}^{(3)}[1,1,2,2]\right) \, ,  \\ 
    \hat{C}^{4f}_2 &= C_{ll}[1,2,2,1]+C_{ll}[2,1,1,2] \, ,  \\
    \hat{C}^{4f}_3 &= C_{ll}[1,1,2,2]+C_{ll}[2,2,1,1] \, ,  \\
    \hat{C}^{4f}_4 &= C_{ll}[1,3,3,1]+C_{ll}[3,1,1,3] \, ,  \\
    \hat{C}^{4f}_{5} &=C_{ee}[1,1,2,2]+C_{ee}[1,1,3,3]+C_{ee}[2,2,1,1]+C_{ee}[3,3,1,1] \\
    &\hspace{10pt}- C_{qe}[1,1,1,1]-C_{qe}[2,2,1,1]-2\left(C_{eu}[1,1,1,1]+C_{eu}[1,1,2,2]\right) \\ 
    &\hspace{10pt}+ C_{ed}[1,1,1,1] + C_{ed}[1,1,2,2]+C_{ed}[1,1,3,3]+C_{le}[2,2,1,1]+C_{le}[3,3,1,1] \, ,  \\
    \hat{C}^{4f}_{6} &= C_{ee}[1,2,2,1]+C_{ee}[2,1,1,2]+C_{ee}[1,3,3,1]+C_{ee}[3,1,1,3] \, .
\end{split}
\end{align}

\section{Numerical Results and Supplemental Plots}
\label{sec:numbers}

\begin{table}[H]
    \centering
    \caption{SM results for the total cross-section for $e^+e^-\rightarrow ZH$ for unpolarized beams and left- and right-handed electron polarizations.  The contributions are separated as defined in Eq.~(\ref{eq:sigpar}). }
    \vspace{7pt}
\begin{tabular}{ccccccc}
        \hline\hline\\[-2.5ex]
        & $\sqrt{s}$ [GeV]
        & $\sigma^\textrm{LO}_\textrm{SM}$ [fb]
        & $\Delta\sigma^\textrm{NLO}_\textrm{SM,weak}$ [fb]
        & $\Delta\sigma^\textrm{NLO}_\textrm{SM,QED}$ [fb]
        & $\sigma^\textrm{NLO}_\textrm{SM,weak}$ [fb]
        & $\sigma^\textrm{NLO}_\textrm{SM}$ [fb] \\
        \hline
        & $240$ & 239.2 & -7.1 & -37.7 & 232.1 & 194.4  \\
        unpolarized & $365$ & 116.9 & -3.7& +7.9 & 113.2 & 121.1 \\
        & $500$ & 57.1  & -2.2 & +8.2& 54.9 & 63.1  \\
        \hline
        & $240$ & 580.7 & -73.0 & -91.5 & 507.6 & 416.1 \\
        left & $365$ & 283.9 & -35.9 & 19.1 & 247.9 & 267.0 \\
        & $500$ & 138.5 & -19.7 & 19.9 & 118.8 & 138.7 \\
        \hline
        & $240$ & 376.1 & 44.6 & -59.3 & 420.7 & 361.4 \\
        right & $365$ & 183.8 & 21.1 & 12.3 & 205.0 & 217.3 \\
        & $500$ & 89.7 & 11.1 & 12.9 & 100.8 & 113.7 \\
        \hline\hline
    \end{tabular}
    \label{tab:results:sm}
\end{table}

\begin{table}[H]
    \centering
    \caption{Effects on the total cross-section for $e^+e^- \rightarrow ZH$  from operators that occur at tree level.  The IR finite contribution must be added to these numbers to get the full NLO cross-sections. The notation is defined in Eq.~(\ref{eq:sigpar}).}
    \vspace{7pt}
    \begin{tabular}{cccccccc}
        \hline\hline\\[-2.5ex]
        & \hspace{-7pt}\parbox[c]{1cm}{\centering $\sqrt{s}$ \\ {[GeV]} \\[-6pt] \hspace{10pt}}\hspace{-7pt}
        & $\Delta_{i,\rm weak}^{\rm (NLO),L} / \Lambda^2$    
        & $\bar\Delta_i^\textrm{L} /\Lambda^2$ 
        & $\Delta_{i,\rm weak}^{\rm (NLO),R} / \Lambda^2$    
        & $\bar\Delta_i^\textrm{R} / \Lambda^2$ 
        & $\Delta^{\rm (NLO)}_{i,\rm weak} / \Lambda^2$    
        & $\bar\Delta_{i} / \Lambda^2$ \\
        \hline 
        & $240$ & $\hphantom{\textrm{-}} 1.81 \cdot 10^{\textrm{-}1}$ & $\textrm{-}9.33 \cdot 10^{\textrm{-}3}$ & $\textrm{-}1.98 \cdot 10^{\textrm{-}1}$ & $\hphantom{\textrm{-}} 8.79 \cdot 10^{\textrm{-}3}$ & $\hphantom{\textrm{-}} 9.27 \cdot 10^{\textrm{-}3}$ & $\textrm{-}1.12 \cdot 10^{\textrm{-}3}$ \\
$C_{\phi D}$ & $365$ & $\hphantom{\textrm{-}} 1.73 \cdot 10^{\textrm{-}1}$ & $\textrm{-}9.19 \cdot 10^{\textrm{-}3}$ & $\textrm{-}1.92 \cdot 10^{\textrm{-}1}$ & $\hphantom{\textrm{-}} 8.53 \cdot 10^{\textrm{-}3}$ & $\hphantom{\textrm{-}} 7.75 \cdot 10^{\textrm{-}3}$ & $\textrm{-}1.17 \cdot 10^{\textrm{-}3}$ \\
& $500$ & $\hphantom{\textrm{-}} 1.68 \cdot 10^{\textrm{-}1}$ & $\textrm{-}9.13 \cdot 10^{\textrm{-}3}$ & $\textrm{-}1.88 \cdot 10^{\textrm{-}1}$ & $\hphantom{\textrm{-}} 8.03 \cdot 10^{\textrm{-}3}$ & $\hphantom{\textrm{-}} 4.83 \cdot 10^{\textrm{-}3}$ & $\textrm{-}1.25 \cdot 10^{\textrm{-}3}$ \\
\hline 
& $240$ & $\hphantom{\textrm{-}} 1.21 \cdot 10^{\textrm{-}1}$ & $\textrm{-}6.45 \cdot 10^{\textrm{-}3}$ & $\hphantom{\textrm{-}} 1.22 \cdot 10^{\textrm{-}1}$ & $\textrm{-}4.16 \cdot 10^{\textrm{-}3}$ & $\hphantom{\textrm{-}} 1.22 \cdot 10^{\textrm{-}1}$ & $\textrm{-}5.41 \cdot 10^{\textrm{-}3}$ \\
\hline\hline
  
    \end{tabular}
    \label{tab:0fops}
\end{table}

\begin{table}[H]
    \centering
    \caption{Effects on the unpolarized total cross-section for $e^+e^- \rightarrow ZH$ from operators that occur at tree level. The notation is defined in Eq.~(\ref{eq:sigpar}).}
    \vspace{7pt}
    \begin{tabular}{cccccccc}
        \hline\hline\\[-2.5ex]
        Unpolarized
        & \hspace{-7pt}\parbox[c]{1cm}{\centering $\sqrt{s}$ \\ {[GeV]} \\[-6pt] \hspace{10pt}}\hspace{-7pt}
        & $\Delta_i^{\rm (LO)} / \Lambda^2$    
        & $\Delta_{i,\rm weak}^{(\delta \rm NLO)} /\Lambda^2$ 
        & $\Delta_{i,\rm QED} / \Lambda^2$    
        & $\Delta_{i}^{\rm (NLO)} / \Lambda^2$ 
        & $\bar\Delta_{i} / \Lambda^2$ \\
        \hline 
        & $240$ & $\hphantom{\textrm{-}} 2.08 \cdot 10^{\textrm{-}2}$ & $\textrm{-}1.15 \cdot 10^{\textrm{-}2}$ & $\textrm{-}2.22 \cdot 10^{\textrm{-}3}$ & $\hphantom{\textrm{-}} 7.06 \cdot 10^{\textrm{-}3}$ & $\textrm{-}1.12 \cdot 10^{\textrm{-}3}$ \\
$C_{\phi D}$ & $365$ & $\hphantom{\textrm{-}} 2.08 \cdot 10^{\textrm{-}2}$ & $\textrm{-}1.31 \cdot 10^{\textrm{-}2}$ & $\hphantom{\textrm{-}} 2.59 \cdot 10^{\textrm{-}3}$ & $\hphantom{\textrm{-}} 1.03 \cdot 10^{\textrm{-}2}$ & $\textrm{-}1.17 \cdot 10^{\textrm{-}3}$ \\
& $500$ & $\hphantom{\textrm{-}} 2.09 \cdot 10^{\textrm{-}2}$ & $\textrm{-}1.61 \cdot 10^{\textrm{-}2}$ & $\hphantom{\textrm{-}} 4.42 \cdot 10^{\textrm{-}3}$ & $\hphantom{\textrm{-}} 9.25 \cdot 10^{\textrm{-}3}$ & $\textrm{-}1.25 \cdot 10^{\textrm{-}3}$ \\
\hline 
& $240$ & $\hphantom{\textrm{-}} 1.25 \cdot 10^{\textrm{-}1}$ & $\textrm{-}3.44 \cdot 10^{\textrm{-}3}$ & $\textrm{-}1.97 \cdot 10^{\textrm{-}2}$ & $\hphantom{\textrm{-}} 1.02 \cdot 10^{\textrm{-}1}$ & $\textrm{-}5.41 \cdot 10^{\textrm{-}3}$ \\
$C_{\phi \square}$ & $365$ & $\hphantom{\textrm{-}} 1.25 \cdot 10^{\textrm{-}1}$ & $\textrm{-}1.19 \cdot 10^{\textrm{-}2}$ & $\hphantom{\textrm{-}} 8.41 \cdot 10^{\textrm{-}3}$ & $\hphantom{\textrm{-}} 1.22 \cdot 10^{\textrm{-}1}$ & $\textrm{-}5.75 \cdot 10^{\textrm{-}3}$ \\
& $500$ & $\hphantom{\textrm{-}} 1.26 \cdot 10^{\textrm{-}1}$ & $\textrm{-}1.84 \cdot 10^{\textrm{-}2}$ & $\hphantom{\textrm{-}} 1.81 \cdot 10^{\textrm{-}2}$ & $\hphantom{\textrm{-}} 1.26 \cdot 10^{\textrm{-}1}$ & $\textrm{-}6.30 \cdot 10^{\textrm{-}3}$ \\
\hline\hline
 
    \end{tabular}
    \label{tab:0fopsfin} 
\end{table}

\begin{table}[H]
    \centering
     \caption{Effects on the left-polarized cross-section for $e^+e^- \rightarrow ZH$  from operators that occur at tree level. The notation is defined in Eq.~(\ref{eq:sigpar}).}  
    \vspace{7pt}
    \begin{tabular}{cccccccc}
        \hline\hline\\ [-2.5ex]
        Left-polarized
        & \hspace{-7pt}\parbox[c]{1cm}{\centering $\sqrt{s}$ \\ {[GeV]} \\[-6pt] \hspace{10pt}}\hspace{-7pt}
        &  $\Delta_i^{\rm (LO)} / \Lambda^2$    
        & $\Delta_{i,\rm weak}^{(\delta \rm  NLO)} /\Lambda^2$ 
        & $\Delta_{i,\rm QED}^{\rm (NLO)} / \Lambda^2$    
        & $\Delta_{i}^{\rm (NLO)} / \Lambda^2$ 
        & $\bar\Delta_{i} / \Lambda^2$ \\
        \hline
        & $240$ & $\hphantom{\textrm{-}} 1.94 \cdot 10^{\textrm{-}1}$ & $\textrm{-}1.36 \cdot 10^{\textrm{-}2}$ & $\textrm{-}2.95 \cdot 10^{\textrm{-}2}$ & $\hphantom{\textrm{-}} 1.51 \cdot 10^{\textrm{-}1}$ & $\textrm{-}9.33 \cdot 10^{\textrm{-}3}$ \\
$C_{\phi D}$ & $365$ & $\hphantom{\textrm{-}} 1.95 \cdot 10^{\textrm{-}1}$ & $\textrm{-}2.16 \cdot 10^{\textrm{-}2}$ & $\hphantom{\textrm{-}} 1.44 \cdot 10^{\textrm{-}2}$ & $\hphantom{\textrm{-}} 1.88 \cdot 10^{\textrm{-}1}$ & $\textrm{-}9.19 \cdot 10^{\textrm{-}3}$ \\
& $500$ & $\hphantom{\textrm{-}} 1.98 \cdot 10^{\textrm{-}1}$ & $\textrm{-}2.98 \cdot 10^{\textrm{-}2}$ & $\hphantom{\textrm{-}} 3.01 \cdot 10^{\textrm{-}2}$ & $\hphantom{\textrm{-}} 1.99 \cdot 10^{\textrm{-}1}$ & $\textrm{-}9.13 \cdot 10^{\textrm{-}3}$ \\
\hline 
& $240$ & $\hphantom{\textrm{-}} 1.39 \cdot 10^{\textrm{-}1}$ & $\textrm{-}1.77 \cdot 10^{\textrm{-}2}$ & $\textrm{-}2.19 \cdot 10^{\textrm{-}2}$ & $\hphantom{\textrm{-}} 9.91 \cdot 10^{\textrm{-}2}$ & $\textrm{-}6.45 \cdot 10^{\textrm{-}3}$ \\
$C_{\phi \square}$ & $365$ & $\hphantom{\textrm{-}} 1.39 \cdot 10^{\textrm{-}1}$ & $\textrm{-}2.73 \cdot 10^{\textrm{-}2}$ & $\hphantom{\textrm{-}} 9.32 \cdot 10^{\textrm{-}3}$ & $\hphantom{\textrm{-}} 1.21 \cdot 10^{\textrm{-}1}$ & $\textrm{-}6.82 \cdot 10^{\textrm{-}3}$ \\
\hline\hline
 
    \end{tabular}
    \label{tab:0fopsL}	
\end{table}

\begin{table}[H]
    \centering
    \caption{\label{tab:0fopsR} Effects on the right-polarized cross-section for $e^+e^- \rightarrow ZH$ from operators that occur at tree level. The notation is defined in Eq.~(\ref{eq:sigpar}).}  
    \vspace{7pt}
    \begin{tabular}{cccccccc}
        \hline\hline\\ [-2.5ex]
        Right-polarized
        & \hspace{-7pt}\parbox[c]{1cm}{\centering $\sqrt{s}$ \\ {[GeV]} \\[-6pt] \hspace{10pt}}\hspace{-7pt}
        & $\Delta_i^{\rm (LO)} / \Lambda^2$    
        & $\Delta_{i,\rm weak}^{(\delta \rm NLO)} /\Lambda^2$ 
        & $\Delta_{i,\rm QED}^{\rm (NLO)} / \Lambda^2$    
        & $\Delta_{i}^{\rm (NLO)} / \Lambda^2$ 
        & $\bar\Delta_{i} / \Lambda^2$ \\
        \hline 
        & $240$ & $\textrm{-}1.89 \cdot 10^{\textrm{-}1}$ & $\textrm{-}8.94 \cdot 10^{\textrm{-}3}$ & $\hphantom{\textrm{-}} 3.07 \cdot 10^{\textrm{-}2}$ & $\textrm{-}1.67 \cdot 10^{\textrm{-}1}$ & $\hphantom{\textrm{-}} 8.79 \cdot 10^{\textrm{-}3}$ \\
$C_{\phi D}$ & $365$ & $\textrm{-}1.90 \cdot 10^{\textrm{-}1}$ & $\textrm{-}2.77 \cdot 10^{\textrm{-}3}$ & $\textrm{-}1.17 \cdot 10^{\textrm{-}2}$ & $\textrm{-}2.04 \cdot 10^{\textrm{-}1}$ & $\hphantom{\textrm{-}} 8.53 \cdot 10^{\textrm{-}3}$ \\
& $500$ & $\textrm{-}1.88 \cdot 10^{\textrm{-}1}$ & $\hphantom{\textrm{-}} 3.22 \cdot 10^{\textrm{-}5}$ & $\textrm{-}2.59 \cdot 10^{\textrm{-}2}$ & $\textrm{-}2.14 \cdot 10^{\textrm{-}1}$ & $\hphantom{\textrm{-}} 8.03 \cdot 10^{\textrm{-}3}$ \\
\hline 
& $240$ & $\hphantom{\textrm{-}} 1.08 \cdot 10^{\textrm{-}1}$ & $\hphantom{\textrm{-}} 1.38 \cdot 10^{\textrm{-}2}$ & $\textrm{-}1.71 \cdot 10^{\textrm{-}2}$ & $\hphantom{\textrm{-}} 1.05 \cdot 10^{\textrm{-}1}$ & $\textrm{-}4.16 \cdot 10^{\textrm{-}3}$ \\
$C_{\phi \square}$ & $365$ & $\hphantom{\textrm{-}} 1.09 \cdot 10^{\textrm{-}1}$ & $\hphantom{\textrm{-}} 6.81 \cdot 10^{\textrm{-}3}$ & $\hphantom{\textrm{-}} 7.30 \cdot 10^{\textrm{-}3}$ & $\hphantom{\textrm{-}} 1.23 \cdot 10^{\textrm{-}1}$ & $\textrm{-}4.46 \cdot 10^{\textrm{-}3}$ \\
\hline\hline
 
    \end{tabular}
\end{table}

\begin{table}[H]
    \centering
    \caption{  \label{tab:2fops} Effects on the total cross-section for $e^+e^- \rightarrow ZH$  from 2-fermion and scalar operators. The operators in the top portion of the table involve the top quark. The notation is defined in Eq.~(\ref{eq:sigpar}).}
    \vspace{7pt}
    \begin{tabular}{cccccccc}
        \hline\hline\\ [-2.5ex]
        & \hspace{-7pt}\parbox[c]{1cm}{\centering $\sqrt{s}$ \\ {[GeV]} \\[-6pt] \hspace{10pt}}\hspace{-7pt}
        &  $\Delta_{i,\rm weak}^{\rm (NLO),L} / \Lambda^2$    
        & $\bar\Delta_i^\textrm{L} /\Lambda^2$ 
        & $\Delta_{i,\rm weak}^{\rm (NLO),R} / \Lambda^2$    
        & $\bar\Delta_i^\textrm{R} / \Lambda^2$ 
        & $\Delta^{\rm (NLO)}_{i,\rm weak} / \Lambda^2$    
        & $\bar\Delta_i / \Lambda^2$ \\
        \hline
        & $240$ & $\textrm{-}1.38 \cdot 10^{\textrm{-}2}$ & $\textrm{-}1.69 \cdot 10^{\textrm{-}2}$ & $\hphantom{\textrm{-}} 1.36 \cdot 10^{\textrm{-}2}$ & $\hphantom{\textrm{-}} 1.14 \cdot 10^{\textrm{-}2}$ & $\textrm{-}1.34 \cdot 10^{\textrm{-}3}$ & $\textrm{-}4.09 \cdot 10^{\textrm{-}3}$ \\
$C_{\phi q}^{(1)}[3,3]$ & $365$ & $\textrm{-}2.13 \cdot 10^{\textrm{-}2}$ & $\textrm{-}1.64 \cdot 10^{\textrm{-}2}$ & $\hphantom{\textrm{-}} 1.61 \cdot 10^{\textrm{-}2}$ & $\hphantom{\textrm{-}} 1.03 \cdot 10^{\textrm{-}2}$ & $\textrm{-}4.38 \cdot 10^{\textrm{-}3}$ & $\textrm{-}4.30 \cdot 10^{\textrm{-}3}$ \\
& $500$ & $\textrm{-}2.87 \cdot 10^{\textrm{-}2}$ & $\textrm{-}1.57 \cdot 10^{\textrm{-}2}$ & $\hphantom{\textrm{-}} 2.16 \cdot 10^{\textrm{-}2}$ & $\hphantom{\textrm{-}} 8.46 \cdot 10^{\textrm{-}3}$ & $\textrm{-}5.57 \cdot 10^{\textrm{-}3}$ & $\textrm{-}4.63 \cdot 10^{\textrm{-}3}$ \\
\hline 
& $240$ & $\textrm{-}9.38 \cdot 10^{\textrm{-}3}$ & $\hphantom{\textrm{-}} 2.73 \cdot 10^{\textrm{-}3}$ & $\hphantom{\textrm{-}} 1.24 \cdot 10^{\textrm{-}2}$ & $\hphantom{\textrm{-}} 5.79 \cdot 10^{\textrm{-}3}$ & $\hphantom{\textrm{-}} 5.08 \cdot 10^{\textrm{-}4}$ & $\hphantom{\textrm{-}} 4.12 \cdot 10^{\textrm{-}3}$ \\
\hline\hline
 
    \end{tabular}
\end{table}

\begin{table}[H]
    \centering
    \caption{    \label{tab:4fops} Effects on the total cross-section for $e^+e^- \rightarrow ZH$  induced by 4-fermion operators.  The operators in the upper portion of the table involve a top. The contribution to the right polarization proportional to $C_{lq}^{(3)}[1,1,3,3]$ originates from the redefinition of the vev in $\Delta r$, Eq.~\eqref{eq:gmu:smeft}. The notation is defined in Eq. \eqref{eq:sigpar}.}
    \vspace{7pt}
    \begin{tabular}{cccccccc} 
        \hline\hline \\ [-2.5ex]
        & \hspace{-7pt}\parbox[c]{1cm}{\centering $\sqrt{s}$ \\ {[GeV]} \\[-6pt] \hspace{10pt}}\hspace{-7pt}
        & $\Delta_{i,\textrm{weak}}^{(\textrm{NLO}),\textrm{L}} / \Lambda^2$    
        & $\bar\Delta_i^\textrm{L} /\Lambda^2$ 
        & $\Delta_{i,\textrm{weak}}^{(\textrm{NLO}),\textrm{R}} / \Lambda^2$    
        & $\bar\Delta_i^\textrm{R} / \Lambda^2$ 
        & $\Delta^{(\textrm{NLO})}_{i,\textrm{weak}} / \Lambda^2$    
        & $\bar\Delta_i / \Lambda^2$ \\
        \hline
        & $240$ & 0 & 0 & $\hphantom{\textrm{-}} 2.82 \cdot 10^{\textrm{-}5}$ & $\textrm{-}3.07 \cdot 10^{\textrm{-}2}$ & $\hphantom{\textrm{-}} 1.28 \cdot 10^{\textrm{-}5}$ & $\textrm{-}1.39 \cdot 10^{\textrm{-}2}$ \\ 
$C_{eu}[1,1,3,3]$ & $365$ & 0 & 0 & $\textrm{-}8.23 \cdot 10^{\textrm{-}2}$ & $\textrm{-}7.11 \cdot 10^{\textrm{-}2}$ & $\textrm{-}3.72 \cdot 10^{\textrm{-}2}$ & $\textrm{-}3.22 \cdot 10^{\textrm{-}2}$ \\ 
& $500$ & 0 & 0 & $\textrm{-}2.77 \cdot 10^{\textrm{-}1}$ & $\textrm{-}1.32 \cdot 10^{\textrm{-}1}$ & $\textrm{-}1.27 \cdot 10^{\textrm{-}1}$ & $\textrm{-}6.08 \cdot 10^{\textrm{-}2}$ \\ 
\hline 
& $240$ & $\textrm{-}2.90 \cdot 10^{\textrm{-}5}$ & $\hphantom{\textrm{-}} 3.16 \cdot 10^{\textrm{-}2}$ & 0 & 0 & $\textrm{-}1.59 \cdot 10^{\textrm{-}5}$ & $\hphantom{\textrm{-}} 1.73 \cdot 10^{\textrm{-}2}$ \\ 
$C_{lu}[1,1,3,3]$ & $365$ & $\hphantom{\textrm{-}} 8.45 \cdot 10^{\textrm{-}2}$ & $\hphantom{\textrm{-}} 7.31 \cdot 10^{\textrm{-}2}$ & 0 & 0 & $\hphantom{\textrm{-}} 4.63 \cdot 10^{\textrm{-}2}$ & $\hphantom{\textrm{-}} 4.00 \cdot 10^{\textrm{-}2}$ \\ 
& $500$ & $\hphantom{\textrm{-}} 2.92 \cdot 10^{\textrm{-}1}$ & $\hphantom{\textrm{-}} 1.40 \cdot 10^{\textrm{-}1}$ & 0 & 0 & $\hphantom{\textrm{-}} 1.58 \cdot 10^{\textrm{-}1}$ & $\hphantom{\textrm{-}} 7.56 \cdot 10^{\textrm{-}2}$ \\
\hline\hline 
    \end{tabular}
\end{table}

\begin{table}[H]
    \centering
    \caption{Effects on the total cross-section for $e^+e^- \rightarrow ZH$  induced by CP violating operators.}
    \vspace{7pt}
    \begin{tabular}{ccccc} 
        \hline\hline \\ [-2.5ex]
        & $\sqrt{s}$ [GeV]
        & $\Delta_{i,\textrm{weak}}^{(\textrm{NLO}),\textrm{L}} / \Lambda^2$    
        & $\Delta_{i,\textrm{weak}}^{(\textrm{NLO}),\textrm{R}} / \Lambda^2$    
        & $\Delta_{i,\textrm{weak}}^{(\textrm{NLO})} / \Lambda^2$ \\
        \hline 
        & $240$ & $\hphantom{\textrm{-}} 2.78 \cdot 10^{\textrm{-}3}$ & $\hphantom{\textrm{-}} 5.51 \cdot 10^{\textrm{-}3}$ & $\hphantom{\textrm{-}} 9.77 \cdot 10^{\textrm{-}4}$ \\ 
$C_{\phi \tilde{B}}$ & $365$ & $\hphantom{\textrm{-}} 8.03 \cdot 10^{\textrm{-}3}$ & $\hphantom{\textrm{-}} 1.40 \cdot 10^{\textrm{-}2}$ & $\hphantom{\textrm{-}} 1.94 \cdot 10^{\textrm{-}3}$ \\ 
& $500$ & $\hphantom{\textrm{-}} 1.03 \cdot 10^{\textrm{-}2}$ & $\hphantom{\textrm{-}} 2.59 \cdot 10^{\textrm{-}2}$ & $\hphantom{\textrm{-}} 6.31 \cdot 10^{\textrm{-}3}$ \\ 
\hline 
& $240$ & $\hphantom{\textrm{-}} 2.26 \cdot 10^{\textrm{-}2}$ & $\hphantom{\textrm{-}} 8.97 \cdot 10^{\textrm{-}4}$ & $\hphantom{\textrm{-}} 1.28 \cdot 10^{\textrm{-}2}$ \\ 
$C_{\phi \tilde{W}}$ & $365$ & $\hphantom{\textrm{-}} 7.88 \cdot 10^{\textrm{-}2}$ & $\hphantom{\textrm{-}} 8.06 \cdot 10^{\textrm{-}4}$ & $\hphantom{\textrm{-}} 4.35 \cdot 10^{\textrm{-}2}$ \\ 
& $500$ & $\hphantom{\textrm{-}} 1.12 \cdot 10^{\textrm{-}1}$ & $\hphantom{\textrm{-}} 3.77 \cdot 10^{\textrm{-}4}$ & $\hphantom{\textrm{-}} 6.09 \cdot 10^{\textrm{-}2}$ \\
\hline\hline 
    \end{tabular}
    \label{tab:cp}
\end{table}

\begin{figure}[H]
    \centering	
    \includegraphics[width=.32\textwidth]{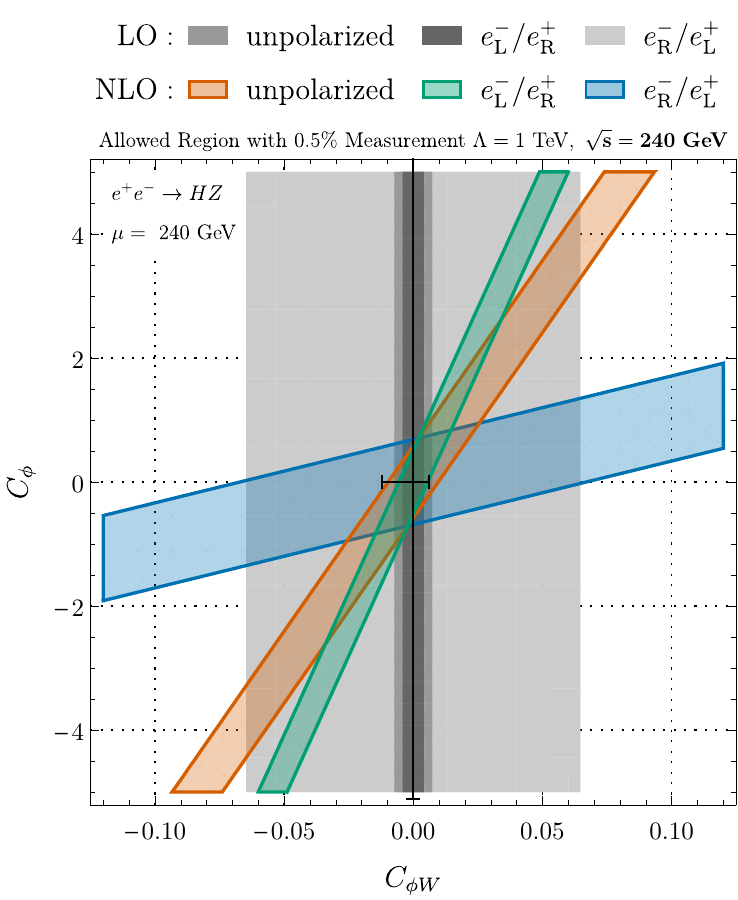}\hspace{5pt}
    \includegraphics[width=.32\textwidth]{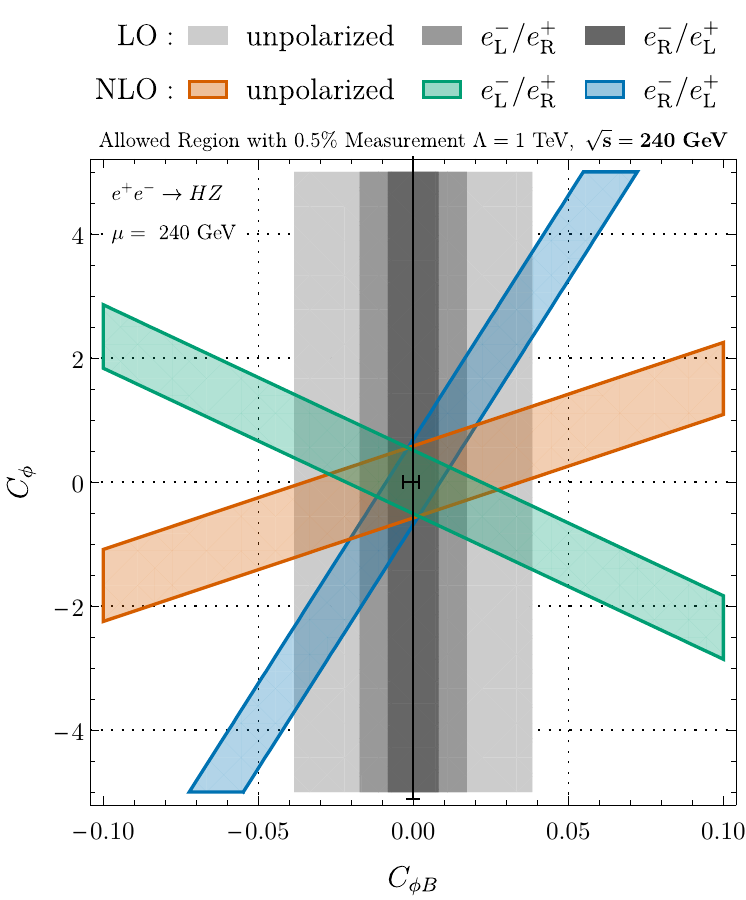}\hspace{5pt}
    \includegraphics[width=.32\textwidth]{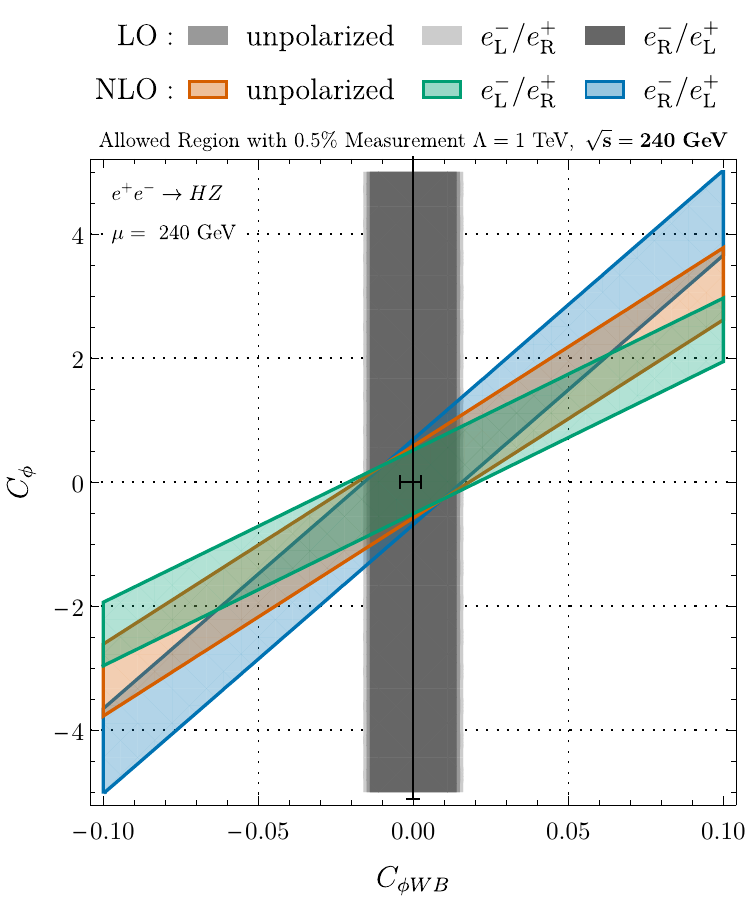}
    \caption{Contributions from modifications of the Higgs tri-linear coupling $C_\phi$ to the cross-section for $e^+e^-\rightarrow ZH$ correlated with those from operators occurring at tree level.  The sensitivity to a $0.5\%$ measurement at $\sqrt{s}=240~\textrm{GeV}$ is shown. Note that there is no sensitivity to $C_\phi$ at tree level. Data points show single parameter limits and are taken from Ref.~\cite{Ellis:2020unq}. We note that the positive limit on $C_\phi$ is much larger then the range of the plot.}
    \label{fg:tree1}
\end{figure}

\begin{figure}[H]
    \centering	 \includegraphics[width=.32\textwidth]{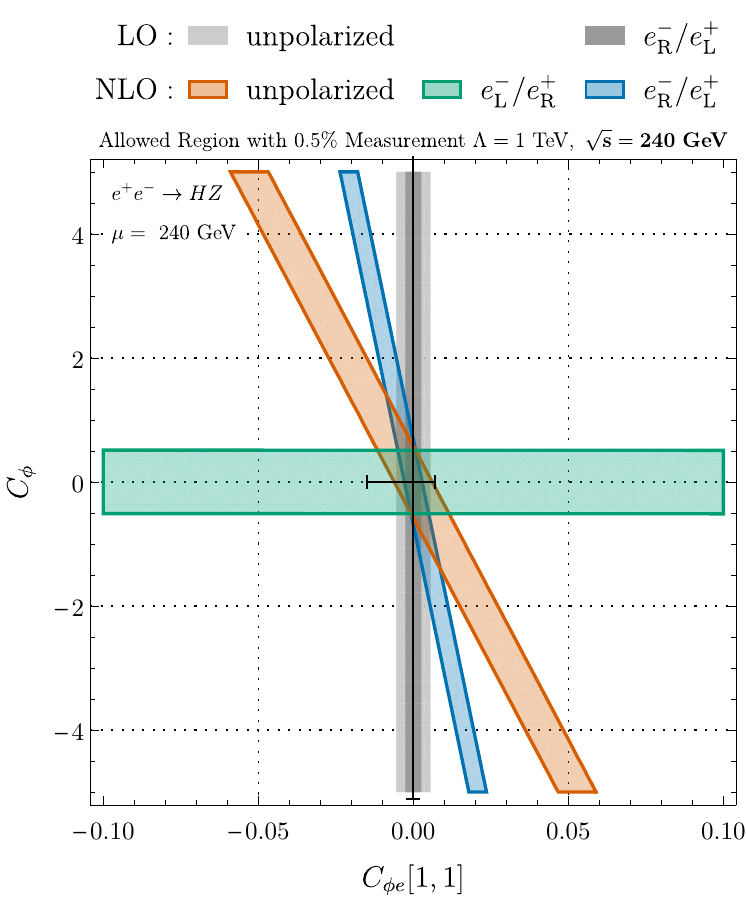}\hspace{5pt}
    \includegraphics[width=.32\textwidth]{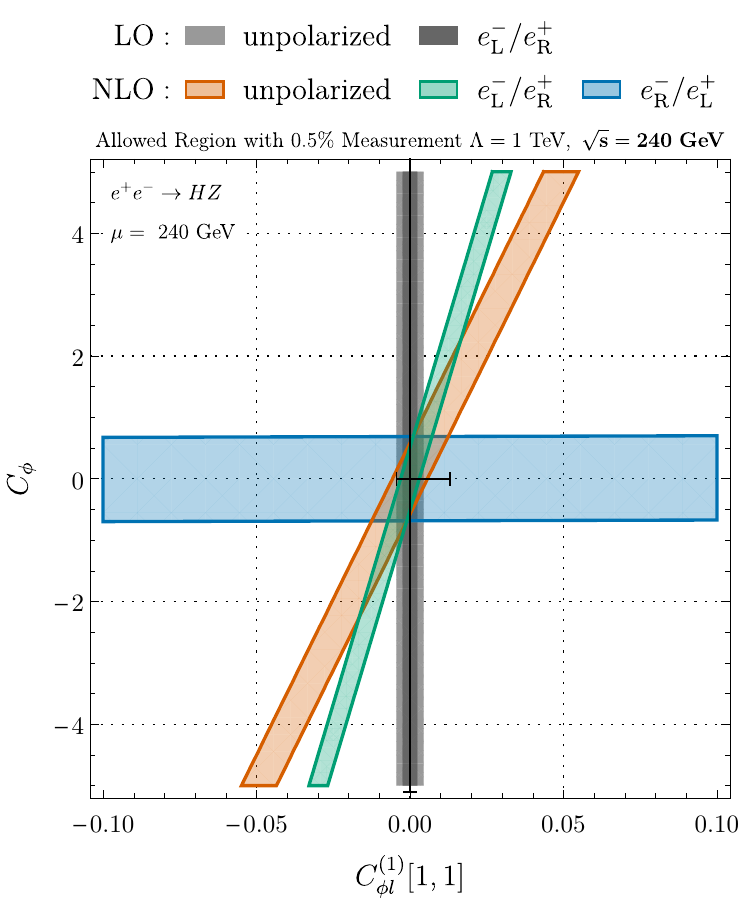}\hspace{5pt}
    \includegraphics[width=.32\textwidth]{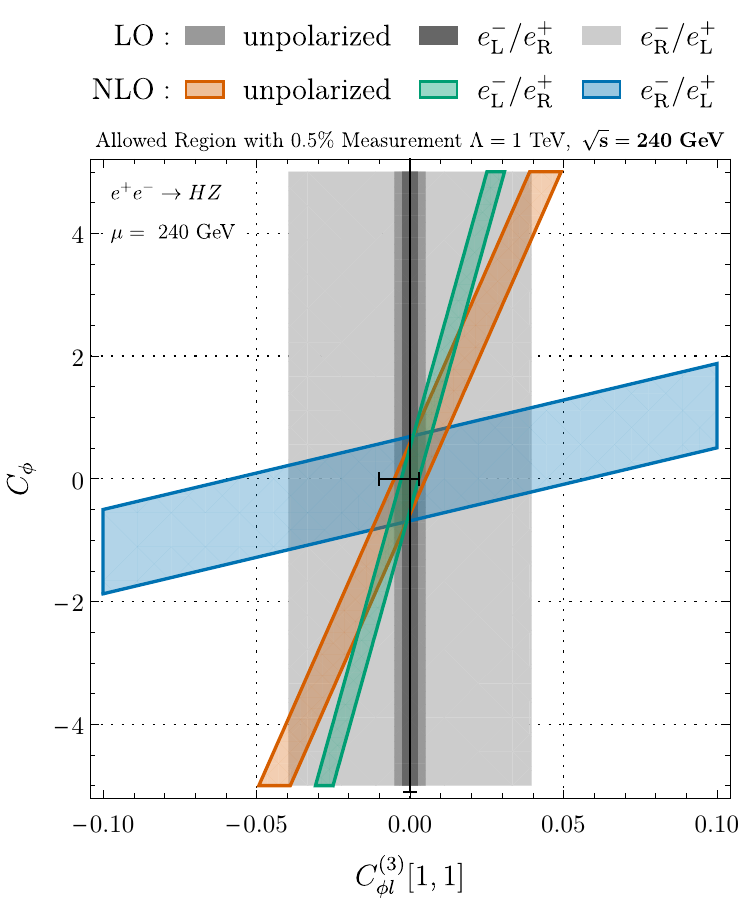}\\[8pt]
    \includegraphics[width=.32\textwidth]{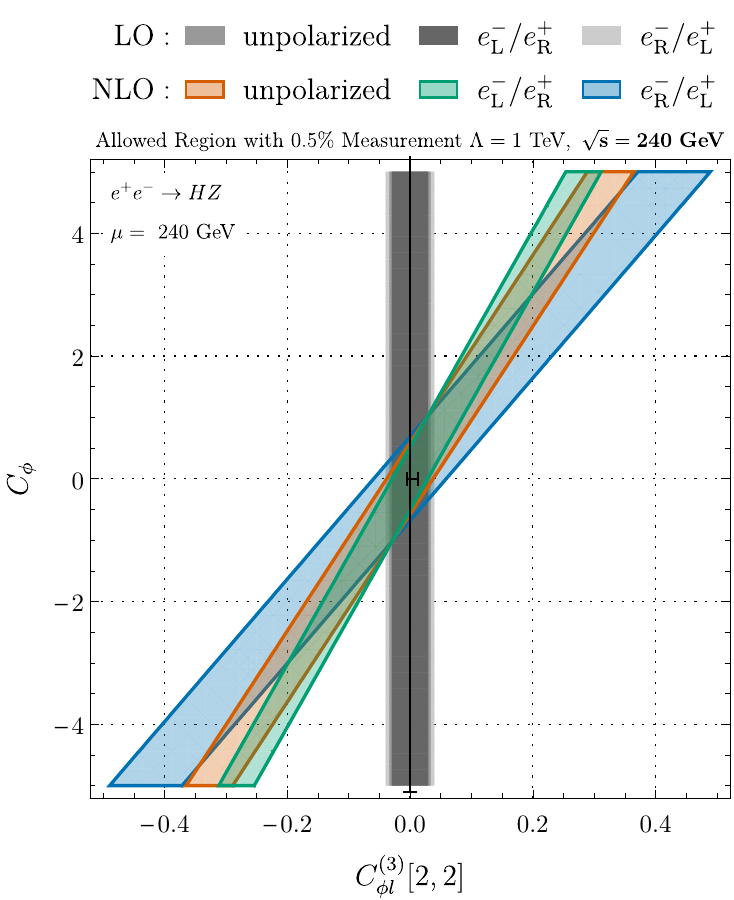}\hspace{5pt}
    \includegraphics[width=.32\textwidth]{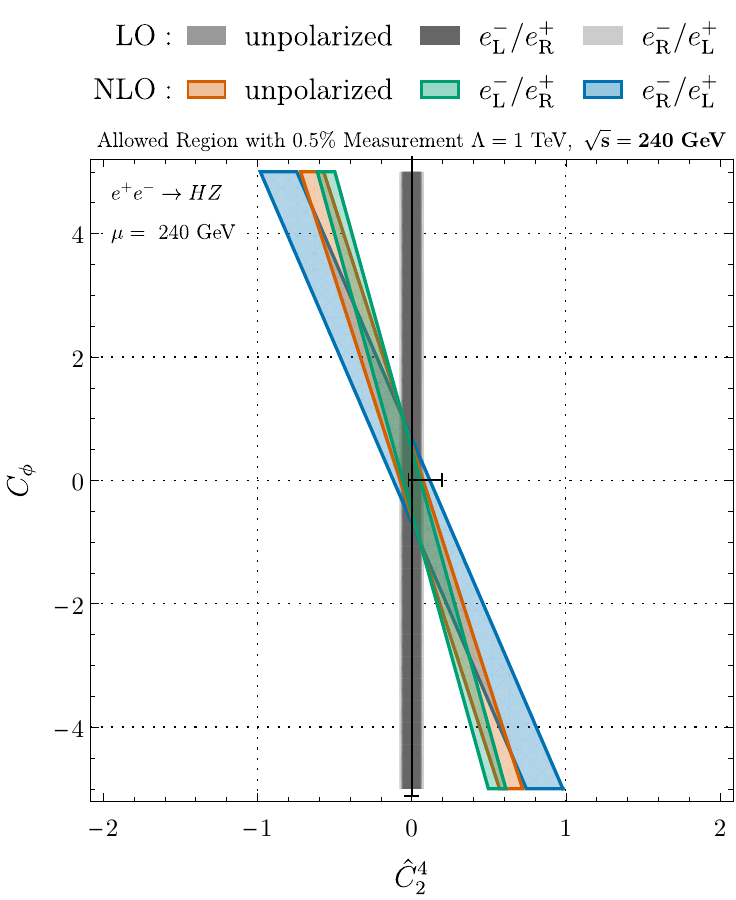}
    \caption{Contributions from modifications of the Higgs tri-linear coupling $C_\phi$ to the cross-section for $e^+e^-\rightarrow ZH$ correlated with those from operators occurring at tree level.  The sensitivity to a $0.5\%$ measurement at $\sqrt{s}=240~\textrm{GeV}$ is shown. Note that there is no sensitivity to $C_\phi$ at tree level. Data points show single parameter limits and are taken from Ref.~\cite{Ellis:2020unq}. We note that the positive limit on $C_\phi$ is much larger then the range of the plot.}
    \label{fg:figmore}
\end{figure}

\begin{figure}[H]
    \centering	
    \includegraphics[width=.32\textwidth]{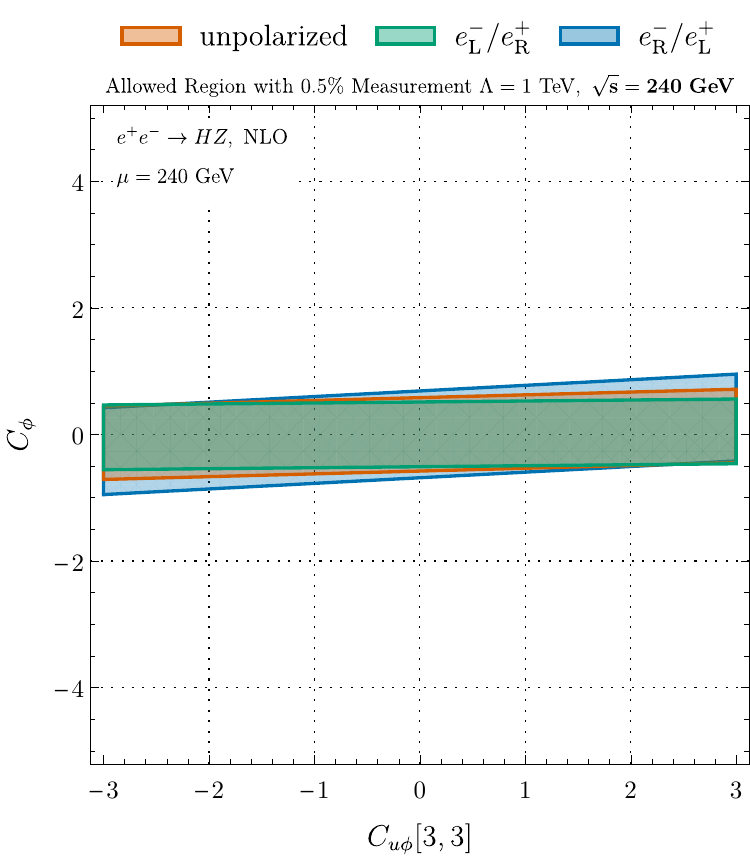}\hspace{5pt}
    \includegraphics[width=.32\textwidth]{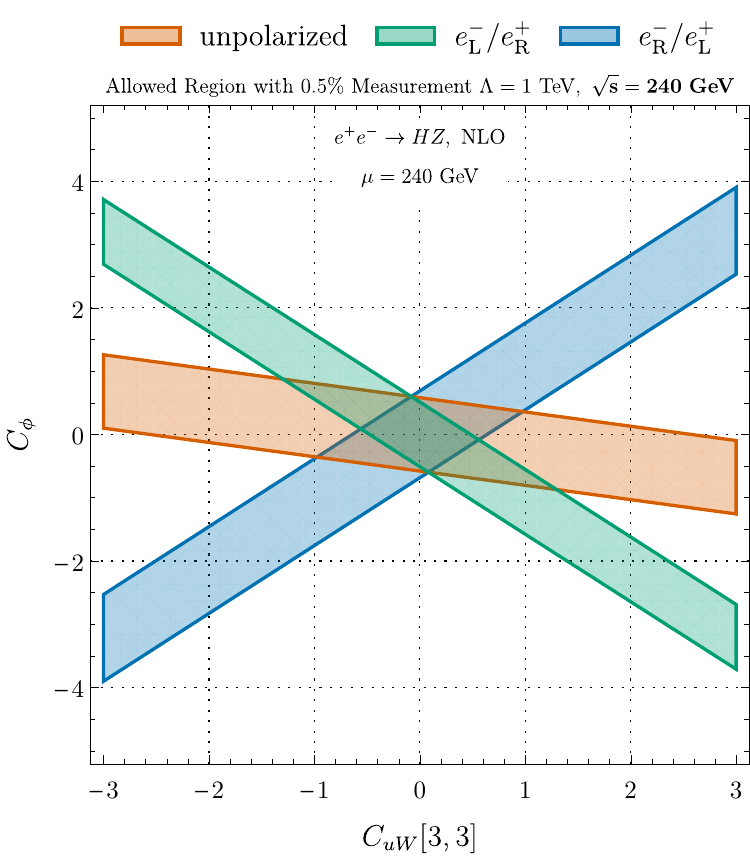}\hspace{5pt}
    \includegraphics[width=.32\textwidth]{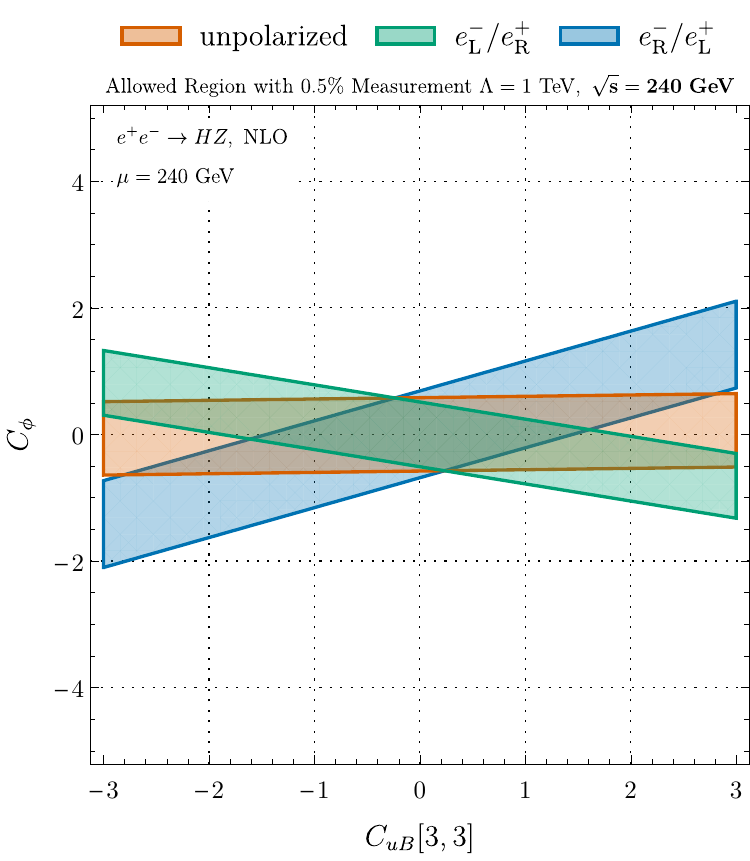}\\[8pt]
    \includegraphics[width=.32\textwidth]{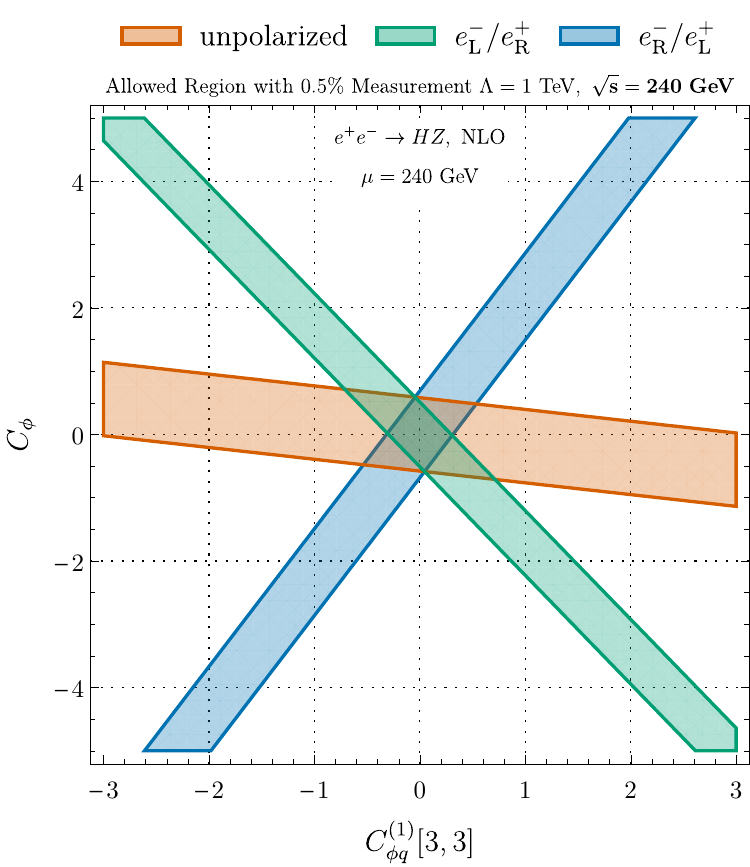}\hspace{5pt}
    \includegraphics[width=.32\textwidth]{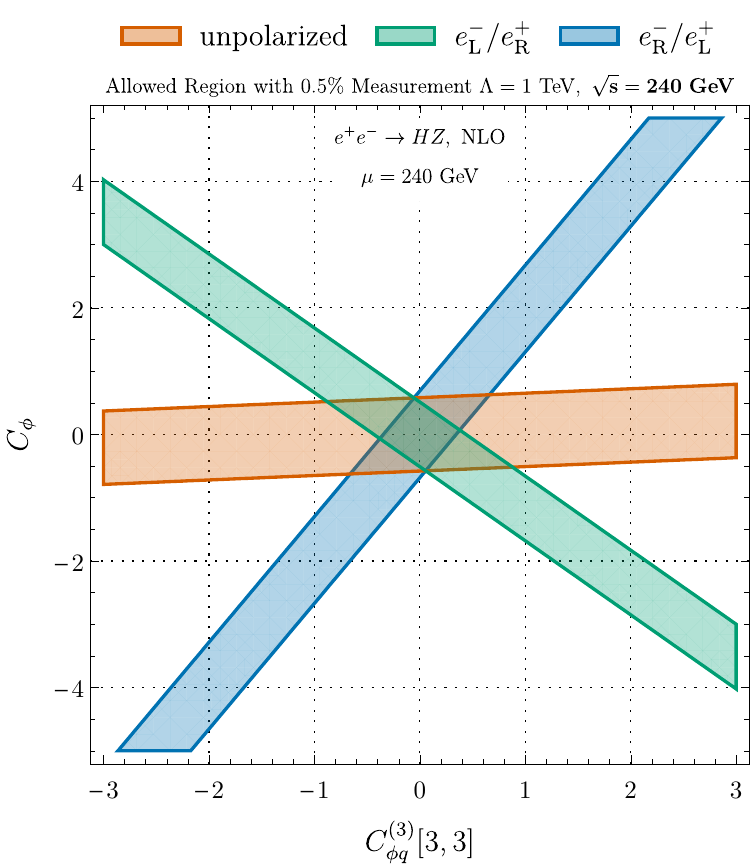}
    \caption{Contributions from modifications of the Higgs tri-linear coupling $C_\phi$ on the cross-section for $e^+e^-\rightarrow ZH$ correlated with those from 2-top quark operators. The sensitivity to a $0.5\%$ measurement at $\sqrt{s}=240~\textrm{GeV}$ is shown. Note that there is no sensitivity to $C_\phi$ or to the 2-top operators at tree level. }
    \label{fg:chtop2}
\end{figure}

\begin{figure}[H]
    \centering	
    \includegraphics[width=.32\textwidth]{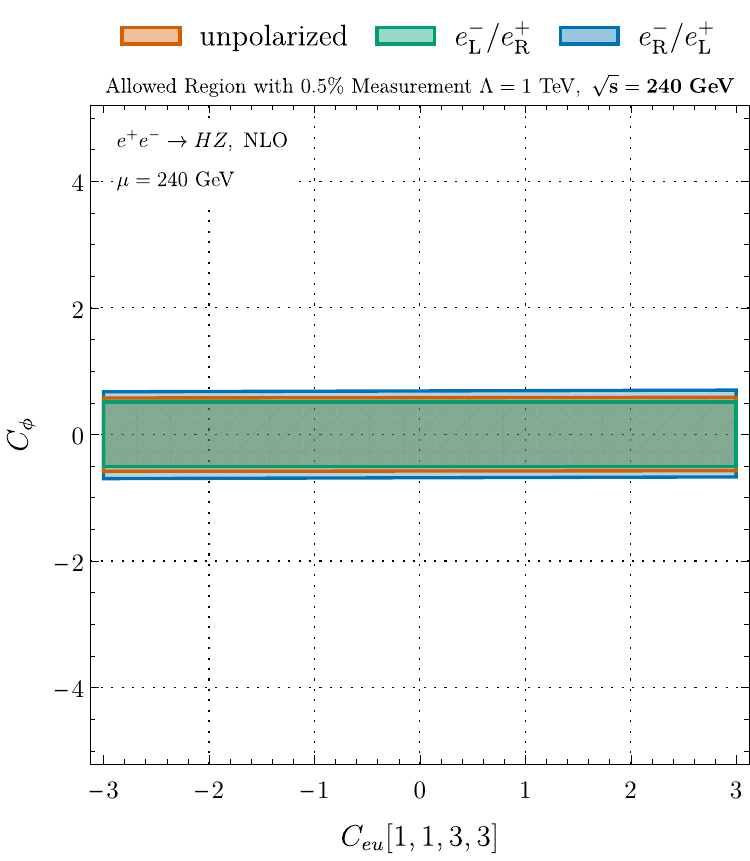}\hspace{5pt}
    \includegraphics[width=.32\textwidth]{figures/cphi_vs_4f/Clq11133_Cphi.pdf}\hspace{5pt}
    \includegraphics[width=.32\textwidth]{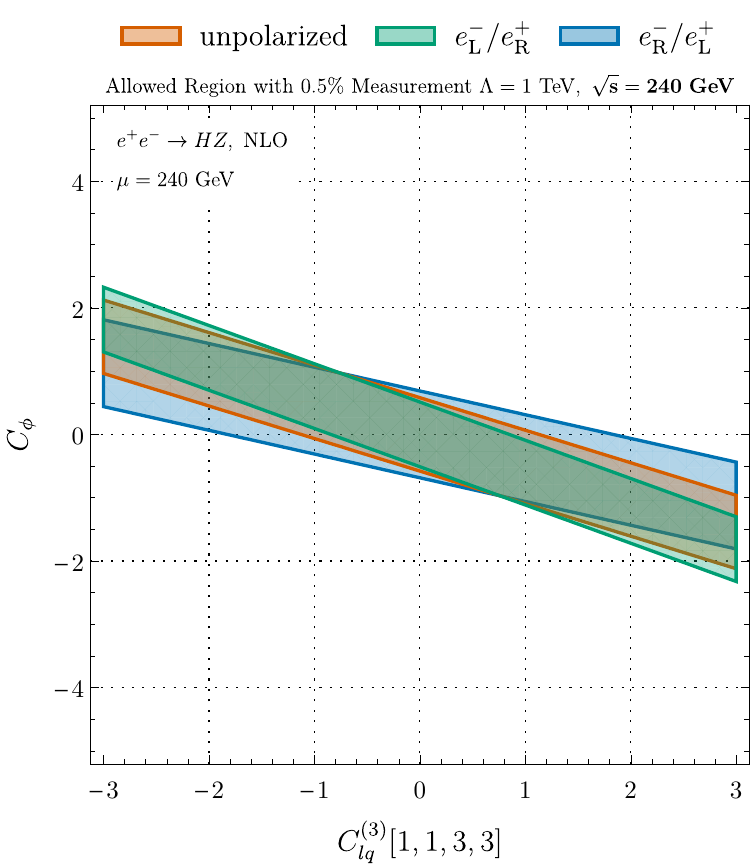}\\[8pt]
    \includegraphics[width=.32\textwidth]{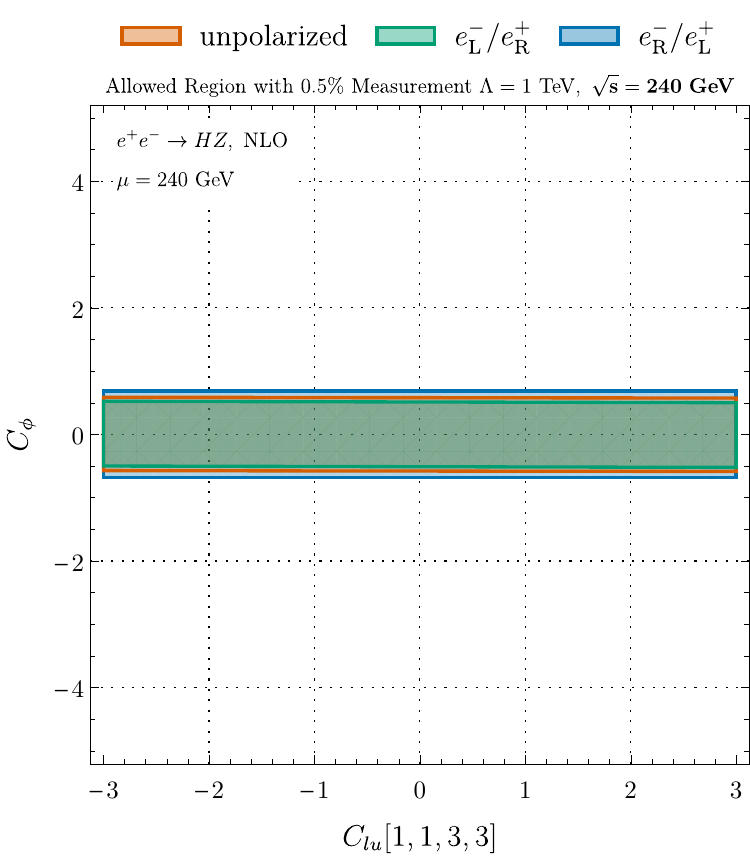}\hspace{5pt}
    \includegraphics[width=.32\textwidth]{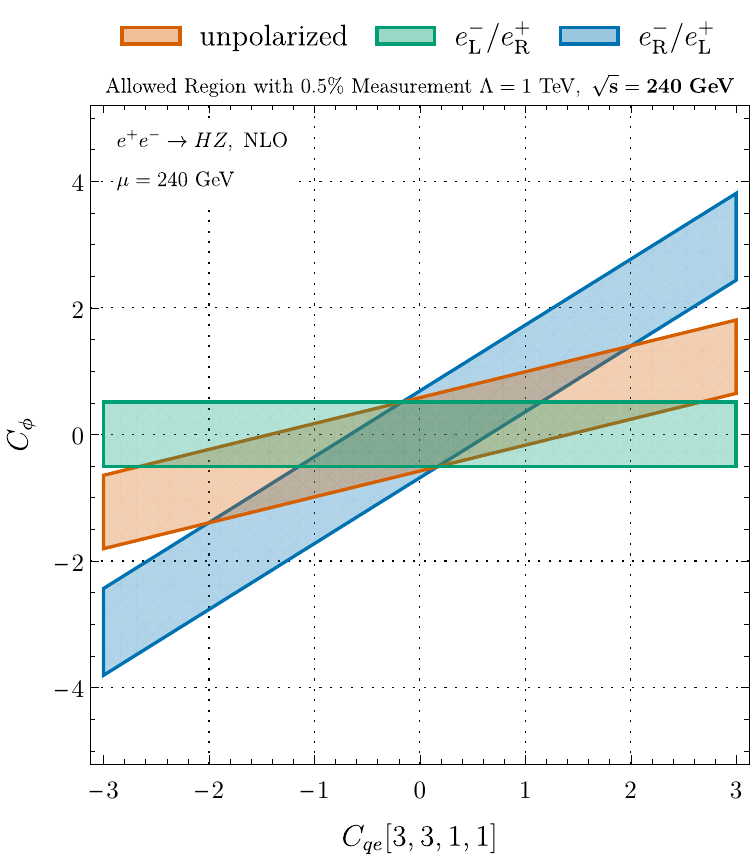}
    \hspace{20pt}
    \caption{ Sensitivity to a $0.5\%$ measurement of $e^+e^-\rightarrow ZH$  at $\sqrt{s}=240~\textrm{GeV}$ to the 4-fermion $eett$ operators that first appear at NLO.}
    \label{fg:4fch}
\end{figure}

\begin{figure}[H]
    \centering	
    \includegraphics[width=.32\textwidth]{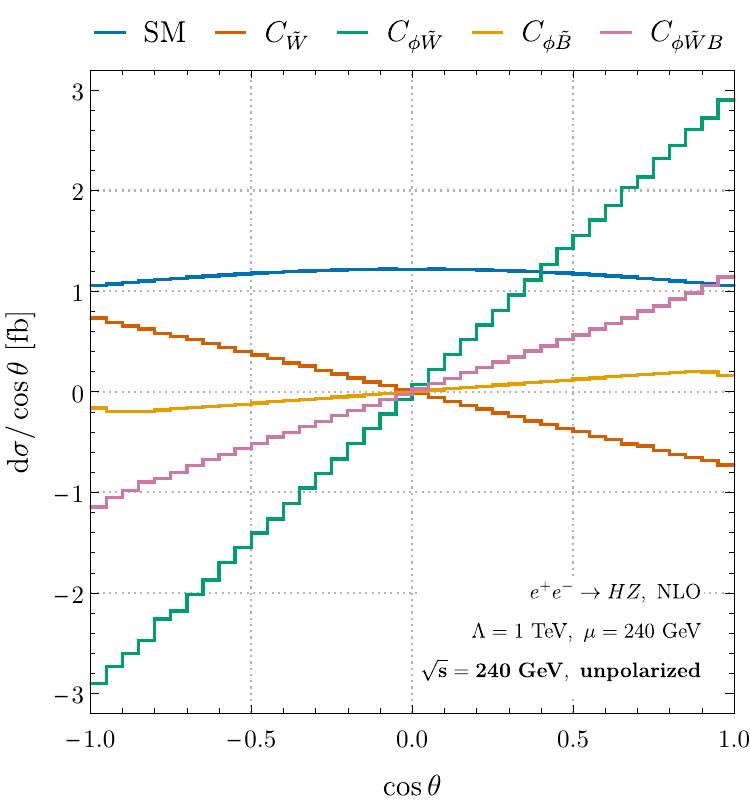}\hspace{5pt}
    \includegraphics[width=.32\textwidth]{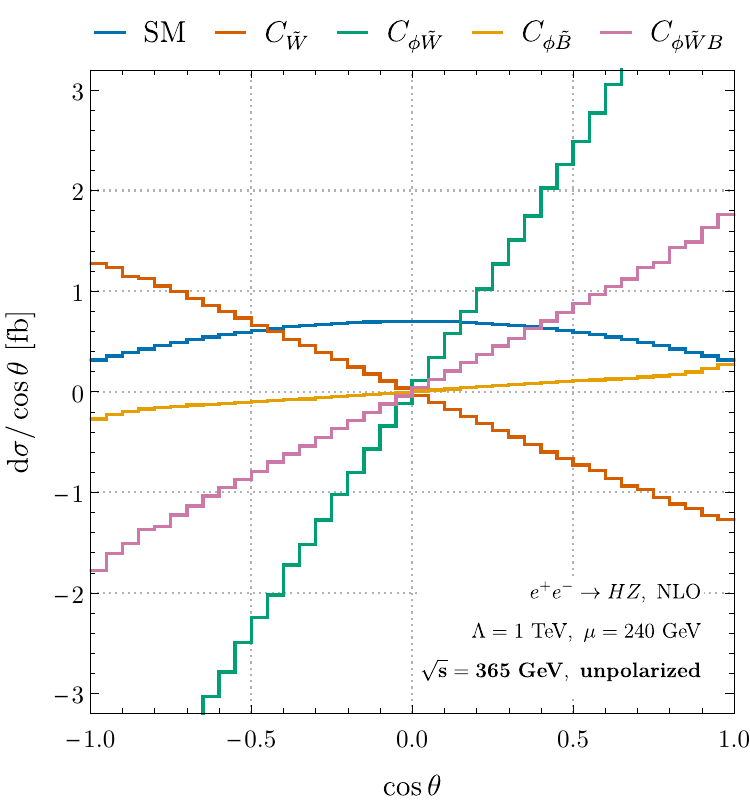}\hspace{5pt}
    \includegraphics[width=.32\textwidth]{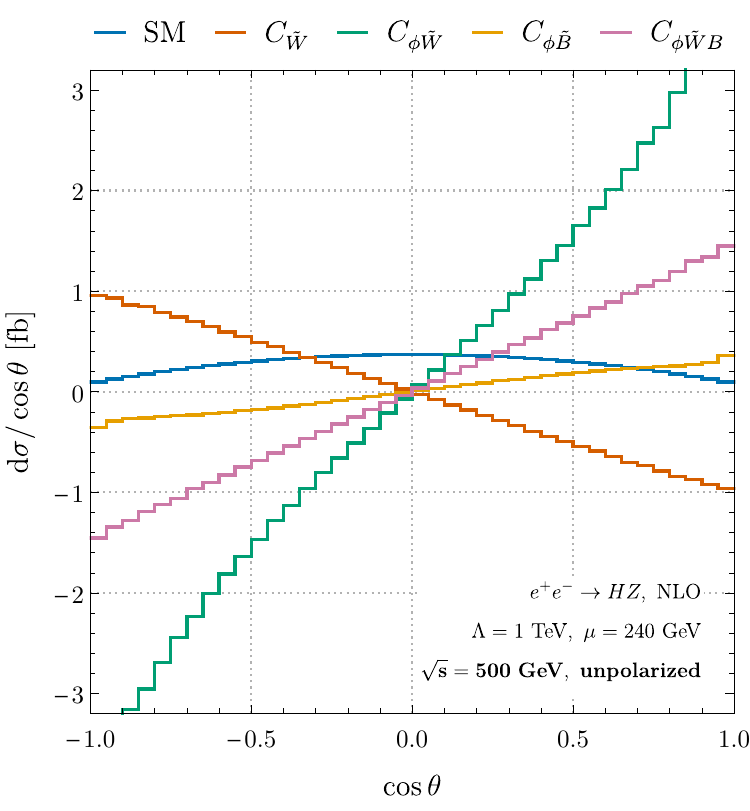}
    \caption{CP violating effects with unpolarized beams for purely bosonic operators.}
	\label{histo:th_h_cp_violation_summary}
\end{figure}

\bibliographystyle{utphys}

%%%%%%%%%%%%%%%%%%%%%%%%%%%%%%%%%%%%%%%%%%%%%%%%%%

\end{document}